\documentclass[twocolumn]{aastex631}

\usepackage{graphicx}
\usepackage{graphics}
\usepackage{amsmath}
\usepackage{amssymb}
\usepackage{url}
\usepackage{hyperref}
\usepackage{xcolor}
\usepackage{physics}
\usepackage{booktabs}

\defcitealias{peeples19}{Paper~I}
\defcitealias{corlies20}{Paper~II}
\defcitealias{zheng20}{Paper~III}
\defcitealias{simons20}{Paper~IV}
\defcitealias{lochhaas21}{Paper~V}
\defcitealias{lochhaas23}{Paper~VI}
\defcitealias{wright24}{Paper~VII}

\newcommand{\hii}{\ion{H}{2}}
\newcommand{\red}[1]{{\color{black}#1}} 

\newcommand{\s}{s}

\newcommand{\Msun}{$\mathrm{M}_\odot$}

\newcommand{\z}{$z$}

\newcommand{\jwst}{\textit{JWST}}
\newcommand{\hst}{\textit{HST}}

\newcommand{\U}{$\left \langle U \right \rangle$}

\usepackage[normalem]{ulem}

\newcommand{\aref}[1]{\hyperref[#1]{Appendix~\ref{#1}}}

\defcitealias{Rigby:2011aa}{R11}
\defcitealias{Rigby:2018aa}{R18a}
\defcitealias{Rigby:2018ab}{R18b}
\defcitealias{Kewley:2002fk}{KD02}
\defcitealias{Whitaker:2014aa}{W14}
\defcitealias{Blanc:2015aa}{B15}
\defcitealias{Pettini:2004qe}{PP04}
\defcitealias{Shi:2007aa}{S07}
\defcitealias{Dopita:2016aa}{D16}
\defcitealias{Kobulnicky:2004aa}{KK04}
\defcitealias{Izotov:2006ab}{I06}
\defcitealias{Garnett:1995aa}{G95a}
\defcitealias{Garnett:1995ab}{G95b}
\defcitealias{Levesque:2014aa}{LR14}
\defcitealias{Kewley:2008aa}{KE08}
\defcitealias{Bian:2018aa}{BKD18}
\defcitealias{Sanders:2016aa}{S16}
\defcitealias{Strom:2018aa}{S18}
\defcitealias{Kewley:2019aa}{K19a}
\defcitealias{Kewley:2019ab}{K19b}
\defcitealias{Goldbaum:2016aa}{G16}
\defcitealias{Carton:2017aa}{C17}
\defcitealias{Acharyya:2019aa}{A19}
\defcitealias{Acharyya:2020aa}{A20}
\defcitealias{Acharyya:2020ab}{A20b}
\defcitealias{Sanchez:2018ab}{S18}
\defcitealias{Wylezalek:2018aa}{W18}
\defcitealias{Wang:2022aa}{W22}

\shorttitle{FOGGIE VIII: Gas phase metallicity gradient evolution}
\shortauthors{Acharyya et al.}

\begin{document}
 
\title{Figuring Out Gas \& Galaxies In Enzo (FOGGIE) VIII: \\ Complex and Stochastic Metallicity Gradients at $z > 2$}

\author[0000-0003-4804-7142]{Ayan Acharyya}
\affiliation{Department of Physics \& Astronomy, Johns Hopkins University, 3400 N.\ Charles Street, Baltimore, MD 21218}
\affiliation{INAF - Astronomical Observatory of Padova, vicolo dell’Osservatorio 5, IT-35122 Padova, Italy}
\correspondingauthor{Ayan Acharyya}
\email{ayan.acharyya@inaf.it}

\author[0000-0003-1455-8788]{Molly S.\ Peeples}
\affiliation{Space Telescope Science Institute, 3700 San Martin Dr., Baltimore, MD 21218}
\affiliation{Department of Physics \& Astronomy, Johns Hopkins University, 3400 N.\ Charles Street, Baltimore, MD 21218}

\author[0000-0002-7982-412X]{Jason Tumlinson}
\affiliation{Space Telescope Science Institute, 3700 San Martin Dr., Baltimore, MD 21218}
\affiliation{Department of Physics \& Astronomy, Johns Hopkins University, 3400 N.\ Charles Street, Baltimore, MD 21218}

\author[0000-0002-2786-0348]{Brian W.\ O'Shea}
\affiliation{Department of Computational Mathematics, Science, and Engineering, 
Department of Physics and Astronomy, 
National Superconducting Cyclotron Laboratory,  
Michigan State University}

\author[0000-0003-1785-8022]{Cassandra Lochhaas}
\affiliation{Space Telescope Science Institute, 3700 San Martin Dr., Baltimore, MD 21218}
\affiliation{Center for Astrophysics, Harvard \& Smithsonian, 60 Garden St., Cambridge, MA 02138}
\affiliation{NASA Hubble Fellow}

\author[0000-0002-1685-5818]{Anna C.\ Wright}
\affiliation{Department of Physics \& Astronomy, Johns Hopkins University, 3400 N.\ Charles Street, Baltimore, MD 21218}

\author[0000-0002-6386-7299]{Raymond C.\ Simons}
\affiliation{Space Telescope Science Institute, 3700 San Martin Dr., Baltimore, MD 21218}
\affiliation{Department of Physics, University of Connecticut, 196A Auditorium Road Unit 3046, Storrs, CT 06269 USA}


\author[0000-0001-7472-3824]{Ramona Augustin}
\affiliation{Space Telescope Science Institute, 3700 San Martin Dr., Baltimore, MD 21218}
\affiliation{Leibniz Institute for Astrophysics Potsdam (AIP), An der Sternwarte 16, Potsdam 14482, Germany}

\author[0000-0002-6804-630X]{Britton D. Smith}
\affiliation{Institute for Astronomy, University of Edinburgh, Royal Observatory, EH9 3HJ, UK}

\author[0009-0001-8366-7606]{Eugene Hyeonmin Lee}
\affiliation{Space Telescope Science Institute, 3700 San Martin Dr., Baltimore, MD 21218}
\affiliation{University of Texas at Austin, Physics, Math, and Astronomy Building, 2515 Speedway, Austin, TX 78712}

\begin{abstract}

Gas-phase metallicity gradients are a crucial element in understanding the chemical evolution of galaxies. We use the FOGGIE simulations to study the metallicity gradients ($\nabla Z$) of six Milky Way-like galaxies throughout their evolution. FOGGIE galaxies generally exhibit steep negative gradients for most of their history, with only a few short-lived instances reaching positive slopes that appear to arise mainly from interactions with other galaxies. FOGGIE concurs with other simulation results but disagrees with the robust observational finding that flat and positive gradients are common at $z>1$. By tracking the metallicity gradient at a rapid cadence of simulation outputs ($\sim 5$--10 Myr), we find that theoretical gradients are highly stochastic: the FOGGIE galaxies spend $\sim 30-50$\% of their time far away from a smoothed trajectory inferred from analytic models or other, less high-cadence simulations. This rapid variation makes instantaneous gradients from observations more difficult to interpret in terms of physical processes. Because of these geometric and stochastic complications, we explore non-parametric methods of quantifying the evolving metallicity distribution at $z > 1$. We investigate how efficiently non-parametric measures of the 2-D metallicity distribution respond to metal production and mixing. Our results suggest that new methods of quantifying and interpreting gas-phase metallicity will be needed to relate trends in upcoming high-$z$ {\it JWST} observations with the underlying physics of gas accretion, expulsion, and recycling in early galaxies. \end{abstract}

\section{Introduction}
\label{sec:intro}

The spatial distribution of metals in the gas phase of the interstellar medium is a crucial diagnostic for the chemical evolution of galaxies. Conventionally, the distribution of metals is quantified in terms of the slope of the azimuthally-averaged metallicity profile, the ``metallicity gradient'', defined as
\begin{equation}\label{eqn:nablaz}
    \nabla Z = \dv{\log{Z/Z_{\odot}}}{r}.
\end{equation}
A negative $\nabla Z$ implies that the metallicity decreases radially, whereas a positive $\nabla Z$ denotes a metal-poor center of the disk relative to the outskirts. Metallicity gradient studies have helped us make considerable progress in understanding galaxy evolution at low-redshift. Thanks to spatially resolved spectroscopy and advanced models, the last $\sim$two decades has witnessed a plethora of such studies.

Steep negative metallicity gradients have improved our understanding of the inside-out star-formation scenario whereby stars are formed first in the centers of galaxies and subsequently at outer radii \citep{Marino:2016aa, Belfiore:2017aa}, allowing more time for metal enrichment in galaxy centers than galaxy outskirts. \citet{Mingozzi:2020aa}  attribute the steepening of the metallicity gradient with increasing stellar mass to increasing star-formation activity, and the flattening of gradient at the highest stellar masses to the increasing prevalence of mergers \citep[see also][]{Belfiore:2017aa}. Thus, negative gradients can teach us about the star-formation history and mass assembly of galaxies. 

Flat gradients, on the other hand, imply efficient, radially outward mixing of metals via gas flows \citep[e.g.,][]{Sanchez-Menguiano:2018aa, Simons:2021aa}. \citet{Sanchez-Menguiano:2018aa} suggest that, for galaxies (or regions of galaxies) with flat metallicity gradients, this could be an indicator that the metal budget in these regions is dominated by gas flows, rather than in situ production. Moreover, interactions and merger events can lead to efficient gas mixing, and consequently, flat or positive gradients \citep[e.g.,][]{Krabbe:2008aa, Kewley:2010aa, Rich:2012aa, Miralles:2014aa, Vogt:2015aa, Munoz:2018aa}. Thus, flat or positive gradients help us learn about gas flows and interactions in the local Universe.

Typically, a mix of negative \citep[e.g.][]{Vila-Costas:1992aa, Garnett:1997aa, Bresolin:2002aa, Kennicutt:2003aa, Bresolin:2007aa, Moustakas:2010aa, Cecil:2014aa, Sanchez:2014aa} and flat or positive metallicity gradients \citep[e.g.][]{Sanchez-Menguiano:2016aa, Belfiore:2017aa, Molina:2017aa, Poetrodjojo:2018aa, Sanchez-Menguiano:2018aa} are observed in the local Universe. However, a significant fraction of the high-redshift ($z>1$) measurements of metallicity gradients are flat or positive \citep{Wuyts:2016aa, Wang:2017aa, Wang:2019aa, Curti:2020aa, Simons:2021aa, Li:2022aa}. This is in contrast with most recent simulations with a wide range of numerical methods and input physics \citep{Gibson:2013a, Ma:2017aa, Hemler:2021a}, which generally show negative gradients at all redshifts. \red{However, recent FIRE-2 simulations \citep{Bellardini:2021aa, Graf:2024aa} have demonstrated systematically flat radial gradients at high-redshifts ($z\sim$1-2), though not significantly positive gradients.}

With its high angular resolution imaging spectroscopy and IFU capabilities, the {\em James Webb Space Telescope}  ({\jwst}) is able to  map spatially-resolved metallicity gradients for statistically large samples of galaxies at $z > 2$ \citep{Wang:2022aa, Venturi:2024aa}. For instance, \citet{Wang:2022aa} measured the first metallicity gradient with the {\jwst} as part of the \textit{GLASS-JWST-ERS} survey (Grism Lens Amplified Survey from Space). They reported a significantly positive metallicity gradient for a 10$^9$~M$_{\odot}$ galaxy at $z=3$. This single case already intensifies the tension between observed and theoretical metallicity gradients, particularly at high redshift. Therefore, it is timely that we employ theoretical models to aid the interpretation of the forthcoming high-$z$ metallicity gradient observations.

In this work, we use simulations from the FOGGIE project (``Figuring Out Gas \& Galaxies in Enzo''). FOGGIE ran high spatial resolution ($\sim 270$ comoving pc) cosmological zoom-in simulations of six Milky-Way mass galaxies. In addition to very high resolution in the CGM and the disk-halo interface, FOGGIE generates a high cadence of outputs (every $\sim$ 5 Myr), which allows us to assess the stochastic variation of metallicity gradients over time. As we show in this paper, these attributes of FOGGIE demonstrate that the traditional approach of quantifying spatial distribution of metals via a single, azimuthally-averaged slope is inadequate in capturing the full distribution of metals in high-$z$ galaxies. We therefore call for a more informative measure of the spatial distribution of ISM metallicity. 

This paper is organised as follows. In \autoref{sec:methods} we describe the FOGGIE simulations and the methods we apply in our analysis. We present the corresponding results in \autoref{sec:results}. We discuss the implications of positive metallicity gradients in the context of the literature in \autoref{sec:posgrad}. In \autoref{sec:what_drives_grad} we discuss what current studies, including our work, hint at the physics governing the metallicity gradient evolution. We note the caveats and challenges for this topic of study, and put them in the context of the upcoming {\jwst} observations in \autoref{sec:caveats}, followed by the summary in \autoref{sec:sum}.

\section{Analysis methods}
\label{sec:methods}

This section details our simulations (\S \ref{sec:sims}) and how we select out the galactic disk (\S \ref{sec:disk_criteria}). \autoref{sec:Zgrad_method} lays out a detailed description of the radial metallicity profiles, Section \ref{sec:proj_comp} considers the effects of mapping 3-D simulations to 2-D projections, and \autoref{sec:Zscat_method} describes the non-parametric quantification of the metallicity distribution in detail.

\subsection{The FOGGIE simulations}
\label{sec:sims}

\begin{figure}
    \centering
    \includegraphics[scale=0.2]{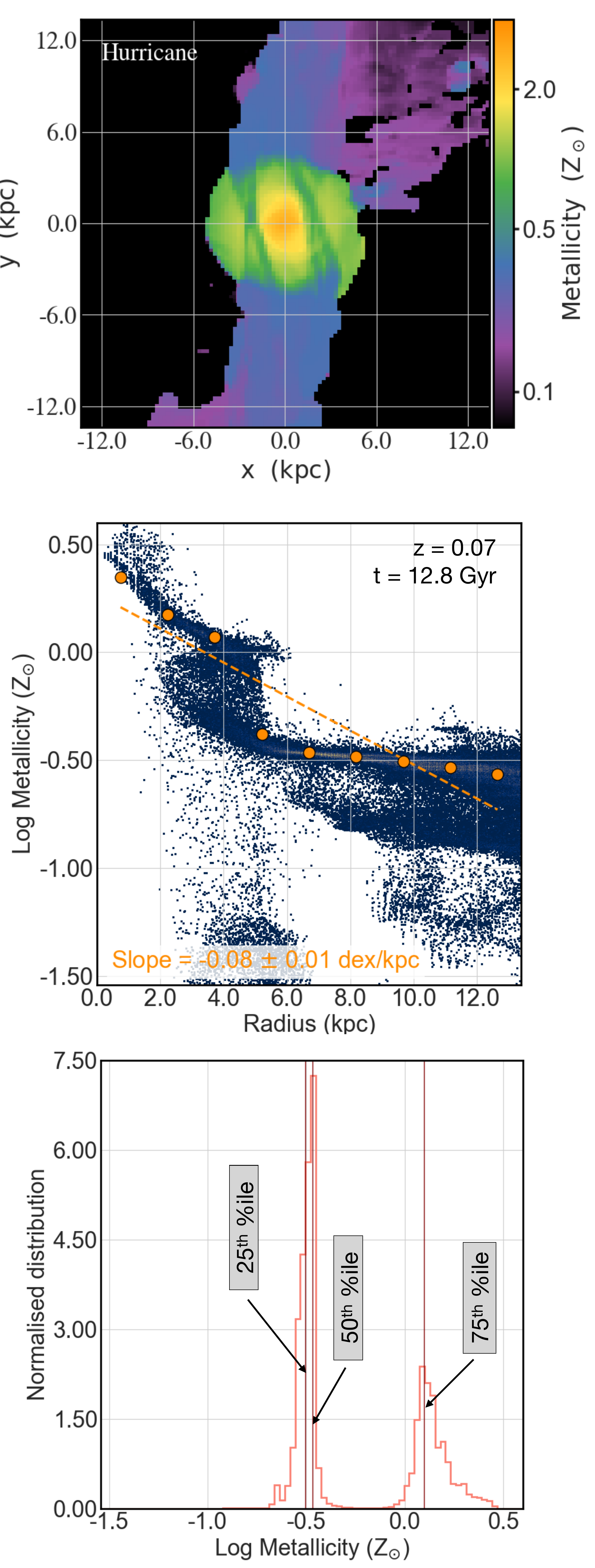}
    \caption{\textit{Top:} Mass-weighted, projected metallicity field of one of the FOGGIE halos, Hurricane, at $z = 0.07$. \textit{Middle:} Corresponding radial metallicity profile out to 10 $h^{-1}$ckpc. The color-coding in the background represents a 2-D histogram, where a high concentration of gas cells at a given metallicity-radius bin is denoted by yellow and a low concentration is denoted by dark blue. The orange circles denote mass-weighted mean metallicity in radial bins of size 1 $h^{-1}$ckpc. The dashed line is a linear fit to the orange circles, the slope of which we define as the metallicity gradient.
    \textit{Bottom:} Histogram of the mass-weighted metallicity out to 10 $h^{-1}$ckpc is denoted in orange. The vertical lines denote various percentiles as indicated in the text boxes.
    }
    \label{fig:Zgrad_snap}
\end{figure}

We use cosmological zoom-in simulations from the Figuring Out Gas \& Galaxies In Enzo (FOGGIE) project\footnote{\url{https://foggie.science/index.html}} \citep{Peeples:2019aa, Corlies:2020aa, Zheng:2020aa, Simons:2020aa, Lochhaas:2021aa, Lochhaas:2023aa, Wright:2024aa}. The simulations have been described in detail in previous FOGGIE papers; however, we summarize them briefly here. 

Firstly, a large cosmological box of 100 $h^{-1}$ comoving Mpc was run with Enzo at relatively low resolution to provide halo catalogs from which to select for zooms. \red{A flat $\Lambda$CDM cosmology \citep{Planck:2014aa} was used for these simulations, with $1 - \Omega_\lambda = \Omega_m = 0.285$, $\Omega_b = 0.0461$, and $h=0.695$}. Six halos were picked such that they are Milky-Way mass at $z=0$. Then, these six halos were re-simulated with Enzo (\citealp{Bryan:2014aa}; last described in \citealp{Brummel-Smith:2019aa}) within a ``zoom in'' region. FOGGIE is distinguished from most other cosmological simulations by a unique adaptive mesh refinement (AMR) scheme that allows ``forced refinement'' within a refine box that follows the zoomed galaxy through the domain. Inside this moving box, 200 $h^{-1}$ comoving kpc on a side, there is a minimum level of refinement, corresponding to a maximum cell size, regardless of density or other physical properties. As first presented in FOGGIE Paper IV \citep{Simons:2020aa}, the current production FOGGIE simulations impose a \textit{minimum} level of refinement $n_{\rm ref}=9$ ($1{,}100$ comoving pc cells) everywhere within the ``refined box". In order to ensure that the cooling length is resolved, we also employ a ``cooling refinement'' prescription that refines cells larger than the cooling length---determined by the local sound speed and cooling time---up to a \textit{maximum} refinement level $n_{\rm ref}=11$ (274 comoving pc cells). Since both forced and cooling refinement are active in the moving box, all of the ISM and almost all of the CGM ($\gtrsim 95\%$) are resolved at the highest level, i.e., 91 pc at $z=2$ \citep{Simons:2020aa}, and the median cell mass in the CGM and disk-halo interface is $<200$ \Msun \citep{Lochhaas:2023aa}.  The high resolution enables FOGGIE to capture small scale dynamic processes such as metal mixing and transport, thereby making these simulations well suited for our study \citep[see also][]{vandeVoort:2019aa, Hummels:2019aa}. The six halos were selected to have no major mergers after $z=2$. This serves as the ideal ``control'' scenario for our study because we can compare the behavior of the metallicity gradient with ($z\gtrsim2$) and without ($z\lesssim2$) major mergers. We refer the reader to Table 1 for a detailed list of global properties of all the halos.

We use the $z=0.07$ snapshot of the Hurricane halo to demonstrate our methods. We present our main results from all six halos---Tempest, Maelstrom, Squall, Blizzard, Hurricane, Cyclone---for the first part of our analysis where we compare the cosmic evolution of metallicity gradients of FOGGIE galaxies in the context of current observations. Thereafter, we focus on one halo, Hurricane, for the rest of this work where we investigate potential factors that impact the metallicity gradient evolution seen in the FOGGIE simulations. We choose this halo as it has the most interactions out of all the FOGGIE halos, and therefore has a diverse range of gradients. While we present detailed analysis for only Hurricane in the main body of the paper for simplicity, we show in \autoref{sec:ap} that our main conclusions remain qualitatively unchanged for the other halos as well.

\subsection{Selection criteria for the gas disk}
\label{sec:disk_criteria}

Observed metallicity gradients are limited to the gaseous, star-forming regions of galaxies and their immediate environment. We therefore limit our analysis to within 10~$h^{-1}$ comoving kpc of the galaxy center. We choose this extent because it is equivalent to $\approx 2.5\times r_e$, where $r_e \sim 4$\,kpc is a typical effective radius of the gas disk of star-forming galaxies at $z\sim0$. Additionally, we impose a simple redshift-dependent density criterion such that $\rho_{\rm cut} = 2\times 10^{-26}$ g cm$^{-3}$ if $z > 0.5$, $\rho_{\rm cut} = 2\times 10^{-27}$ g cm$^{-3}$ if $z < 0.25$, and a linear interpolation with respect to physical time (not redshift) between $0.25 \leq z \leq 0.5$. Note that this density criterion is based on proper density, and not comoving density. We select only the material with density above $\rho_{\rm cut}$, and apply our analysis methods on this dense material. The choice of this density criterion was used by \citet{Lochhaas:2021aa} to capture the dense ISM material. We deliberately do not impose any cylindrical or geometrical criteria to capture the ISM disk, because the ``disk'' is likely to have irregular, clumpy morphology, particularly at high redshifts (see \autoref{sec:Zscat_method}) which would not be captured by a well-behaved cylindrical geometry. It is worth noting that even at low-$z$ FOGGIE galaxies have significant warps \citep{Lochhaas:2023aa, Simons:2024aa}.

\subsection{Radial metallicity gradient}
\label{sec:Zgrad_method}
\autoref{fig:Zgrad_snap} shows (a) the projected gas phase metallicity at the top, projected along the arbitrary $y$ axis, (b) the radial mass-weighted metallicity profile in the middle, and (c) the mass-weighted histogram at the bottom. All the panels correspond to the $z=0.07$ snapshot of the Hurricane halo, and show metallicities within a 10~$h^{-1}$ comoving kpc sphere (corresponds to 13 physical kpc at $z = 0.07$), after employing the density criterion described above. We provide figures similar to \autoref{fig:Zgrad_snap}, but corresponding to the the other FOGGIE halos and several redshift epochs, in Figures \ref{fig:Zgrad_snap_tempest}-\ref{fig:Zgrad_snap_hurricane} in \autoref{sec:ap}.

The top panel reveals distinct ``inner'' and ``outer'' disks in the metallicity space, where there is a steep drop in metallicity in the outer disk, compared to the inner disk. We note that whenever there is such a steep drop in metallicity between an ``inner'' and ``outer'' disk in the FOGGIE simulations, active star formation is typically confined to the inner disk only.

The color-coding of the middle panel of \autoref{fig:Zgrad_snap} represents a 2-D histogram of the gas cells, with lighter blue implying a higher concentration of cells. In this case there is a notable ``break'' in the metallicity profile between a steeper trend in the central few kpc and a flatter, almost constant, metallicity profile in the outskirts. 

The outer profile is coincident with a diffuse, non-star-forming disk in the outskirts. Although we see this ``break" in the metallicity profile ubiquitously in our simulations, we highlight a particularly prominent example here to aid our explanation of the fitting routine.

The orange circles depict the mass-weighted average metallicity in spherical bins of width $1 h^{-1}$ comoving kpc (ckpc) centered at the galaxy center, and a linear fit to them is shown by the dashed orange line. The resulting slope, along with its uncertainty, is denoted in the orange text in the middle panel of \autoref{fig:Zgrad_snap}. Note that only the gas that passes the disk density cut (\autoref{sec:disk_criteria}) is included here. The line is fitted to the mass-weighted mean for each bin, and the natural, physical variation around each mean value is treated as an ``uncertainty'' in the inverse-square weighting, i.e., each bin receives a weight equal to the inverse square of the ``uncertainty''. Owing to mass-weighting, the central bins dominate the fit because that is where most of the disk mass is concentrated. 

Observed gas-phase metallicity profiles are primarily derived from emission line maps tracing the ionized ISM. The primary drivers of the ionization in such \hii\ regions are young, massive stars. Therefore, observed profiles preferentially trace a luminosity-weighted metallicity. The central regions of Milky Way type galaxies typically not only have more stars but also more massive bright stars, leading to a more luminous center than the outskirts \citep{Minniti:2008aa}. Therefore, the observed metallicity profile has a higher weight-contribution from the center. In our case, the fit to the mass-weighted radial bins preferentially traces the central part of the galaxy, which is similar in effect to a fit to an observed metallicity profile.

We repeat this radial fitting routine for each snapshot of each halo, from $z = 4$ to the latest snapshot available. We start our analysis at $z = 4$ because at higher redshifts the ``disk'' morphology is not apparent and mergers are too frequent and rapid to maintain a radial profile.

\subsection{Evaluation of projection effects}
\label{sec:proj_comp}
\begin{figure*}
    \centering
    \includegraphics[width=\linewidth]{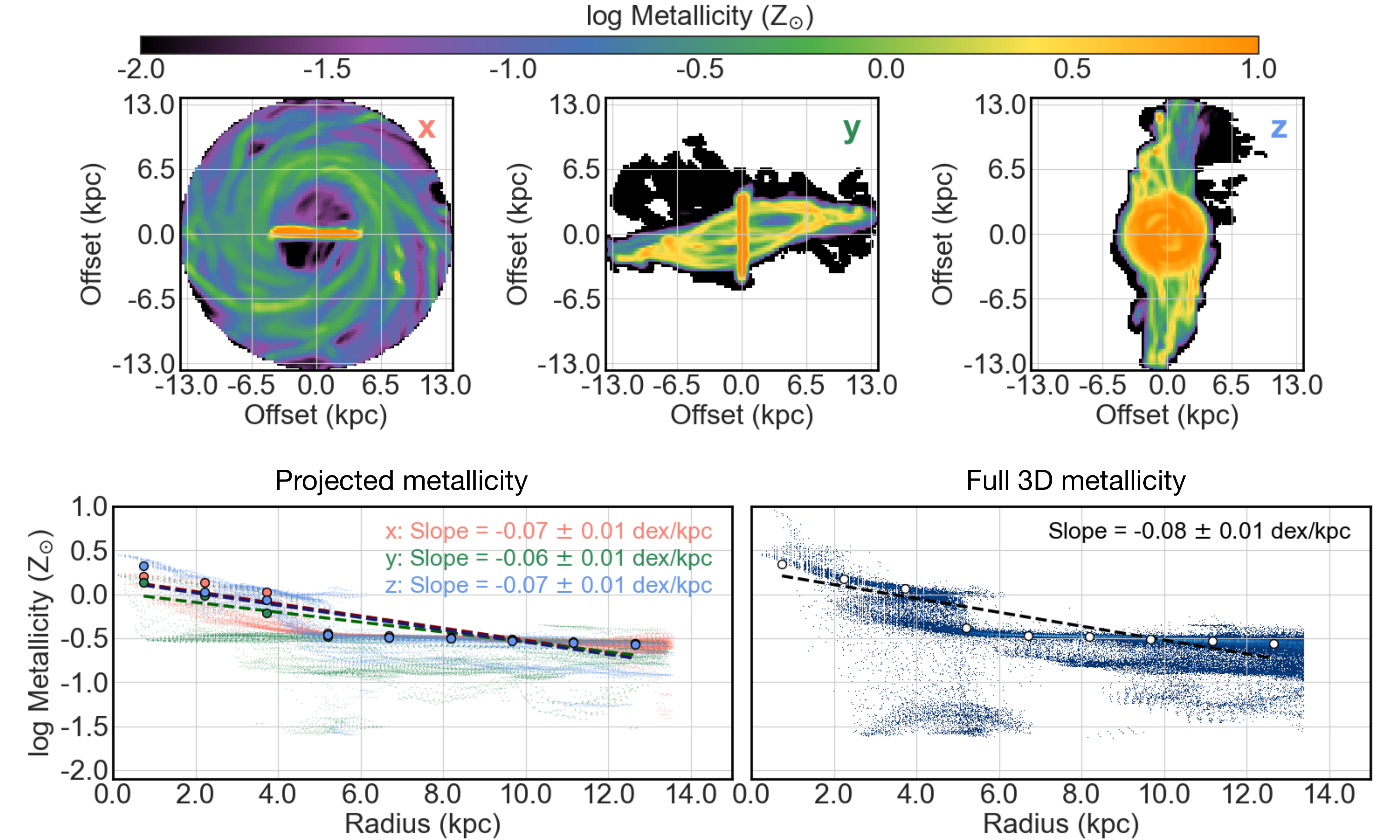}
    \caption{Comparison of metallicity distribution as seen in projection vs the full 3-D distribution. \textit{Top:} Mass-weighted, projected metallicity of Hurricane at $z = 0.07$, as seen from three different lines of sight -- $x$, $y$, and $z$ axes, with respect to the simulation box. \textit{Bottom left:} Corresponding radial profile out to $10 h^{-1}$ckpc for the mass-weighted metallicity projected along the three axes, with each color corresponding to an axis of projection. The fitting process is outlined in \autoref{sec:Zgrad_method}. \textit{Bottom right:} Same as the left panel but for the full 3-D metallicity in the simulation, rather than projected metallicity. The projected and 3-D metallicity distributions look qualitatively similar.}
    \label{fig:proj_vs_3d}
\end{figure*}

Observed metallicity distributions are, necessarily, projected quantities along the line of sight of the observation. Simulations, on the other hand, have access to the full 3-D distribution of the physical property concerned. Therefore, it is important to compare the full 3-D distribution of metals in the FOGGIE simulations to its projected counterpart in order to assess the potential impact of projection effects on simulated gradients. Our analysis for the rest of this paper is based on the intrinsic 3-D metallicities in the FOGGIE galaxies, but here we show that our qualitative results are not expected to change significantly when accounting for projection effects.

In \autoref{fig:proj_vs_3d} we demonstrate the effect of projection on the metallicity maps and radial profiles as seen in projection from three different lines of sight (along the cardinal axes), as well as on the full 3-D  metallicity profile. The top row of \autoref{fig:proj_vs_3d} corresponds to the projected 2-D metallicity maps. The bottom-left panel shows the azimuthally-averaged fits to the radial metallicity profile, where each color corresponds to a different projection. The bottom-right panel depicts the radial profile of the full 3-D metallicity distribution, same as the middle panel of \autoref{fig:Zgrad_snap}. We do not see any qualitative difference in the radial profiles due to the different angles of projection. Indeed, the fitted gradients (colored text in \autoref{fig:proj_vs_3d}) obtained in cases of projections along various lines of sight and that from the 3-D distribution all agree with each other within 1$\sigma$ uncertainties. Therefore, we conclude the projection effects do not impact the main science results of this paper. Henceforth we proceed with the full 3-D metal distribution for the rest of the paper, disregarding projection effects.

It is worth noting that forward modeling the simulated data to produce synthetic observables is generally a good approach in comparing observations with simulations. However, that is outside the scope of our current work for the reasons described in \autoref{sec:caveats}. Therefore, all measurements of metallicity discussed in this work refer to the intrinsic metallicity of the gas cells in the FOGGIE simulations.

\subsection{Non-parametric characterization of the metallicity distribution}
\label{sec:Zscat_method}

Fitting a single, smooth gradient to the entire radial metallicity profile of the disk, while informative, may not necessarily be the best representation of the full 2-D distribution of metallicity. For instance, ``breaks" in radial profiles and azimuthal variations are washed out in the conventional gradient approach. Moreover, defining a gradient makes assumptions about a disk geometry, which may not hold at high-$z$. We therefore explore a non-parametric approach to quantify the distribution of metallicity. We discuss the detailed implications of this approach in \autoref{sec:nparam}. We note that a similar approach of quantifying the metallicity distribution function (MDF) has already been utilised by the stellar metallicity community to great scientific success, ranging from the Milky Way \citep{Hayden:2015aa, Loebman:2016aa} to stellar halos \citep{Durrell:2001aa} and satellites \citep{Kirby:2011aa, Chiti:2020aa}.

We characterize the spatial distribution of metallicity within $10 h ^{-1}$ comoving kpc from the galaxy center by representing it as a histogram and computing the 25$^{\rm th}$, 50$^{\rm th}$ and 75$^{\rm th}$ percentiles. The bottom panel of \autoref{fig:Zgrad_snap} shows the metallicity distribution for Hurricane at $z = 0.07$, along with vertical lines depicting the different percentiles. We compute the inter-quartile range (Z$_{\mathrm{IQR}}$) as the difference between the 75$^{\rm th}$ (Z$_{\mathrm{75}}$) and 25$^{\rm th}$ (Z$_{\mathrm{25}}$) percentiles, and use it, along with the median, to trace the evolution of the metallicity distribution across cosmic time (\autoref{sec:Zscat_res}). All the percentiles are in $\log{(\mathrm Z_{\odot})}$ units, and therefore Z$_{\mathrm{IQR}}$ is the difference between  Z$_{\mathrm{75}}$ and  Z$_{\mathrm{25}}$ in log space.

In the snapshot depicted in \autoref{fig:Zgrad_snap}, we see a bimodal distribution: a narrow peak at low metallicity, followed by a broad distribution at higher metallicity. This is a manifestation of the `break' in the radial metallicity profile, as described in \autoref{sec:Zgrad_method}. We conclude that the low metallicity peak in the distribution is contributed by an extended, metal-poor cold gas disk on the outskirts of the galaxy. Such a bimodal (in metallicity) disk is likely a result of misaligned disks (see \autoref{sec:break}), which survive for a majority of the cosmic time, until very low redshift ($z \lesssim 0.3$). Therefore, the metallicity distribution exhibits a double-peaked nature for most snapshots, except at very low redshift when it becomes single-peaked. 

Our attempts to fit these multi-peaked distributions with skewed Gaussian models have proven to be non-trivial, and not sufficiently informative for tracing the chemical evolution of the galaxies. This further justifies our view that it is difficult to capture the underlying physics leading to the full 3-D distribution of metals in the galaxy disk by parameterizing it with simple 1D models. 

\begin{table*}
	\small
	\centering
\begin{tabular}{llllllll}
\toprule
      Halo &    $z$ & Time (Gyr) & M$_{\star}$/M$_{\odot}$ & $\log Z_{\rm total}$/Z$_{\odot}$ & $\log Z_{\rm median}$/Z$_{\odot}$ & $\log Z_{\rm IQR}$/Z$_{\odot}$ & $\nabla Z$ (dex/kpc) \\
\midrule
   Tempest &  $0.0$ &    $13.62$ &                 $10.72$ &                          $-0.13$ &                           $-0.18$ &                         $0.36$ &     $-0.062\pm0.002$ \\
   Tempest &  $1.0$ &     $5.89$ &                 $10.56$ &                $\phantom{-}0.17$ &                 $\phantom{-}0.17$ &                         $0.12$ &     $-0.191\pm0.022$ \\
   Tempest &  $2.0$ &     $3.32$ &                 $10.19$ &                          $-0.10$ &                           $-0.31$ &                         $0.41$ &     $-0.114\pm0.016$ \\
 Maelstrom &  $0.0$ &    $13.62$ &                 $11.02$ &                $\phantom{-}0.13$ &                 $\phantom{-}0.12$ &                         $0.20$ &     $-0.037\pm0.002$ \\
 Maelstrom &  $1.0$ &     $5.89$ &                 $10.79$ &                          $-0.05$ &                           $-0.09$ &                         $0.47$ &     $-0.136\pm0.003$ \\
 Maelstrom &  $2.0$ &     $3.32$ &                 $10.48$ &                          $-0.20$ &                           $-0.42$ &                         $0.61$ &     $-0.245\pm0.010$ \\
    Squall &  $0.0$ &    $13.62$ &                 $11.07$ &                          $-0.08$ &                           $-0.38$ &                         $0.64$ &     $-0.082\pm0.011$ \\
    Squall &  $1.0$ &     $5.89$ &                 $10.45$ &                $\phantom{-}0.09$ &                 $\phantom{-}0.09$ &                         $0.21$ &     $-0.071\pm0.005$ \\
    Squall &  $2.0$ &     $3.32$ &                  $9.18$ &                $\phantom{-}0.07$ &                 $\phantom{-}0.00$ &                         $0.18$ &     $-0.133\pm0.008$ \\
  Blizzard &  $0.0$ &    $13.62$ &                 $11.14$ &                          $-0.13$ &                           $-0.20$ &                         $0.42$ &     $-0.068\pm0.002$ \\
  Blizzard &  $1.0$ &     $5.89$ &                 $10.91$ &                $\phantom{-}0.16$ &                 $\phantom{-}0.13$ &                         $0.23$ &     $-0.048\pm0.003$ \\
   Cyclone &  $1.0$ &     $5.90$ &                 $11.12$ &                          $-0.04$ &                           $-0.09$ &                         $0.09$ &     $-0.018\pm0.010$ \\
   Cyclone &  $2.0$ &     $3.32$ &                 $10.49$ &                $\phantom{-}0.16$ &                 $\phantom{-}0.19$ &                         $0.26$ &     $-0.125\pm0.047$ \\
 Hurricane &  $0.0$ &    $13.62$ &                 $11.37$ &                          $-0.20$ &                           $-0.45$ &                         $0.54$ &     $-0.076\pm0.009$ \\
 Hurricane &  $1.0$ &     $5.89$ &                 $11.04$ &                $\phantom{-}0.10$ &                 $\phantom{-}0.06$ &                         $0.28$ &     $-0.120\pm0.007$ \\
 Hurricane &  $2.0$ &     $3.32$ &                 $10.42$ &                $\phantom{-}0.08$ &                 $\phantom{-}0.08$ &                         $0.65$ &     $-0.247\pm0.018$ \\
\bottomrule
\end{tabular}

	\caption{A list of the measured quantities within 10 $h^{-1}$ ckpc for each snapshot of each FOGGIE halo -- stellar mass, total metallicity,  median and inter-quartile range (IQR) of the metallicity distribution, and the fitted radial gradient (see \autoref{sec:Zgrad_method}). All the quantities are quoted in log$_{10}$ scale, with the exception of redshift, time (in Gyr) and metallicity gradient (in dex/kpc). This table is intended to merely function as an abridged version of the full table which is available online in a machine-readable format at \url{https://ayanacharyya.github.io/data/master_table_Zpaper_upto10.0ckpchinv_wtby_mass_fitmultiple_wdencut_islog.txt}.}
    \label{tab:fitted}
\end{table*}

In \autoref{tab:fitted} we tabulate the fitted and other measured quantities for all FOGGIE halos, including the total stellar mass and total metallicity. \red{The stellar mass is defined as the total mass in stars within 10 $h^{-1}$ckpc. The total metallicity is the ratio of metal mass to gas mass within 10 $h^{-1}$ckpc, after applying the density criteria to select the disk (see \autoref{sec:disk_criteria}). We then normalize it using solar metallicity of 0.01295, which is the default solar metallicity in \texttt{yt} \citep{Turk:2011aa}, and is consistent with the solar metallicity used in Cloudy photoionization models. In \autoref{tab:fitted} we report this normalized total metallicity in log-scale. The total metallicity of FOGGIE galaxies are higher than those observed for galaxies of similar mass and type, particularly at high-redshift. FOGGIE galaxies exhibit super-solar metallicity $\sim$50 \% of the time. This is due to a combination of the current feedback and star-formation prescription in FOGGIE which leads to centrally concentrated star-formation and too may metals being locked up in the stars. For a detailed explanation see \citet{Wright:2024aa}. Although the absolute metallicity values are somewhat higher than observed, the \textit{relative} variation in metallicity, i.e., metallicity \textit{gradients}, is robust and therefore renders our primary science conclusions unaffected.}

In the table presented here we only include a few snapshots, intended to serve as examples. The full table is available online in a machine readable format at \url{https://ayanacharyya.github.io/data/master_table_Zpaper_upto10.0ckpchinv_wtby_mass_fitmultiple_wdencut_islog.txt}.

\begin{figure*}
    \centering
    \includegraphics[width=\linewidth]{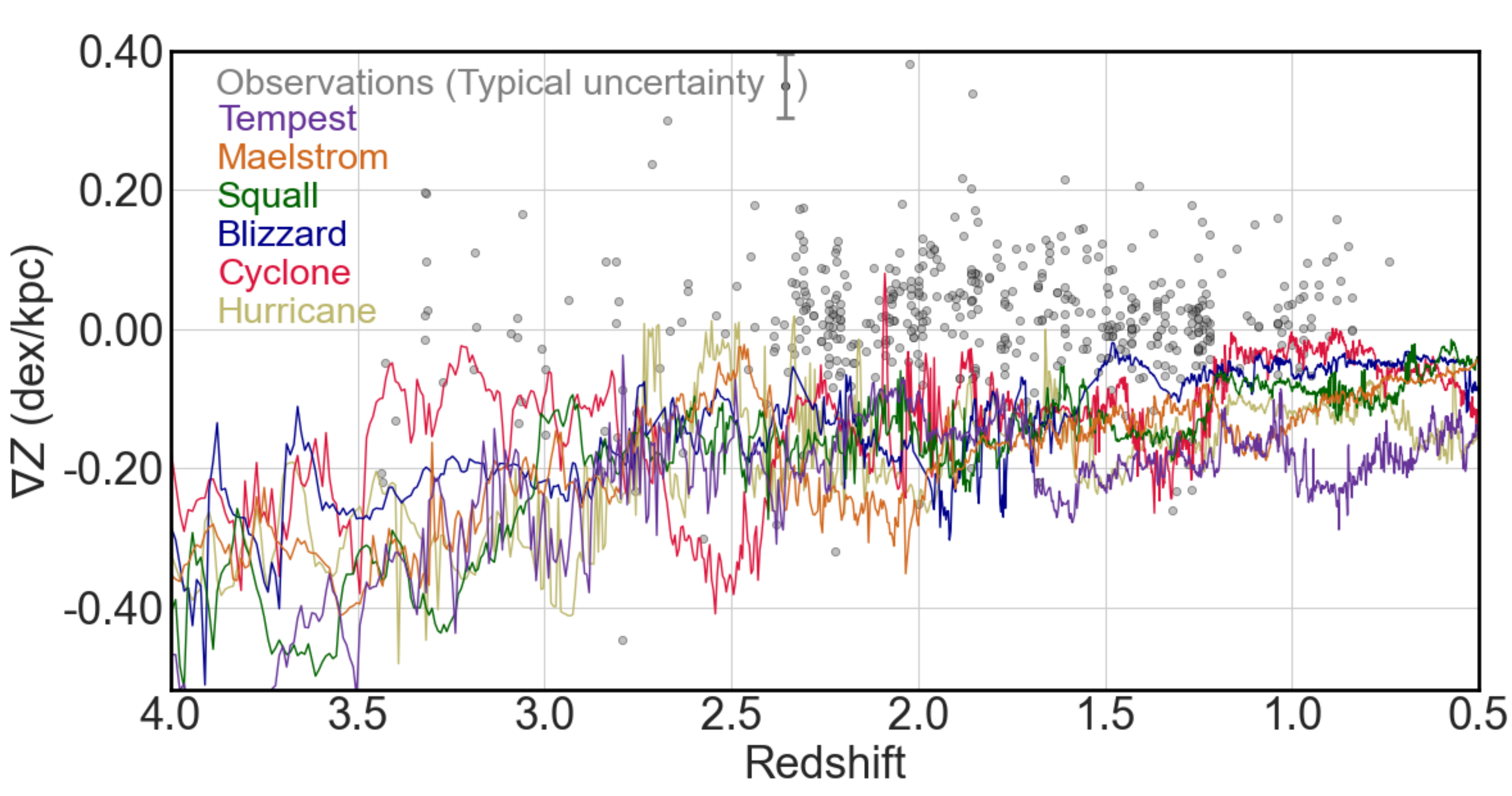}
    \includegraphics[width=\linewidth]{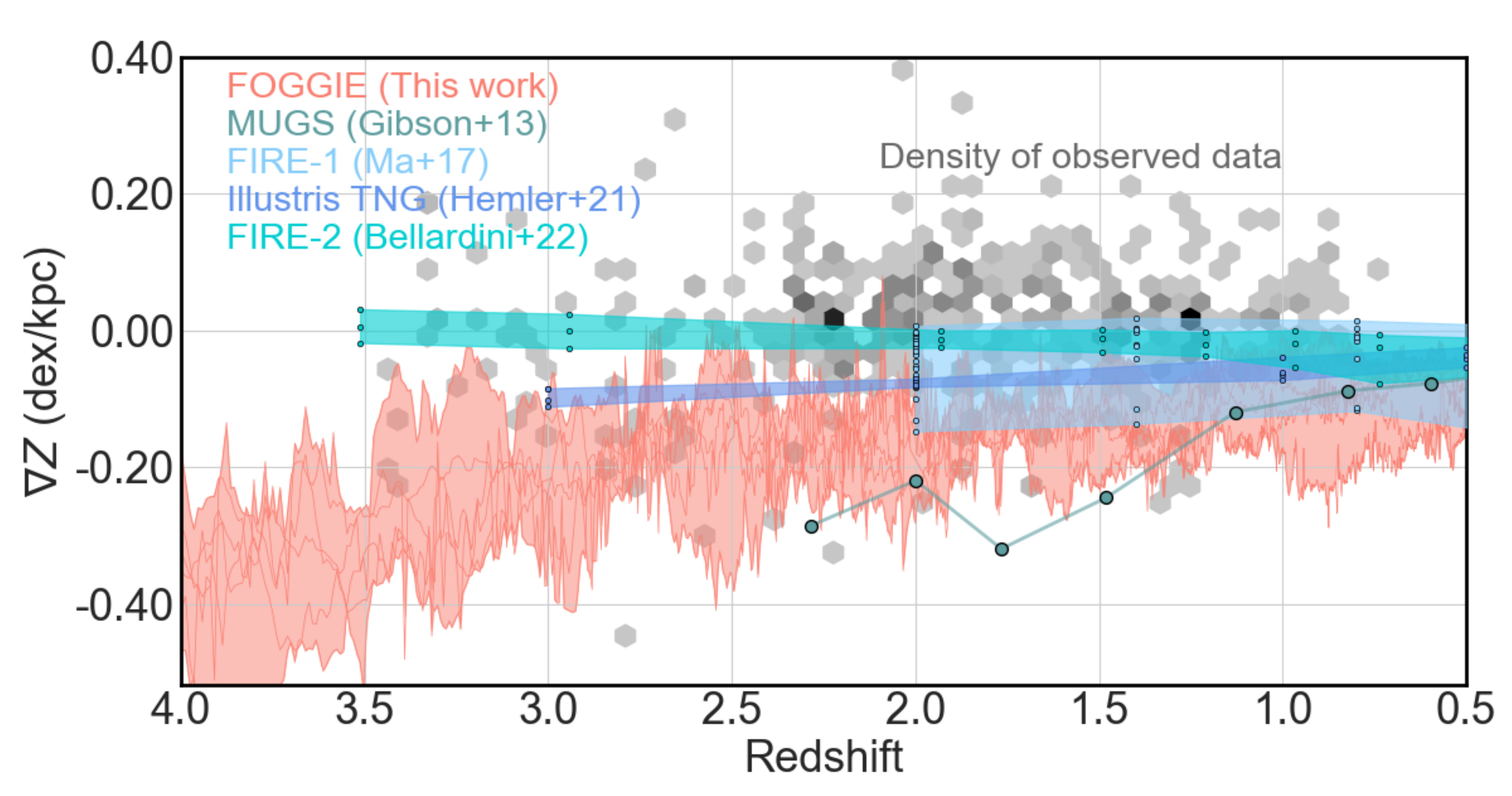}
    \caption{\textit{Top:} Comparison of the evolution of FOGGIE metallicity gradients and observed gradients from the literature. The colored lines show the evolution of metallicity gradients (within 10 $h^{-1}$ckpc) of the six FOGGIE galaxies. The rest of this figure was adapted from Figure 3 of \citet{Wang:2022aa}. The grey circles denote observed gradients from a variety of literature sources -- ground-based integral-field spectroscopy (IFS) with \citep{Swinbank:2012aa, Jones:2013aa, Leethochawalit:2016aa, Forster-Schreiber:2018aa} and without \citep{Bundy:2015aa} AO-assist, as well as space-based slitless spectroscopy with {\hst}/WFC3 \citep{Jones:2015aa, Wang:2020aa, Simons:2021aa, Li:2022aa} \red{and {\jwst}/NIRISS (Li et al. submitted)}. \textit{Bottom:} Comparison between FOGGIE (salmon) and metallicity gradient evolution predicted by other simulations---MUGS \citep[][teal]{Gibson:2013a}, FIRE-1 \citep[][light blue]{Ma:2017aa}, TNG50 \citep[][darker blue]{Hemler:2021a}, \red{and FIRE-2 \citep[turquoise]{Bellardini:2022aa}. Note that here we plot the full extent of the Fe/H gradient of stars at formation time from FIRE-2, corresponding to the top panel of Figure 7 in \citet{Bellardini:2022aa}.} The density of observed gradients is shown by the hexbins, with darker bins implying a higher density of observations.}
    \label{fig:Zlit}
\end{figure*}

\section{Results}
\label{sec:results}

\subsection{General patterns in radial gradients}
\label{sec:Zgrad_res}

Our starting point for results is the traditional radial metallicity gradient as seen in data and in models (top panel of \autoref{fig:Zlit}). The observational data for comparison is shown in the gray circles, as compiled by \citet{Wang:2022aa} in their paper presenting the first high-$z$ gradient measurement with {\jwst}. These data are predominantly ground-based IFU measurements or {\hst} slitless spectroscopy. Notably, the data at $z = 1-2.5$ exhibits just as many flat or positive gradients as negative ones. 

The six FOGGIE galaxies have negative gradients for nearly all of their history, with only one (Hurricane) briefly approaching or touching a flat gradient (see \autoref{sec:posgrad}). There is also a broad trend for the FOGGIE galaxies to start with steeper negative gradients ($\sim -0.5$ to $-0.1$ dex/kpc) and evolve toward shallower negative gradients ($\sim -0.3$ to $-0.1$ dex/kpc). The bottom panel of \autoref{fig:Zlit} depicts the \textit{spread} in metallicity gradients predicted by different simulations---MUGS \citep{Gibson:2013a}, FIRE\red{-1} \citep{Ma:2017aa}, \red{FIRE-2 \citep{Bellardini:2022aa}} and Illustris TNG \citep{Hemler:2021a} in comparison to that of FOGGIE \citep[see also Figure 3 of][]{Wang:2022aa}. This basic comparison between observed and simulated gradients, in both the panels, shows a strong tension; none of these state-of-the-art simulation suites yields gradients that are positive as often as they are negative. This systematic offset is robust: the different observational techniques employed here generally agree with one another about a distribution scattered around a flat gradient, while the simulations all occupy the negative range. This inconsistency poses a challenge to our interpretation of metallicity gradients in terms of the production and distribution of metals in galaxies. 

FOGGIE generates snapshots every $\sim$5.3 Myr, which makes the rapid stochastic fluctuations of the gradients readily apparent. \red{FIRE-2 \citep{Bellardini:2021aa, Bellardini:2022aa, Graf:2024aa} simulations have an output cadence of 20-25 Myr, making these the most suitable comparison sample to FOGGIE (detailed in \autoref{sec:lit_obs}). However, both these simulations contrast} with the $50-100$ Myr cadence of most previous simulations \citep[e.g.,][]{Gibson:2013a, Ma:2017aa, Hemler:2021a}, as shown in the bottom panel of \autoref{fig:Zlit}. The considerable degree of stochastic fluctuation in the FOGGIE metallicity gradient evolution demonstrates that metallicity gradients indeed vary significantly over short timescales, suggesting that trends deduced from outputs several Gyr apart may not be robust. 

The abundance of observed positive gradients at all redshifts is in contrast with simulated ones. On par with other simulations, FOGGIE too does not generally reproduce the observed significantly positive metallicity gradients. Notably, we do measure positive gradients in FOGGIE on a few occasions---particularly for Hurricane, as shown in the top panel of \autoref{fig:Zlit}. However, such instances in FOGGIE are short-lived, and stem from \textit{external} processes such as accretion or mergers, and as such are highly subject to rapid fluctuations (see \autoref{sec:posgrad} for detailed discussion). 

We do not see long-lived positive gradients in settled and well-behaved disks. This could imply that the observed positive gradients are also instantaneous states in the galaxies' lifetimes. However, the ubiquity of the observed positive gradients could potentially imply that the simulations are missing some key physical process. It is worth noting that the metallicity observations themselves potentially suffer from limited spatial resolution, signal-to-noise issues, scatter between different metallicity diagnostics used, and uncertainties in the underlying photoionization models. Although these uncertainties can lead to artificially flatter gradients \citep[e.g.,][]{Yuan:2013aa, Acharyya:2020aa}, none of these effects have yet been demonstrated to bias the observed gradients towards \textit{positive} values.

Our results unambiguously show significant scatter in the metallicity gradient evolution, particularly at high-redshift. Not only is there galaxy-to-galaxy scatter of $\sim0.1$--0.4 dex but there also is considerable stochasticity (variations of $\sim 0.1$--0.2 dex in slope over a $\sim$ few Myr) in the evolution of an individual galaxy. Therefore, the short timescale variations that FOGGIE's high time cadence data outputs allow us to study are not negligible. Lack of a high time-cadence would lead to sparser sampling of this stochastic behavior, and therefore such low-cadence studies would be less likely to capture the large excursions the metallicity gradient makes along its evolutionary course. We discuss this further in \autoref{sec:scatter}. Given the stochastic behavior of the models, it would be challenging to interpret the metallicity gradient evolution via forthcoming high-$z$ observations with {\jwst} and large ground based telescopes (e.g., the Giant Magellan Telescope; GMT).

\subsubsection{Evolution of the radial gradient}
\label{sec:evolution}

The top panel of \autoref{fig:Zgrad_smooth} shows the metallicity gradient evolution of one FOGGIE halo, Hurricane, as a function of time. The circles mark the times corresponding to whole number redshift values i.e., $z=0$, 1, 2, and so on, from right to left. The faint line in the background shows the evolution sampled at a lower cadence of $\sim$ 500 Myr, whereas the thicker line in the foreground represents the intrinsic cadence ($\sim 5$ Myr) of FOGGIE simulations. We see an overall flattening of the metallicity gradient after $z\sim1$. We provide figures similar to the top panel of \autoref{fig:Zgrad_smooth}, but corresponding to the other FOGGIE halos, in \autoref{sec:ap}.

While the lower-cadence curve captures the overall general trend in metallicity gradient, it misses out on the short timescale variations, particularly at high-$z$ ($z \geq 1$). We therefore quantify the fraction of evolution that would be otherwise ``missed'' by studies with such lower cadence. The shaded area represents a typical observational uncertainty of $\pm 0.03$ dex/kpc around the lower-cadence evolution. We emphasise that the shaded area does not represent the uncertainty in the simulated gradient, but the usual amount of uncertainty seen in observational studies. Every time the solid line exceeds the shaded region, we consider that as a significant excursion of the metallicity gradient evolution away from the ``general'' (smoothed) behavior. This behavior suggests that lower-cadence simulations or static models are missing a significant degree of the stochastic variation. But observations are snapshots in time and can easily catch a real galaxy in one of these excursions. We find that this galaxy is more than the typical observational uncertainty away from its general behavior for $\sim 50$ \% of the time at $z > 1$. In other words, a cadence of $\sim 500$ Myr fails to capture almost half of this galaxy's evolution at high-$z$. Observations, by definition, capture a single snapshot during a galaxy's lifetime. Our work demonstrates that, for a Milky-Way type galaxy, it is challenging to interpret an instantaneous high-$z$ metallicity gradient observation because any snapshot may be a large and/or short-lived excursion away from the general pattern. This further highlights the challenges to interpreting forthcoming high-$z$ metallicity gradient measurements with {\jwst} or GMT.

\subsubsection{Rapid variations in radial gradient}
\label{sec:variations}

\begin{figure*}
    \centering
    \includegraphics[width=\linewidth]{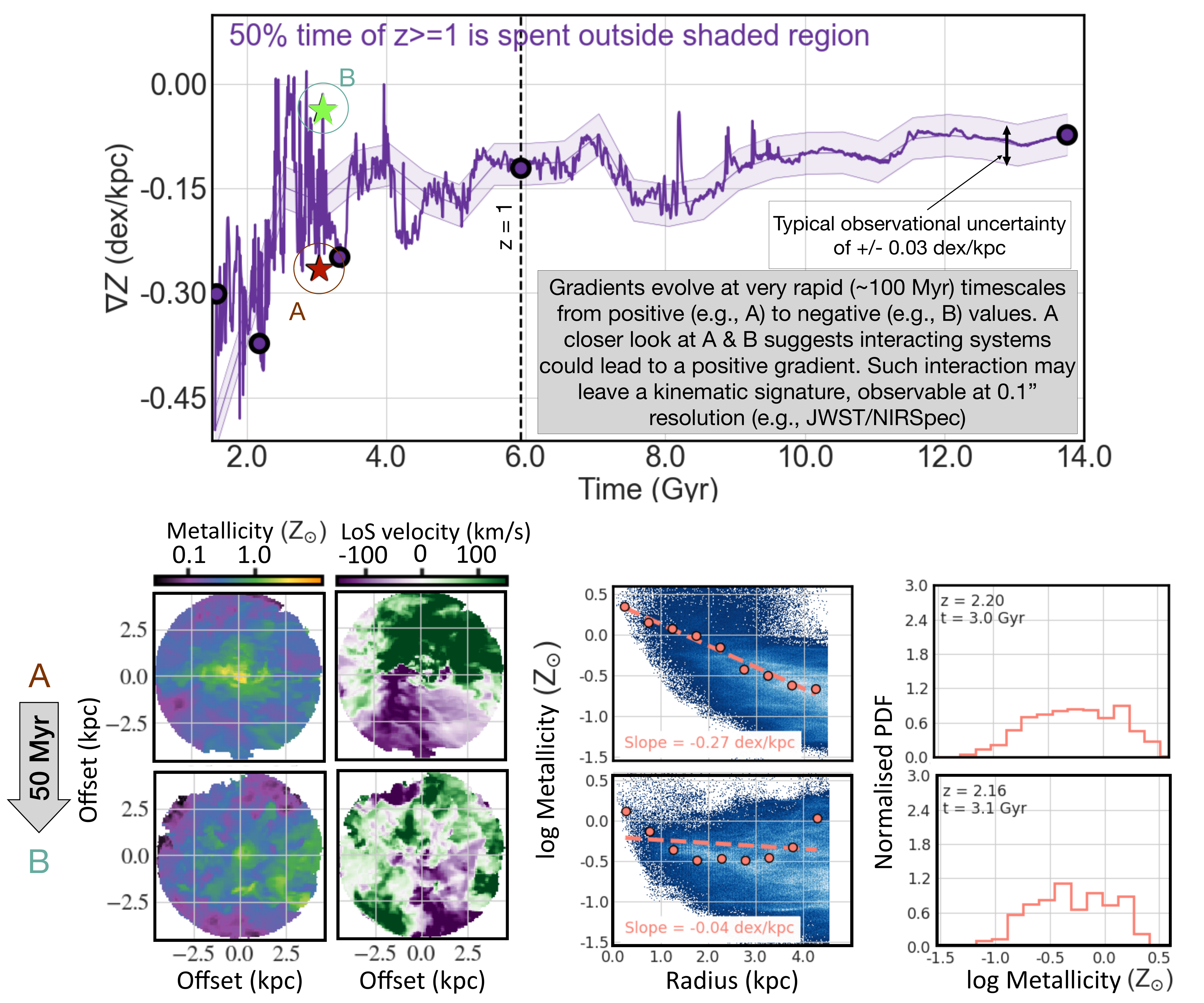}
    \caption{\textit{Top:} The bold purple line shows the full evolution of the metallicity gradient of Hurricane---one of the FOGGIE galaxies---with a cadence of $\approx$ 5 Myr. The thinner purple line in the background denotes the same evolution with a lower cadence of 500 Myr. This thin line is a proxy for the relatively low-stochasticity, average evolution of the gradient over time. The shaded region of $\pm$ 0.03 dex/kpc around the lower-cadence evolution is intended to depict the typical observational uncertainty in gradient measurements. The purple circles denote whole number redshifts $z=$ 0, 1, 2, 3 and 4 (from right to left). This instantaneous gradient for this galaxy is outside this typical 1$\sigma$ uncertainty 50\% of the time up to $z >= 1$, denoted by the black dashed line.\\
    \textit{Middle:} From left to right, the panels denote the projected metallicity map, line of sight velocity map, radial profile and full distribution of mass-weighted metallicity of the time stamp corresponding to A, as denoted in the top panel.
    \textit{Bottom:} Same as above but corresponding to the time stamp B, denoted in the top panel. The time stamps A \& B are only $\sim 50$ Myr apart, although they show a stark contrast ($\sim0.25$ dex) in the gradients. This demonstrates the short timescale variations in metallicity distribution, leading to the large scatter in gradients, particularly at high-$z$.}
    \label{fig:Zgrad_smooth}
\end{figure*}

An example of rapid variation of the radial metallicity gradient at high-redshift is demonstrated in the lower two panels of \autoref{fig:Zgrad_smooth}. The middle and bottom panels of \autoref{fig:Zgrad_smooth} allow a closer look at specific time stamps during the evolution of Hurricane, denoted by the brown and green star-symbols in the top panel, ``A'' and ``B'', respectively. ``A'' displays a strong negative metallicity gradient, whereas ``B''---only $\sim 50$\,Myr later---shows a significantly flatter (but still negative) gradient. Each row of these two bottom panels shows, from left to right, the projected mass-weighted metallicity map, the mass-weighted line-of-sight velocity map, the radial metallicity profile, and the mass-weighted metallicity distribution. The projected metallicity map, radial profile and distribution plots follow the same template as in \autoref{fig:Zgrad_snap}. The line-of-sight velocity map depicts the kinematics of the gas in the disk at each snapshot. Our goal is to investigate the cause of the rapid flattening of the gradient in just $\sim 50$\,Myr by examining the kinematics of the gas. We choose the line-of-sight velocity as a metric because it is a readily accessible quantity to spectroscopic observations.

By comparing the snapshots ``A'' and ``B'' it is clear that the flatter metallicity gradient at the later time is due to the presence of metal-rich clumps on the outskirts of the central disk in ``B''. This is clearly not a resolution-effect, although resolution can play a crucial role in such measurements (see \autoref{sec:posgrad}). In this case, the extent of our analysis---dense gas within $10 h^{-1}$ ckpc of the disk center---serendipitously includes interacting systems which leads to high metallicity on the outskirts, thereby flattening the overall radial profile\footnote{We have repeated our analysis by considering a smaller disk, down to $3 h^{-1}$ ckpc, and arrived at the same qualitative conclusions.}. It appears that these interacting systems are impacting the line-of-sight velocity map too, making it less smooth and organised compared to that in ``A''. Therefore, disturbed kinematics could potentially be correlated with rapid interactions and therefore flatter gradients. Such interactions are more common at high-$z$ when the galaxies are undergoing more frequent mergers and violent starbursts and only last for a short time. Indeed, $\lesssim$100\,Myr after ``B'' the gradient is strongly negative again, as seen in the top panel of \autoref{fig:Zgrad_smooth}.

\subsection{Non-parametric characterization}
\label{sec:Zscat_res}
We investigate a non-parametric approach to quantify the distribution of metallicity within a certain distance from the center, which we describe in \autoref{sec:Zscat_method}. Our new approach focuses on the distribution of different metallicity values present within the disk, rather than \textit{where} in the disk the different metallicities occur.

The top panel of \autoref{fig:time_series_hurricane} shows the time evolution of the metallicity gradient. This is the same as the top panel of \autoref{fig:Zgrad_smooth}. We reproduce it here again for ease of comparison against the time evolution of other quantities derived from the non-parametric approach, as discussed in the following paragraphs. In each panel, the circles represent whole number redshifts, i.e., $z=0$, 1, 2, 3 and 4 from right to left.

In the middle panel of \autoref{fig:time_series_hurricane} we show the evolution of the median (brown) and inter-quartile range (IQR, blue) of the full metallicity distribution (see \autoref{fig:Zgrad_snap}). Both of these quantities are independent of geometric assumptions or fitting the data to a model. Therefore, the median and the IQR represent a non-parametric, quantitative description of the distribution of the mass-weighted gas-phase metallicity. \red{We do not see any significant \textit{overall} evolution from $z=4$ to $z=0$ in the IQR. The median metallicity shows a slight decline over this redshift range, discussed in detail in \autoref{sec:sf_response}.} However, both exhibit significant scatter, particularly \textit{downward} detours at several epochs throughout the lifetime of the galaxy.

The bottom panel of \autoref{fig:time_series_hurricane} shows the star formation rate of Hurricane as a function of time \citep{Lochhaas:2021aa}. We provide figures similar to \autoref{fig:time_series_hurricane}, but corresponding to the the other FOGGIE halos, in \autoref{sec:ap}.

\subsubsection{How does star-formation shape the metallicity distribution?}
\label{sec:sf_response}
\begin{figure*}
    \centering
    \includegraphics[width=0.99\linewidth]{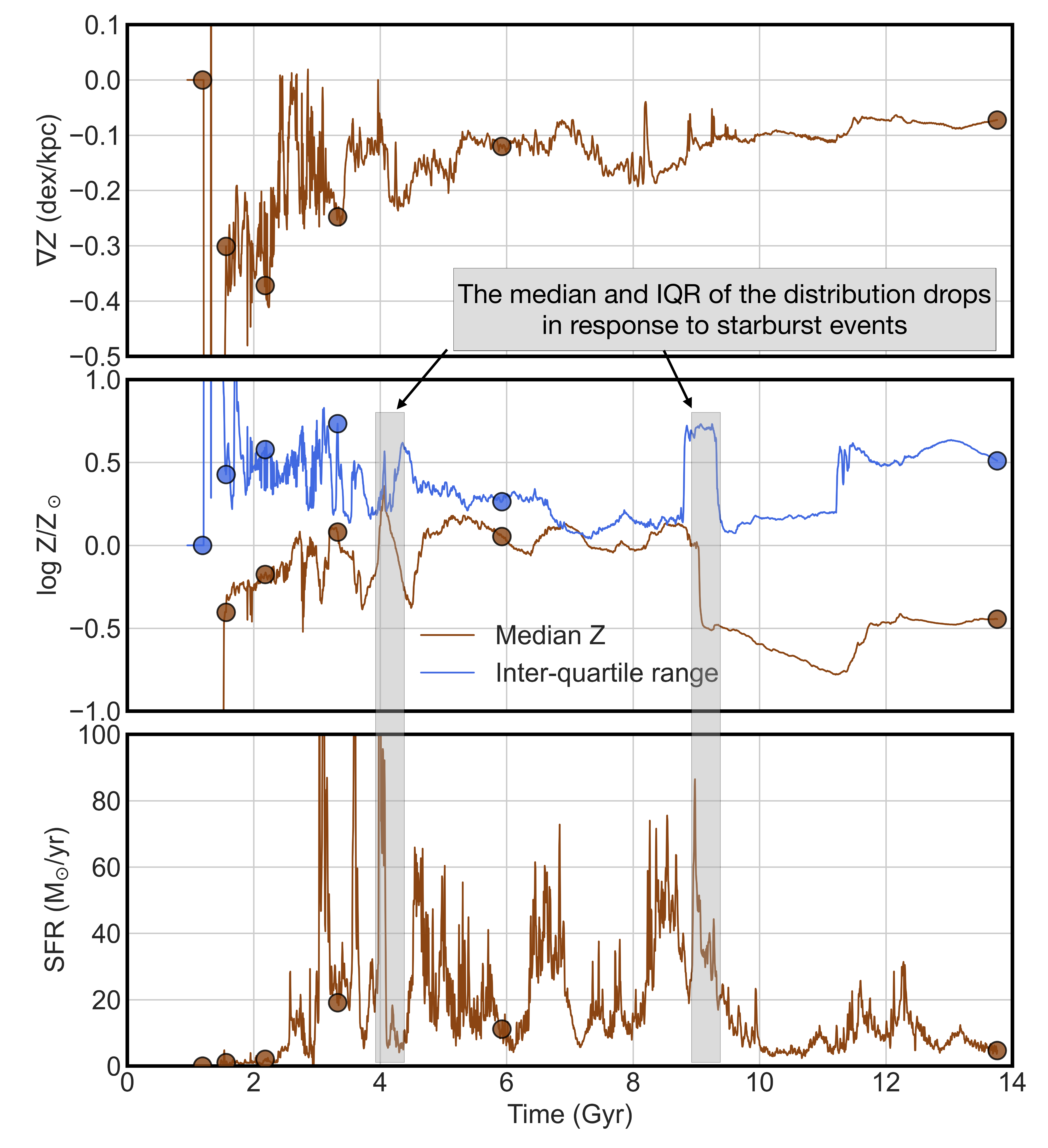}
    \caption{Each panel shows the time evolution of various quantities of Hurricane. Circles in each panel denote whole number redshifts $z=$ 0, 1, 2, 3 and 4 (from right to left). The \textit{first panel} shows metallicity gradient, the same as top panel of \autoref{fig:Zgrad_smooth}. The \textit{second panel} shows the median and inter-quartile range (in log-space) of the full distribution of the metallicity within 10 $h^{-1}$ ckpc, in brown and blue respectively. 
    The \textit{bottom panel} denotes the star formation rate history. The highlighted regions in gray show that the inter-quartile range of the metallicity distribution decreases in response to star formation episodes.}
    \label{fig:time_series_hurricane}
\end{figure*}

Our goal is to investigate the changes in the metallicity distribution in response to star formation. \red{We therefore compare the middle and the bottom panel of \autoref{fig:time_series_hurricane}, as well as similar figures corresponding to the other FOGGIE halos provided in \autoref{sec:ap}, in order to draw the following conclusions. Overall, the median metallicity (brown line in the middle panel) declines in the last 5--6 Myr when star-formation slows down. The magnitude of this decline ranges from $\sim0.1$ to $-0.5$ dex, depending on by \textit{how much} the galaxy's SFR declines, i.e., a sharper fall in the SFR is accompanied by a larger drop in median metallicity. For instance, Blizzard sees the sharpest drop in SFR---$\sim$100 \Msun/yr at $\sim$9 Gyr and $\lesssim$1 \Msun/yr afterwards (\autoref{fig:time_series_blizzard}). The median gradient drops steadily by $\sim$0.5 dex in the same span of time. The SFR of Tempest, on the other hand, drops from $\sim$10 \Msun/yr at $\sim$10 Gyr to $\lesssim$1 \Msun/yr thereafter. Consequently, the median metallicity drops by only $\sim$0.1 dex for Tempest over this time (\autoref{fig:time_series_tempest}). Maelstrom (\autoref{fig:time_series_maelstrom}) is an exception, where we see the median metallicity \textit{increase} steadily for the last $\sim$10 Gyr. Overall, the median of the metallicity distribution seems to tend to drop in absence of star-formation feedback, while every starburst boosts it up. The small but numerous starbursts in Maelstrom, right up to $z\sim0$ possibly explains the lack of drop in the metallicity.}

It appears that the width (IQR) of the metallicity distribution tends to decrease following prominent starburst events, as highlighted in the figure. Rapid star formation epochs are closely ($\sim$ few Myr) followed by supernovae events, which leads to deposition of metals as well as momentum on to the gas. \red{This sudden injection of metals leads to a narrower distribution (blue curve in the middle panel) and a momentary increase in median metallicity (brown curve). Thereafter, the momentum feedback expels metal-rich gas (from the vicinity of the starburst) before it can mix with the ISM, leading to a gradual decrease in median metallicity.} It takes the IQR $\sim2$\,Gyr to ``recover'' to its initial value at the start of the starburst, hinting at the timescale for mixing of metals throughout the disk.

It is worth noting that the radial gradient measurement (top panel) does not exhibit any strong response to the star formation history. A slight flattening (upward excursion) is occasionally seen in radial gradient due to mixing of gas following starbursts, but other factors, such as the rapid stochasticity in the gradient and interactions with nearby galaxies, wash out any reliable signature in response to starbursts, particularly at high redshift.

\section{What flattens or inverts the metallicity gradients?}
\label{sec:posgrad}
In this section we elaborate on the ubiquity of positive and flat metallicity gradients in observations, and their scarcity in simulations. We demonstrate that although FOGGIE galaxies sometimes exhibit positive gradients, it is a short-lived phenomenon---and does not occur sufficiently often to be able to explain the observations. We emphasise that the inability to reproduce long-lived positive gradients is common to all simulations, and therefore indicates a major gap in our understanding of chemical evolution of galaxies. 

\subsection{Comparison with literature}
\label{sec:lit_obs}

Comparison with existing observational studies shows tensions in cases where positive metallicity gradients have been observed. Observations of flat or positive metallicity gradients, particularly at $1<z<3$, is a robust finding given it has been reported by various studies using a multitude of techniques \citep{Wuyts:2016aa, Wang:2017aa, Wang:2019aa, Wang:2020aa, Carton:2018aa, Curti:2020aa, Simons:2021aa, Li:2022aa}. Recently \citet{Wang:2022aa} and \citet{Venturi:2024aa} pushed the redshift-boundary to $z\gtrsim3$ and $z\sim8$, respectively, using {\jwst} observations. \citet{Vallini:2024aa} measured metallicity gradients at $z\sim7$ based on ALMA observations. All of these recent measurements report flat/positive gradients all the way out to $z\sim8$. This is something that theoretical models need to be able to explain. Simulation-based studies (including this work) have attempted to do so, but without any conclusive results yet. \red{Note that directly comparing observed and simulated gradients implicitly assumes a flat mass-to-light ratio profile, which is a caveat we discuss further in \autoref{sec:caveats}.}

The evolution of metallicity gradients of FOGGIE galaxies over long timescales broadly agrees with existing simulations. Multiple theory-based studies---including MUGS \citep{Gibson:2013a}, FIRE-1 \citep{Ma:2017aa}, \red{FIRE-2 \citep{Bellardini:2021aa, Bellardini:2022aa, Graf:2024aa}} and Illustris \citep{Hemler:2021a}---have addressed this problem using a variety of different input physics. All these simulations have reported high-$z$ metallicity gradients to be usually negative, only sometimes flat, and almost never positive. \red{Moreover, there is a clear broad agreement between all these simulations at low-redshift.} This is consistent with FOGGIE as well. This broad agreement, in spite of the diversity among these simulation suites in numerical treatments of hydrodynamics, input physics, and analysis methods is significant. It might imply that the simulations are underpredicting the timescale of certain processes, such as mixing of metal rich gas accreted on the disk outskirts, leading to very short-lived positive gradients. Alternatively, it is possible that poor numerical resolution leads to inaccurate mixing of metals through numerical viscosity \citep{Wadsley:2008aa}.

While the instantaneous metallicity gradients of FOGGIE galaxies do sometimes achieve positive values,  these are the result of the highly stochastic behavior of the gradient during the evolution of a given galaxy (\autoref{sec:mergers}). The gradient quickly settles back to being negative on short timescales ($\sim$ 50 Myr). The FOGGIE galaxies do not develop a stable, long-lived positive metallicity gradient at any point during their lifetime. 

\subsection{Positive gradients from interactions}
\label{sec:mergers}
\begin{figure*}
    \centering
    \includegraphics[width=\linewidth]{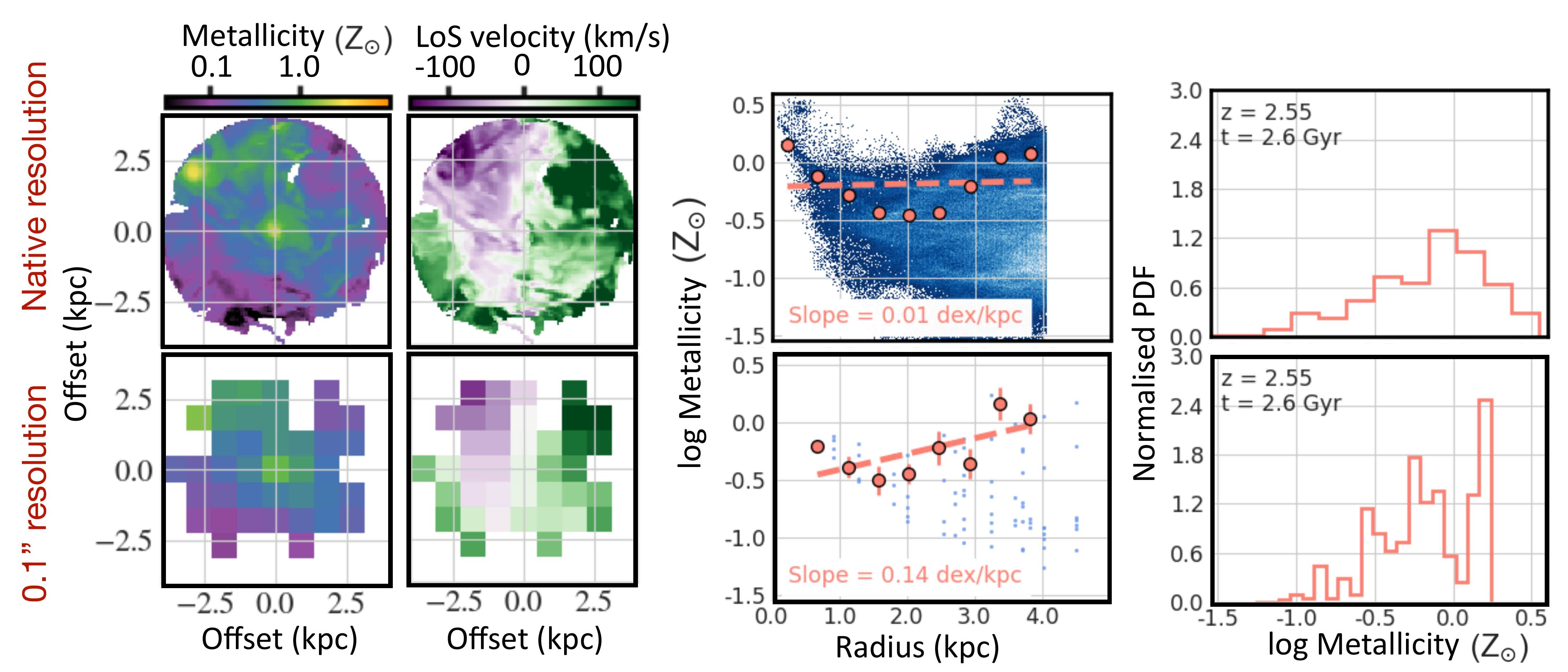}
    \caption{\textit{Top:} From left to right, the panels denote the projected metallicity map, line of sight velocity map, radial profile and full distribution of mass-weighted metallicity at a specific time stamp during the evolution of one of the FOGGIE halos, Hurricane. These panels are shown at the native resolution of the simulation. \textit{Bottom:} Same as above but after binning the quantities at $0.1^{\prime\prime}$/pixel resolution. The resolution comparison demonstrates the detectability of such positive gradients with new observatories e.g., {\jwst}.
    }
    \label{fig:Zgrad_pos}
\end{figure*}

\autoref{fig:Zgrad_pos} depicts one of the rare instances of positive metallicity being seen in the FOGGIE (Hurricane) simulations, and how it would appear at a lower spatial resolution. The quantities depicted are the same as in \autoref{fig:Zgrad_smooth}. While the top row depicts the quantities at the native FOGGIE resolution ($\sim 77$\,pc at $z\sim2.55$), the bottom row shows the same quantities at 0.1$^{\prime\prime}$ resolution ($\sim 0.8$\,kpc at $z\sim2.55$). We achieved the coarsening simply by re-binning the intrinsic 2-D projected map of each relevant quantity, without incorporating a full point spread function (PSF). This provides an \textit{approximate} expectation of how this simulated galaxy might appear to {\jwst}/NIRSpec.

The main takeaway from comparing the two rows of \autoref{fig:Zgrad_pos} is that the occurrence of positive gradients is independent of spatial resolution. Examining the higher resolution counterparts---particularly the top left panel---it is clear that the positive gradient is due to a close passage of a metal-rich satellite on the top left of the central disk. In the lower resolution version, however, it would be extremely challenging to distinguish between the two separate disks in the 2-D metallicity map. Hence, data at such a resolution would hint that the observed positive gradient is associated with a \textit{single} system, without accounting for any mergers/interactions. Turning to 2-D kinematic maps, even at low spatial resolution, might prove useful in distinguishing such systems. As an example, let us consider just the low resolution data, as shown in the bottom row of \autoref{fig:Zgrad_pos}. We see a positive metallicity gradient based on a 2-D metallicity map that does not give us any obvious hints about an interacting system. However, the departure from a smooth line-of-sight velocity map (dark purple in the second panel of bottom row) hints at disturbed kinematics, thus pointing us to the cause of the measured positive gradient.

These minor interactions at high-$z$ occur intermittently and only last for a short time ($\lesssim 100$\,Myr), following the generic behavior of minor mergers in $\Lambda$CDM structure formation. Although positive gradients are rare in FOGGIE simulations ($< 5\%$ of the time within $2 < z < 3$ and $< 1\%$ throughout cosmic time), whenever they do occur, it is due to a close passage of a merger, or interaction with a metal-rich companion. While it might be difficult to identify interacting systems in metallicity-space at the resolution of current space-based observations, the line-of-sight kinematics might provide useful data.

Finding flat/positive metallicity gradients in association with mergers and interactions is not unexpected. In some simulations, merger events can lead initially to \textit{steepening} of gradients, owing to delivery of less enriched gas to the disk outskirts \citep{Buck:2023aa} with subsequent mixing later settling the disk to \textit{flat} gradients \citep{Tissera:2016aa}. Indeed \citet{Tissera:2016aa} and \citet{Tissera:2022aa} reported both positive as well as strongly negative gradients for galaxies with disturbed morphology and/or undergoing interactions with a close neighbour. The evolutionary stage of the merging event dictates whether the gradient would be negative or positive.

\cite{Rosas-Guevara:202aa} showed that galaxies in the voids (as opposed to close to the large scale filaments) in the EAGLE simulations are less likely to harbour positive metallicity gradients. Our work with FOGGIE emphasises that environment and interactions play a key role in flattening or inverting the metallicity gradients.

\section{What drives the general evolution of the metallicity gradients?}
\label{sec:what_drives_grad}
Now we discuss the general evolution of the FOGGIE metallicity gradients in the context of other observational as well as simulation studies. In particular, we emphasise the lack of strong correlations between gradients and other global galaxy properties such as SFR, size, and circular velocity. We highlight the challenges in understanding the physics driving the gradients if we reduce each property to a single globally averaged value. We therefore propose to utilise the entire information embedded in the full 2D maps of metallicity and velocity, via a non-parametric approach.

\subsection{Redshift evolution of gradients}
\label{sec:grad_evol}

In terms of tracking the gradients of individual simulated galaxies across time, \cite{Ma:2017aa} find opposite trends compared to this work (FOGGIE) as well as \citet{Gibson:2013a} and \citet{Hemler:2021a}. Figure 9 of \citet{Ma:2017aa}, which shows the evolution of an individual halo of the FIRE-1 simulations, displays a rapid initial \textit{steepening} of the metallicity gradient early on, followed by little evolution in the past $\sim6$\,Gyr owing to a ``quiet'' phase of evolution, while both FOGGIE (\autoref{fig:Zgrad_smooth}) and MUGS \citep[Figure 1 of][]{Gibson:2013a} display a general \textit{flattening} with time. \citet{Hemler:2021a} reported predominantly negative metallicity gradients, with significant scatter, and an overall \textit{flattening} of gradients at low-$z$ in TNG50 simulations. This contrast can not entirely be explained by contrasting merger histories in the different simulations. While FOGGIE galaxies have been chosen to have fewer major mergers at $z\lesssim 2$, \citet{Ma:2017aa} also note the naturally decreasing merger rates in FIRE simulations at low-$z$. The similarity in merger history leads to the gradients not evolving significantly in the last few billion years. Therefore, having \textit{steepened} at earlier times the FIRE gradients stay steep at low-$z$ due to lack of mergers, whereas FOGGIE gradients generally stay weakly negative at low-$z$, having already \textit{flattened} by $z\sim1$. Such flattening can be attributed to an accretion-dominated phase at later times (see below). \citet{Hemler:2021a} note that mergers flatten the gradients at high-$z$, whereas AGN feedback drives the flattening of gradients at low-$z$ in their TNG50 simulations. \red{However, in the improved FIRE-2 simulations \citet{Bellardini:2021aa} \citep[see also][]{Bellardini:2022aa, Graf:2024aa} find a systematic \textit{steepening} of the gradient with time, which is at odds with the behavior of FOGGIE galaxies, although they do agree at low-redshift. The steepening in FIRE-2 is due to the disk settling down at lower-redshift, and consequently the ISM becoming less turbulent, hindering the radial mixing of metals. This qualitative difference between FIRE-2 and FOGGIE is likely due to different numerical techniques (AMR in FOGGIE vs Lagrangian in FIRE) and feedback prescriptions.}

In the analytic models of \citet{Sharda:2021aa} the gradient is primarily governed by four factors: (a) in-situ star formation, (b) accretion of pristine gas, (c) advection of gas, and (d) diffusion of gas. The FOGGIE gradient evolution qualitatively agrees with the \citet{Sharda:2021aa} models, such that both works hint at an \red{initial} steepening of gradient with time, followed by a ``turnaround'' towards flatter metallicity gradients at late times. \red{This ``turnaround'' is attributed to accretion of pristine gas as the galaxies become more massive, and consequent mixing of this pristine gas within the galaxy. We note that without this accretion-dominated phase, the inside-out quenching and inside-out disk assembly that is typical in MW-type galaxies \citep[e.g.,]{Tacchella:2016ab, Avila-Reese:2018aa} would lead to a build-up for negative metallicity gradient in the galaxy. \citet{Simons:2021aa} confirmed this behavior with the help of closed-box, toy models (see below).} 

Although we have higher absolute metallicity values than predicted by \citet{Sharda:2021aa}, the predicted \textit{scatter} in the gradients matches that of FOGGIE. They attributed the scatter to the ``yield reduction factor" $\phi_y$, which determines the fraction of newly produced metals that are mixed with the ISM rather than being ejected as galactic winds. However, their models hint that the scatter in the gradient is low at high-$z$ and increases around $z\sim0.5$ before decreasing again at lower-$z$. This is in contrast to what we observe with FOGGIE. The scatter in FOGGIE is largest at high-$z$ when the galaxy mass is lowest, and the scatter reduces with time and increasing stellar mass. This discrepancy could be attributed to the lack of merger events in the analytic prescription of \citet{Sharda:2021aa}, because mergers seem to be a key driver of the scatter and stochastic variations of the FOGGIE gradients at high-$z$. Moreover, the \citet{Sharda:2021aa} model assumes the ISM to be in steady-state equilibrium, which the FOGGIE simulations suggest, via high stochasticity, is not really being achieved.

The short timescale variations in gradients enabled by the unique high time-cadence of FOGGIE are seen in other cosmological simulations as well. \citet{Ma:2017aa} found that the scatter in gradient of a single galaxy in the FIRE simulations is statistically the same as that in the full ensemble, indicating that at high-$z$ gradients allow us just an instantaneous view rather than the accumulated history of the galaxy evolution. They also find that bursty star formation can lead to flatter gradients on short timescales.

Semi-analytic models of \citet{Fu:2013aa} show that radial gas flows have little effect on setting the gradient at late-times. Instead, the gradient is set by the fraction of metals ejected into the hot halo before being mixed with the cool ISM because it is this pre-enriched gas that accretes on to the outskirts of the disk at later times, flattening the gradient at low-redshift \citep{Werk:2011aa}. Indeed, \citet{Simons:2021aa} show with the help of a toy model in their Figure 11, that in the absence of metal re-distribution, the metallicity gradient of an isolated galaxy will rapidly ($\sim 10-100$\,Myr depending on mass) become strongly negative. Therefore, simulations must be able to be analyzed on timescales significantly shorter than this in order to capture the short time scale variations in metallicity distribution. With an output cadence of $\sim5$\,Myr, our result demonstrates  that in the presence of a realistic environment and feedback to re-distribute metals, gradients become flatter with time. This is in line with \citet{Gibson:2013a}, who found that enhanced feedback was important for reproducing ``flat'' metallicity gradients. The feedback in the FOGGIE simulations is strong enough to launch metal-rich outflows \citep{Lochhaas:2023aa}, and we do see an overall flattening of metallicity gradients.

\subsection{Correlation with other galaxy properties}
\label{sec:corr}

A significant observational finding is that the metallicity gradients show little-to-no correlation with other global galaxy properties. Figure 3 of \citet{Belfiore:2017aa} compared the observed radial gradients in low-$z$ galaxies with galaxy stellar mass and found a correlation, particularly at stellar masses below $\log{\rm M_*/M_{\odot}}<10$. However, at high-$z$ \citep[Figure 7 of][]{Simons:2021aa} the correlation is weak. Moreover, no meaningful correlation has been observed with other galaxy properties such as SFR, sSFR, velocity dispersion or galaxy size \citep[Figure 9 of][]{Simons:2021aa}. The lack of clues as to what might be governing the positive metallicity gradients makes it a challenging task to ascertain its causes. This is perhaps an indication that collapsing the 2-D spatial information of metallicity distribution to one value---the metallicity gradient---has already taught us all it can about galaxy evolution on its own, and going forward we need to combine it with the diagnostic power of the full spatial distribution.

Simulation-based studies have reported a diverse degree of correlation between metallicity gradient and stellar mass. Gradients in Illustris TNG50 galaxies show little correlation with mass (for an ensemble of galaxies at various cosmic times---not the individual galaxies tracked over cosmic time) in Figure 8 of \citet{Hemler:2021a}, while Figure 6 of \citet{Ma:2017aa} shows a weak correlation with mass where lower mass galaxies exhibit flatter gradients. This result of \citet{Ma:2017aa} is perhaps the closest available simulation-based counterpart to the high-redshift observations of \citep[][Figure 9]{Simons:2021aa}.

\red{\citet{Ma:2017aa} and \citet{Bellardini:2022aa} report an anti-correlation between radial gradient and the relative strength of circular velocity to velocity dispersion ($v_c$/$v_\sigma$), such that more rotationally supported galaxies tend to have steeper gradients, due to inefficient radial mixing.} \citet{Hemler:2021a}\red{, on the other hand,} find weak-to-no correlation between metallicity gradient ($\nabla Z$) and galaxy size, SFR or $v_c$. Such poor correlations have been reported by observational studies as well \citep{Simons:2021aa, Sharda:2021ac}. A common link between all such theoretical and observational studies is that they use globally averaged quantities that reduce complexity of chemical evolution in the galaxy disk. For instance, in a plot of $\nabla Z$ vs $v_c$ both quantities have been derived by reducing the spatial information to a global average, discarding a lot of potentially useful information along the way. We show in  \autoref{fig:Zgrad_smooth} that considering the 2-D map of metallicity and velocity can reveal correlations (e.g., disturbed kinematics correlated with positive gradients) that are lost when each galaxy is reduced to one value of $\nabla Z$ and $v_c$.

\subsection{Implications of a non-parametric approach}
\label{sec:nparam}

Given the difficulties with fitted symmetric gradients, we consider a non-parametric approach to quantifying the full distribution of metallicity. Unlike the radial gradient approach that assumes the existence of a well-behaved disk, which may not hold at high-$z$, the non-parametric approach is independent of geometric assumptions. Even if an observed galaxy's center is not well-constrained because of complex morphology, a non-parametric treatment is still feasible in such a scenario because all one needs is a distribution of observed metallicity values.

Moreover, spatially resolved metallicity studies at high-$z$ often involve lensed galaxies and therefore employ lens model reconstruction. Many high-$z$ galaxies have been observed via gravitational lensing, tying in the lens model uncertainties in the gradients \citep[e.g.,][]{Jones:2013ab, Jones:2015aa, Wang:2019aa, Wang:2020aa}. Our non-parametric approach is less susceptible to the systematic uncertainties of lens models because the metallicity distribution is independent of the physical geometry of the galactic disk. In the context of an observational study of a lensed galaxy, accurately estimating the center of the disk in the source-place is sensitive to the lensing model and associated uncertainties. The metallicity \textit{distribution} however, can be recovered without an accurate estimate of the disk center, therefore being less sensitive to lensing models.

The new approach, however, is not intended to \textit{replace} the conventional approach, given that there are scenarios that can be explored by only the latter. For instance, radial transport of metals on short timescales will have its imprint on the metallicity gradient (by changing the location of the `high' and `low' metallicity pockets along the radius) but will not affect the full distribution of metallicity values. Therefore, the new non-parametric approach may not be ideal for studying radial transport of metals. However, the non-parametric approach can serve as a good ``complement'' to the conventional gradient method, because of its ability to study metallicity distribution of galaxies with disturbed morphology.

\subsection{Breaks in radial metallicity profiles}
\label{sec:break}

We observe a ``break" in the radial metallicity profile in the FOGGIE galaxies throughout most of their lifetimes. The FOGGIE galaxies have a relatively steep, high-metallicity radial profile in the central part, out to $\approx$ 4--5 kpc, followed by a relatively flat metallicity profile. At times (typically $z \gtrsim 0.5$), we observe a sharp drop in metallicity between the inner steep profile and the outer flat profile (see the middle panel of \autoref{fig:Zgrad_snap} as an example). This flat metallicity profile on the outskirts arises from a cold, outer gas disk that is not forming stars. This is in agreement with \citet{Garcia:2023aa} who studied the evolution of the location of the break in Illustris-TNG simulations and concluded that a typical disk in their simulations is formed of two parts---an inner star formation dominated part with a steep gradient, and an outer mixing dominated part with a flat metallicity profile. Multiple observational studies have reported a similar break in the radial gas-phase metallicity profile \citep[e.g.,][]{Sanchez-Menguiano:2018aa}. Although limited spatial resolution could be a potential contributor to the ``flattening" of observed metallicity profiles \citep[e.g.][]{Yuan:2013ab, Carton:2018aa, Acharyya:2020aa}, the ``flattened" outskirts have recently been observed, even with good spatial resolution. This calls for an explanation of possible reasons behind the flat, low-metallicity outskirts such as those observed in the FOGGIE simulations. 

In FOGGIE galaxies the sharp drop in metallicity from the inner to the outer region, when it occurs, is at least partially due to misalignment of the angular momentum vector of the inner and outer gas disks \red{\citep{Simons:2024aa}}. Such a misalignment acts as a barrier to efficient mixing of gas across the inner and outer disks and therefore hinders metal transport. The inner disk continues to be metal enriched owing to local star formation whereas the non-star-forming outer disk remains metal poor. With time, the inner and outer disk orientations align, which eventually leads to one continuous, co-rotating gas disk. Based on animations of gas density projections of the disks of FOGGIE galaxies, this is a slow process and takes $\sim$ 1 Gyr. But once the inner and outer disks are merged into a contiguous structure, it facilitates metal mixing across the inter-disk boundary and makes the radial metallicity profile smoother, and consequently, the gradient flatter.

\subsection{Galaxy-to-galaxy scatter vs time evolution scatter}
\label{sec:scatter}
Individual FOGGIE galaxies exhibit significant scatter in the metallicity gradient, particularly at high-$z$. Following \citet{Simons:2021aa} we compared the scatter in the gradient as a function of galaxy stellar mass. The resulting relation was considerably noisy, though, overall we found the scatter in the gradient to be decreasing with stellar mass, in agreement with \citet{Simons:2021aa}. 

The inter-galaxy variation of metallicity gradient between the six FOGGIE galaxies is of the same order as the variation seen in a given halo's gradient during its lifetime. Note that the substantial diversity in simulated gradients is not unique to FOGGIE; \citet{Ma:2017aa} and \citet{Hemler:2021a} predict large diversity in gradients as well. With such large scatter in the metallicity gradient evolution, it is challenging to interpret the `instantaneous' observations of gas phase metallicity gradients. This is particularly true for high-$z$ observations with {\jwst} because the stochasticity is larger at high-$z$, when the `disk' of the galaxy has not settled. FOGGIE galaxies spend $\sim$30-50\% of the time up to $z > 1$ more than the typical observational uncertainty ($\sigma\approx$0.03 dex/kpc) away from the mean behavior (see \autoref{fig:Zgrad_smooth}). As observations become more precise in the future, and the typical observational uncertainty decreases, the fraction described above will only increase. We conclude that it would be even less likely for observations to capture the overall evolution in metallicity gradient by measuring an instantaneous gradient.

\red{The scatter in metallicity gradients is highly dependent on the method of measuring the gradient, the metallicity diagnostics used \citep{Kewley:2008aa, Poetrodjojo:2021aa} and whether the gradients have been normalized to a scale-radius. Indeed \citet{Sanchez-Menguiano:2018aa} demonstrated that upon normalizing the gradient to the effective radii, local galaxies have similar gradients, i.e., the scatter reduces. We have attempted to compute normalized gradients for the FOGGIE galaxies, by using the stellar half-mass radius as the effective radius. This led the absolute gradient values in dex/r$_e$ units to become steeper than the pre-normalized values in dex/kpc units (as expected, given the multiplication with r$_e$). However, the overall galaxy-to-galaxy scatter or the inter-galaxy scatter did not change significantly upon normalization.}

\section{Caveats and challenges}
\label{sec:caveats}
We now briefly highlight the several challenges in studying metallicity gradient at high-$z$, both from an observational as well as a simulation standpoint, and discuss the relevant caveats specific to the FOGGIE simulations.

\paragraph{Uncertainties in metallicity diagnostics:} 
Simulations and observations measure metallicity gradients in inherently different ways---one uses intrinsic values accessible in the simulations while the other relies on accurately diagnosing the starlight processed by the (often dusty) ISM via empirical and theoretical models. Therefore, direct comparison between simulated and observed metallicity gradients should be interpreted with caution. It is worth noting that unlike the simulations, the observed gradients involve varied metallicity diagnostics, which in turn depend on diverse photoionization models, leading to a  scatter in the observed gradients that may be larger than in the true population \citep{Kewley:2008aa}.

\paragraph{Challenges in defining the disk center:} 
Another reason to be cautious about interpreting fitted linear gradients is that they refer to a center that is not known precisely. At high-$z$, the morphology is more likely to be irregular and clumpy rather than smooth and disk-like, making it difficult to reliably measure a disk center, and consequently, a reliable radial gradient. In the simulations we have access to the full dark matter distribution which can help find the center. Observations, however, typically do not contain such information and are therefore reliant on the surface brightness map to deduce the center, which is affected by the chaotic morphology of the disk. Metallicity distributions that routinely depart from clean linear gradients, and high time variability of the gradient at high-$z$, warrant a search for a different approach to quantify the spatial distribution of metallicity, as discussed in \autoref{sec:Zscat_res}. 

\red{\paragraph{Mass-to-light ratio profile:}
Observed metallicity gradients are derived from emission line fluxes, and consequently, are light-weighted quantities. Simulated gradients, on the other hand, are typically (including in this work) mass-weighted quantities. Therefore, reliable comparisons between observed and simulated gradients would depend on the implicit assumption that the mass-to-light (M/L) ratio radial profile is flat. However, this is not always the case. Studies have demonstrated a radially declining M/L ratio, i.e., a negative color gradient \citep[e.g.,][]{Suess:2019aa, Suess:2019ab, Avila-Reese:2023aa}. A negative M/L ratio profile would imply simulated gradients preferentially have greater contribution from the central region of galaxies, compared to observed gradients. This would typically lead to steeper simulated gradients than observed ones, as seen in \autoref{fig:Zlit}, but still would not make an intrinsically negative gradient appear positive in observations. \citet{Avila-Reese:2023aa} show that the M/L gradient becomes steeper with redshift for low-$z$ ($z\lesssim0.2$) but becomes flatter with redshift at higher-$z$, and overall the M/L gradient is flatter at lower stellar mass. A flatter M/L ratio profile, would nominally make simulated gradients less steep compared to their observed counterparts. In the FOGGIE simulations (and most simulations in general) a given galaxy has a lower mass at higher $z$, which is the flatter M/L gradient parameter space. Therefore, the effects of M/L ratio gradient should be less pronounced at high-$z$ when comparing evolution of observed and simulated metallicity gradients.}

\paragraph{FOGGIE merger history:} 
The initial bursty star formation episodes followed by quieter phases of star formation history, as seen in the FOGGIE galaxies, could potentially be a selection effect. The FOGGIE galaxies are selected to \textit{not} have major mergers after $z\sim2$, and therefore have fewer external stimuli for rejuvenating star formation at lower redshifts. This works in favour of our current analysis because this gives us an ideal laboratory for studying the evolution of metallicity gradients both during bursty (high-$z$) as well as at quiet (low-$z$) star formation episodes.

\paragraph{Star formation \& supernova feedback in FOGGIE:} 
The supernova feedback prescription in current FOGGIE simulations is overall underpowered, and therefore unable to expel enough metal-rich gas. Moreover, the star formation scheme employed is slightly over- and under-efficient in more and less dense regions respectively. For a detailed description of the central concentration of star formation in FOGGIE, see Section 2.7 of \citet{Wright:2024aa}. The combination of these effects leads to too many metals being locked up in the ISM, resulting in too high \textit{absolute} values of metallicity. 
We consider that the centrally concentrated star formation may lead to a systematic impact on the metallicity gradient. However, the gradients reproduced in our simulations are consistent with that of other state-of-the-art cosmological simulations, wherein we all fail to reproduce the observations. Therefore the issues with the star formation prescription does not impact our key results. The upcoming generation of FOGGIE simulations will have improved feedback models. Although the absolute values of metallicities in the current FOGGIE simulations may not be perfect, the current simulations are sufficient for comparing \textit{relative} metallicity quantities, such as the gradient.

\paragraph{Numerical mixing:} 
Achieving accurate ``mixing'' of gas in simulations through numerical methods is a challenging problem \citep{Wadsley:2008aa}, and yet it impacts the metal mixing and therefore, metallicity distributions. Particle-based simulations often employ a sub-grid ``mixing model''. This necessitates calibration with one or more parameters and therefore introduces potential sources of uncertainties. On the other hand, grid-based simulations, such as FOGGIE, often suffer from the issue of over-mixing owing to the numerical recipes used. However, the speed at which metals can diffuse from their point of origin to larger radii is still limited by physical timescales and processes. Therefore we believe that the mixing problem does not impact our main conclusions.

\paragraph{Challenges of forward modeling:} 
A potential next step to better understand this discrepancy is to produce synthetic Integral Field Spectroscopy (IFS) datacubes from cosmological simulations, with realistic beam smearing and noise. Producing synthetic IFS datacubes is generally a powerful tool, but requires a complete understanding of the underlying ionization conditions that power the \hii\ regions and therefore govern the emitted light. In the absence of a detailed ionization model, which is challenging to do at the resolutions achieved in our cosmological simulations, this approach would require significant assumptions regarding the sub-grid physics. Therefore, even after accounting for beam smearing, noise, and metallicity diagnositcs, some systematic uncertainties stemming from the internal ionization structure of \hii\ regions, and reliability of the diagnostics themselves, will continue to remain and are very challenging to remedy entirely. The fact that no galaxy simulation is yet able to reproduce the occurrence of observed positive and flat metallicity gradients (discussed in detail in \autoref{sec:Zgrad_res}) hints that we might still be missing something fundamental in our models. Therefore, we believe that the forward modeling of existing inadequate simulations, compounded by sub-grid physics assumptions, will not help us understand the discrepancy between observed and simulated gradients. We therefore consider it of higher priority to investigate what underlying physics in the simulations can reproduce the observed positive gradients.  

\section{Summary \& Conclusions}
\label{sec:sum}

Spatially resolved gas-phase metallicity studies will imminently be ubiquitous, especially now that {\jwst} is performing beyond expectations. We expect several forthcoming observational studies to target gravitationally lensed high-redshift galaxies in order to perform spatially resolved studies \citep[e.g.][]{Wang:2022aa}. As and when more such observational measurements become available we will need physical models to help us interpret observations.

We use the FOGGIE simulations---hydrodynamic, zoom-in cosmological simulations with high spatial resolution ($\sim 270$ comoving pc in the ISM) and high output cadence ($\sim 5$ Myr)---to study the evolution of the spatial distribution of gas phase metallicity. Capitalizing on our higher cadence of outputs, we show that gas-phase metallicity gradients of FOGGIE galaxies display substantial variations that look stochastic and can exhibit big swings over short time intervals. Consistent with other simulations, including MUGS \citep{Gibson:2013a}, FIRE \citep{Ma:2017aa}, and Illustris TNG50 \citep{Hemler:2021a}, we find that gradients are usually negative, sometimes flat, and rarely positive. The short-lived positive gradients in FOGGIE result from minor interactions leading to a higher-metallicity patch on the outskirts of the disk, not from a ``real'' positive gradient in the disk itself. We find that non-parametric measures of metal distribution, and 2-D maps of velocity, are more responsive to short-time evolution than simple fitted gradients. This approach has other advantages for lensed galaxy studies (see \autoref{sec:nparam}).

Our main conclusions are as follows.
\begin{itemize}
    \item We corroborate the tension between observed and simulated metallicity gradients in terms of the lack of positive gradients in simulations (\autoref{fig:Zlit}).
    
    \item We find that the conventional radial metallicity gradient measurement exhibits large \red{($\sim 1$--2 dex)} scatter on short timescales, particularly at high-redshift, when the galaxy disk might not have completely settled. The magnitude of the scatter is of the same order as the galaxy-to-galaxy variation, thereby making it hard to disentangle time variability and inter-galaxy variation as the driver of the scatter in global relations such as the mass-metallicity relation (MZR) (\autoref{fig:Zgrad_smooth}).
    
    \item However, the overall gradient evolution of all FOGGIE halos exhibit a general flattening over large timescales. This general behavior can be significantly different from the instantaneous gradient at any given time.
    
    \item In fact, FOGGIE halos spend $\sim 30$--50 \% ($\sim 10$--25 \%) of their lifetime more than the typical observational uncertainty away from the mean behavior in gradient evolution up to $z=1$ ($z=0$). Studies with lower cadence outputs would likely miss such small scale variability and therefore be unable to capture $\sim 30$--50 \% of the gradient evolution at high-redshift (\autoref{fig:Zgrad_smooth}).
    
    \item We propose a non-parametric approach of quantifying the metallicity distribution---by characterizing the full distribution of metallicities in the galaxy disk, without any geometric assumptions. We demonstrate that the median and the IQR of such a distribution is more stable at high-redshift, even when the disk may not have fully formed yet. These metrics, particularly the IQR, exhibit a strong response to star formation and stellar feedback (\autoref{fig:time_series_hurricane}).

    \item We demonstrate that, although rarely occurring, positive gradients in the FOGGIE simulations result from closely interacting or merging systems, which can be challenging to distinguish at the spatial resolution of current space-based observations (\autoref{fig:Zgrad_pos}).
    
\end{itemize}

Our work will pave the way for, and help interpret, upcoming {\jwst} high-redshift metallicity distribution studies.

\section*{Acknowledgments}
AA's efforts for this work were supported by NSF-AST 1910414 and HST AR \#16151. \red{AA is also supported by European Union--NextGenerationEU RFF M4C2 1.1 PRIN 2022 project 2022ZSL4BL INSIGHT.} RA, CL, AA, and MSP were supported in part by NASA via an Astrophysics Theory Program grant 80NSSC18K1105. RA and CL also acknowledge financial support from the STScI Director’s Discretionary Research Fund (DDRF). RA acknowledges funding by the European Research Council through ERC-AdG SPECMAP-CGM, GA 101020943. BWO acknowledges support from NSF grants \#1908109 and \#2106575 and NASA ATP grants NNX15AP39G and 80NSSC18K1105. JT and ACW acknowledge support from the \textit{Nancy Grace Roman Space Telescope} Project, under the Milky Way Science Investigation Team. RA's efforts for this work were additionally supported by HST AR \#15012 and HST GO \#16730. EHL acknowledges support from the 2023 STScI Space Astronomy Summer Program. \red{The authors acknowledge the valuable inputs from Dr. Andrew Wetzel and Mr. Zihao Li which made the presentation of the conclusions in this paper more thorough.} Computations described in this work were performed using the publicly-available \textsc{Enzo} code (\href{http://enzo-project.org}{http://enzo-project.org}), which is the product of a collaborative effort of many independent scientists from numerous institutions around the world. Their commitment to open science has helped make this work possible. The 
python packages {\sc matplotlib} \citep{matplotlib2007}, {\sc numpy} \citep{numpy2011}, \textsc{scipy} \citep{scipy2020}, {\sc yt} \citep{yt2011}, {\sc datashader} \citep{datashader2022}, 
and {\sc Astropy} \citep{astropy2013,astropy2018,astropy2022} were all used in parts of this analysis.

\appendix
\section{Appendix}
\label{sec:ap}
\renewcommand\thefigure{\thesection\arabic{figure}}  
\setcounter{figure}{0}

In the main body of the paper we focused our analysis on one FOGGIE halo---Hurricane. In this section we present our results for the other five FOGGIE halos---Tempest, Maelstrom, Squall, Blizzard and Cyclone. Our main conclusions qualitatively hold true for all these halos.

\begin{figure*}
    \centering
    \includegraphics[page=1,width=\linewidth]{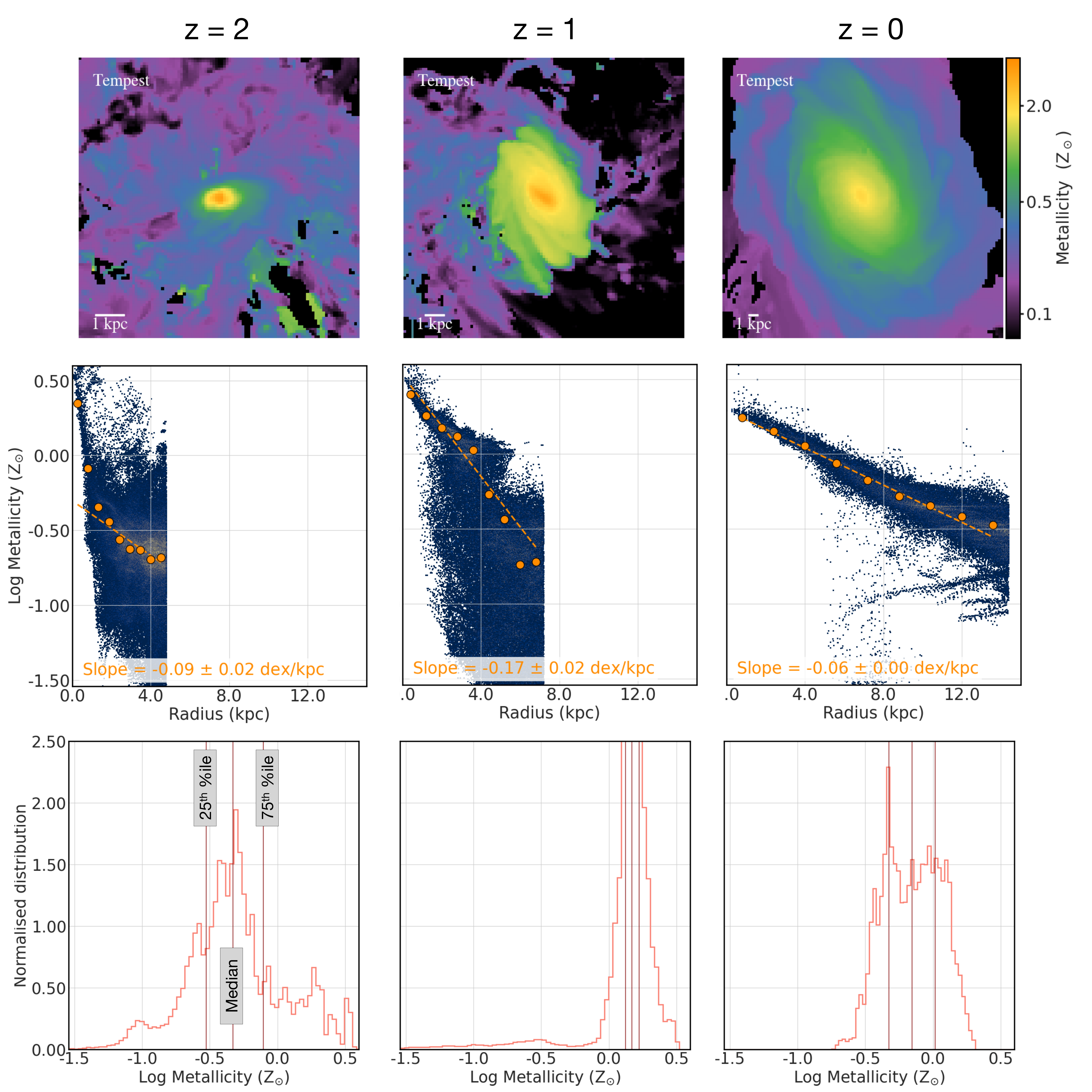}
    \caption{Similar to \autoref{fig:Zgrad_snap}, depicting the mass-weighted projected metallicity (\textit{top row}), radial metallicity profile (\textit{middle row}) and metallicity distribution (\textit{bottom row}) of all gas cells of the halo Tempest, within 10~$h^{-1}$ comoving kpc (ckpc) of the center, and above a certain density threshold (see \autoref{sec:disk_criteria}). The columns denote three different redshifts: $z=2$ (\textit{left}), $z=1$ (\textit{middle}) and $z=0$ (\textit{right}). Limiting our analysis to within 10~$h^{-1}$ ckpc results in a different \textit{physical} extent of our analysis at each epoch. This is demonstrated by (1) the scale bar in the top row (depicting 1 \textit{physical} kpc) being a different length in each panel, and (2) the extent of the radial fit in the middle row being different in each panel.}
    \label{fig:Zgrad_snap_tempest}
\end{figure*}

\begin{figure*}
    \centering
    \includegraphics[page=2,width=\linewidth]{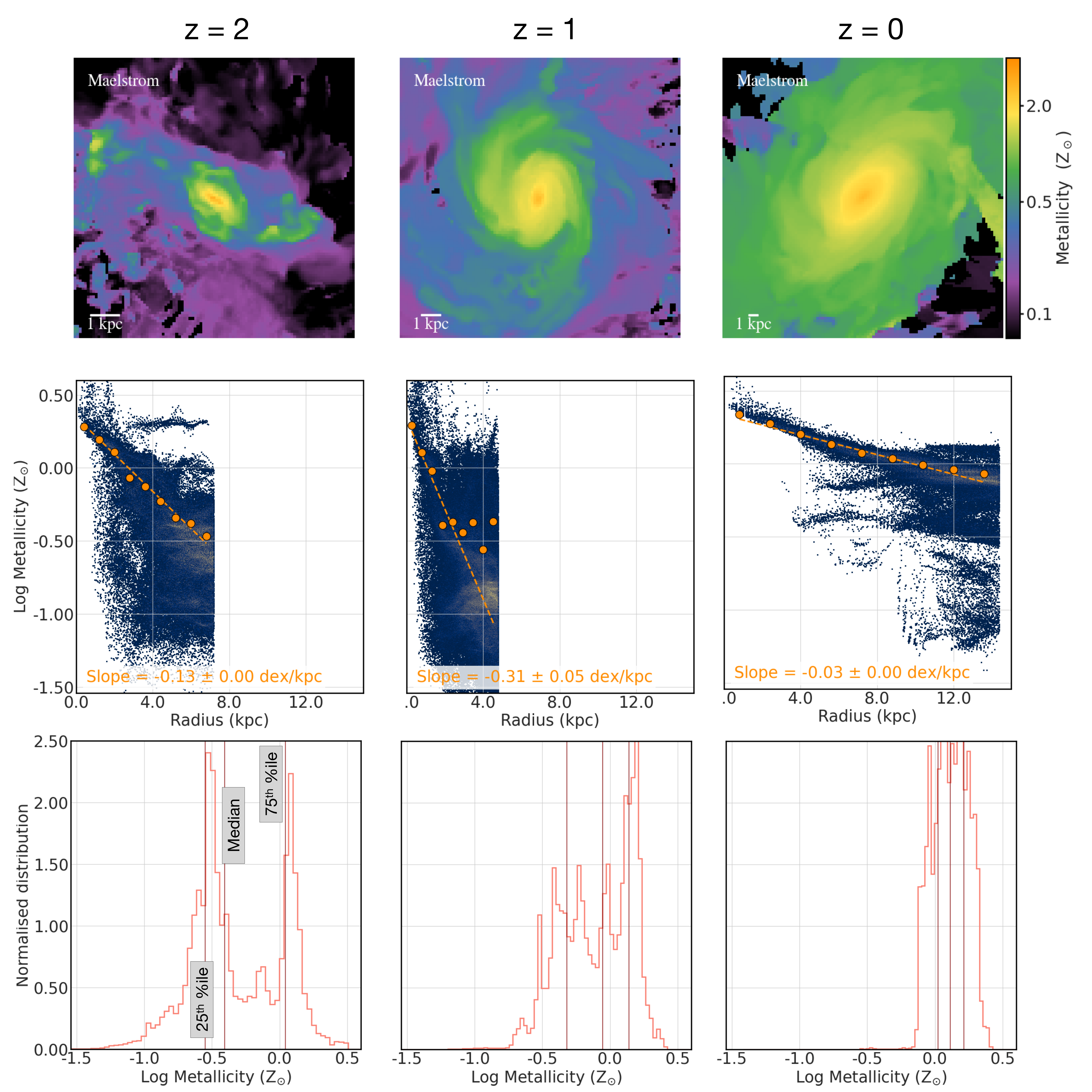}
    \caption{Similar to \autoref{fig:Zgrad_snap_tempest} but for another FOGGIE halo---Maelstrom.}
    \label{fig:Zgrad_snap_maelstrom}
\end{figure*}

\begin{figure*}
    \centering
    \includegraphics[page=3,width=\linewidth]{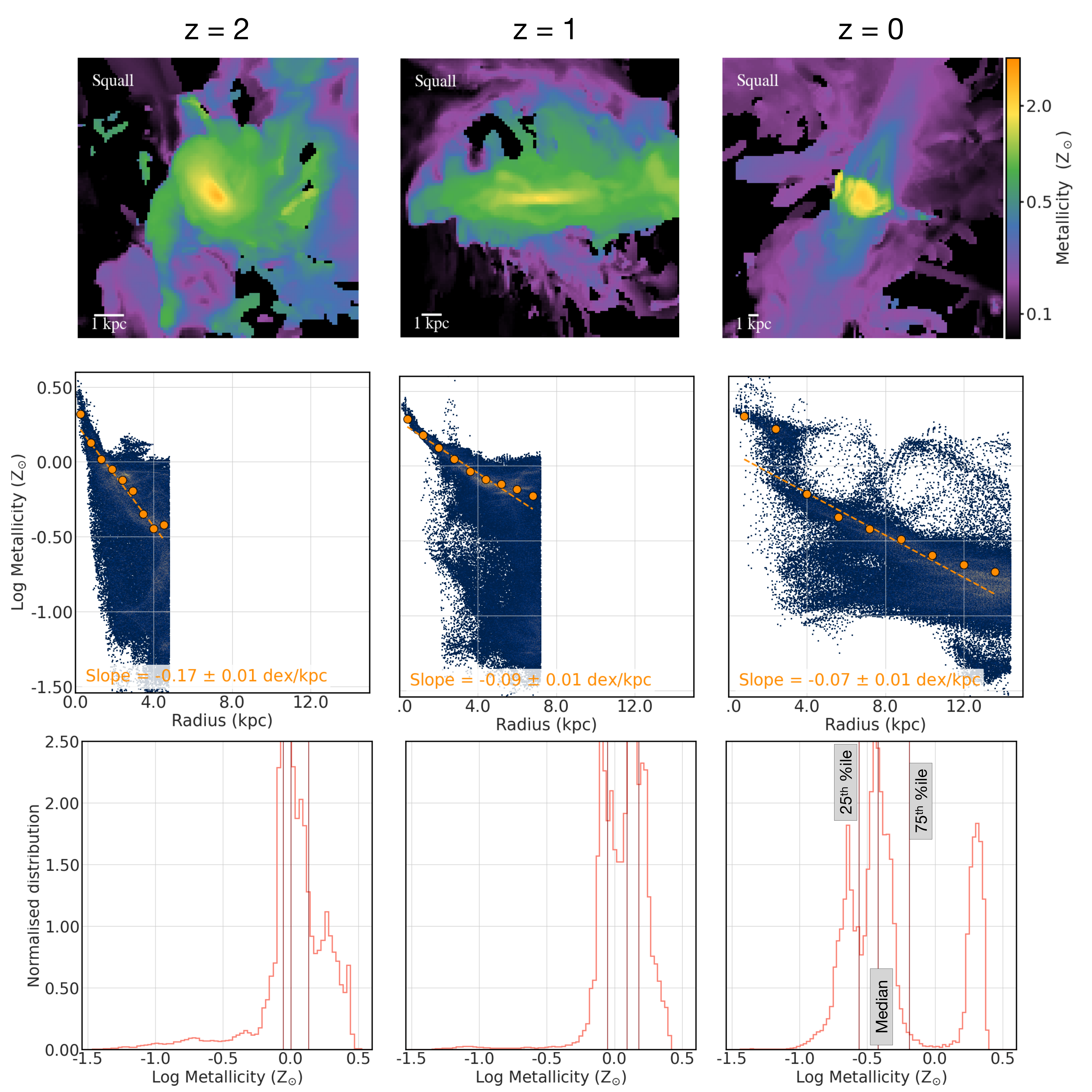}
    \caption{Similar to \autoref{fig:Zgrad_snap_tempest} but for another FOGGIE halo---Squall.}
    \label{fig:Zgrad_snap_squall}
\end{figure*}

\begin{figure*}
    \centering
    \includegraphics[page=4,width=\linewidth]{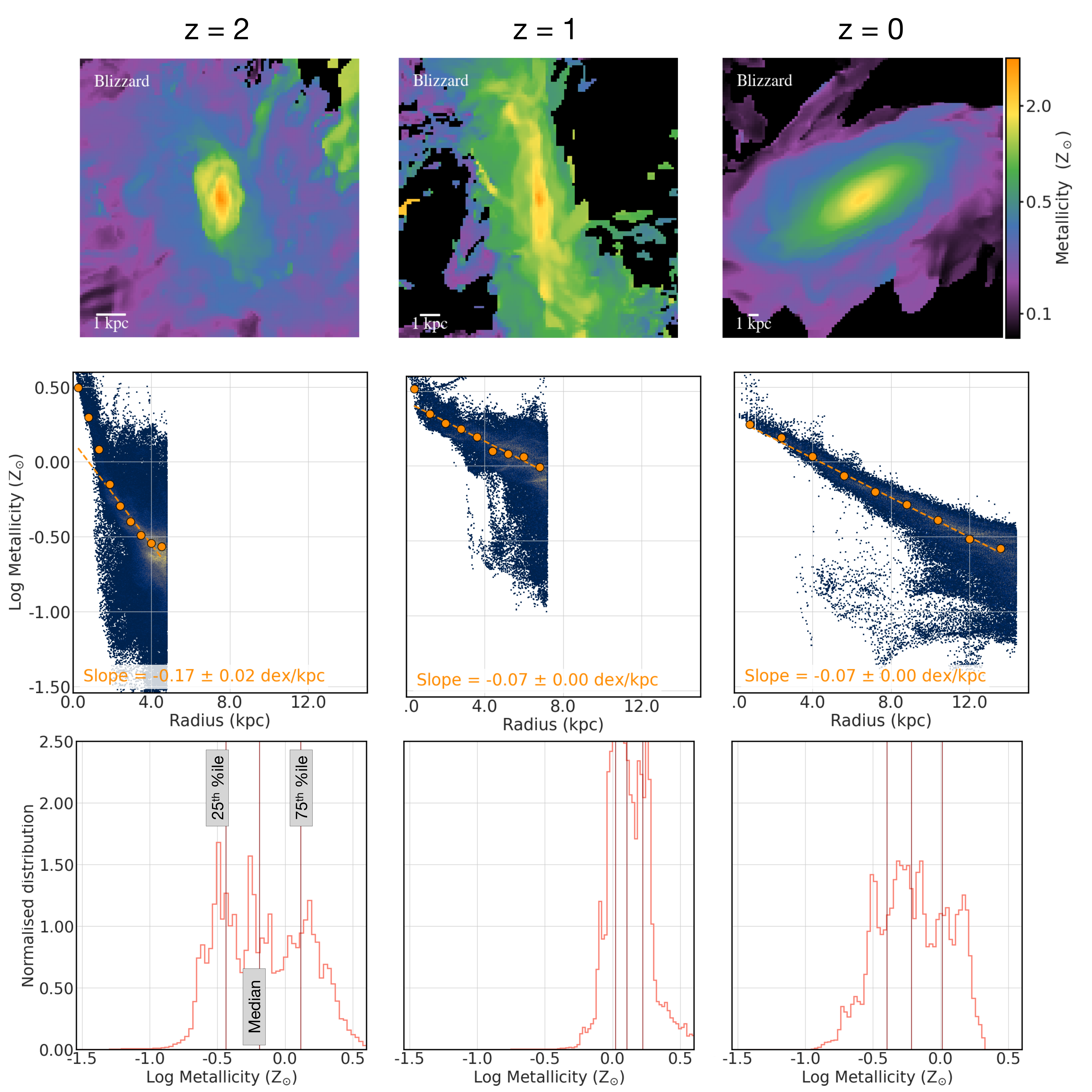}
    \caption{Similar to \autoref{fig:Zgrad_snap_tempest} but for another FOGGIE halo---Blizzard.}
    \label{fig:Zgrad_snap_blizzard}
\end{figure*}

\begin{figure*}
    \centering
    \includegraphics[page=5,width=\linewidth]{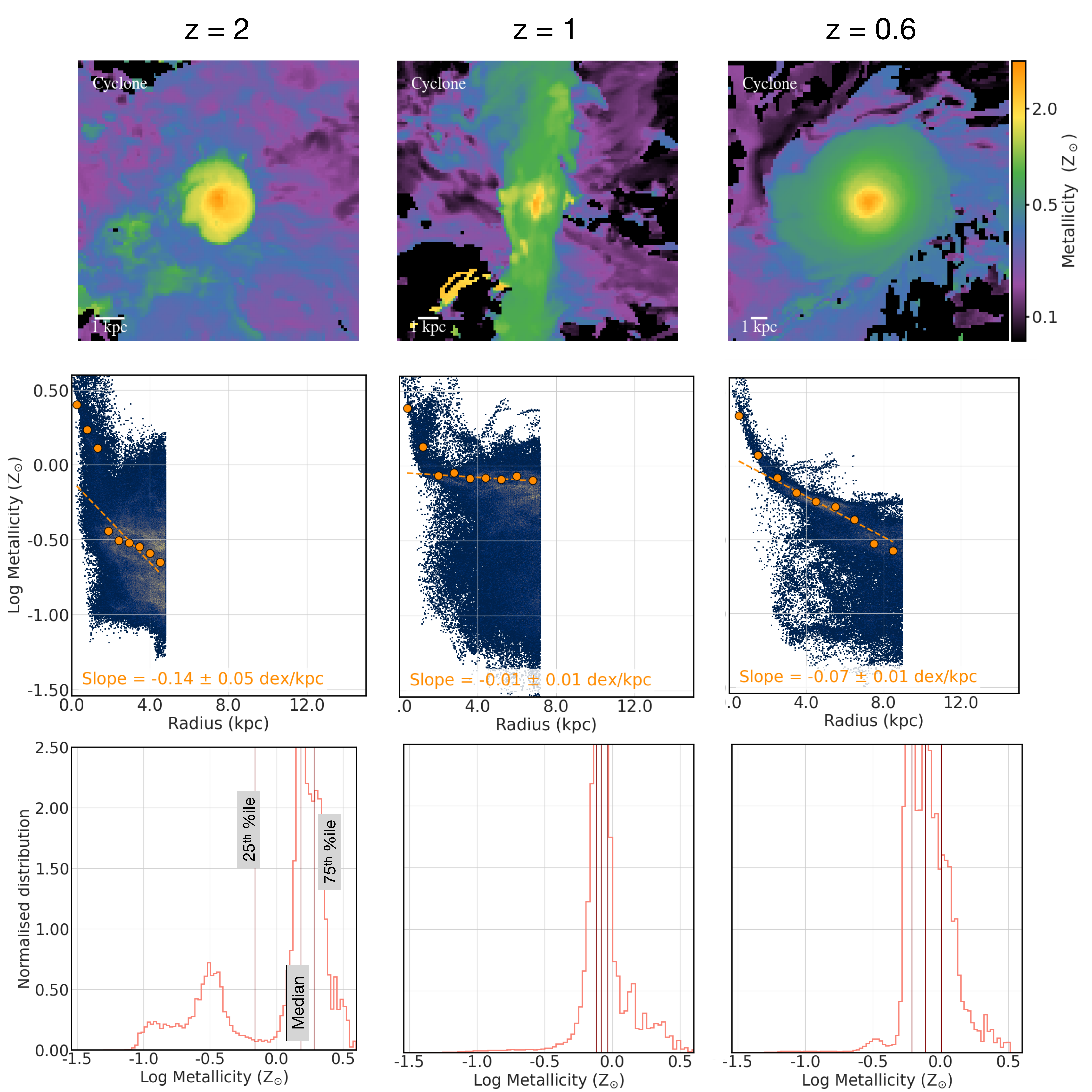}
    \caption{Similar to \autoref{fig:Zgrad_snap_tempest} but for another FOGGIE halo---Cyclone. This FOGGIE halo had been simulated only up to $z=0.5$ at the time of this analysis. In this case the rightmost column corresponds to $z=0.6$ (instead of $z=0$).}
    \label{fig:Zgrad_snap_cyclone}
\end{figure*}

\begin{figure*}
    \centering
    \includegraphics[page=6,width=\linewidth]{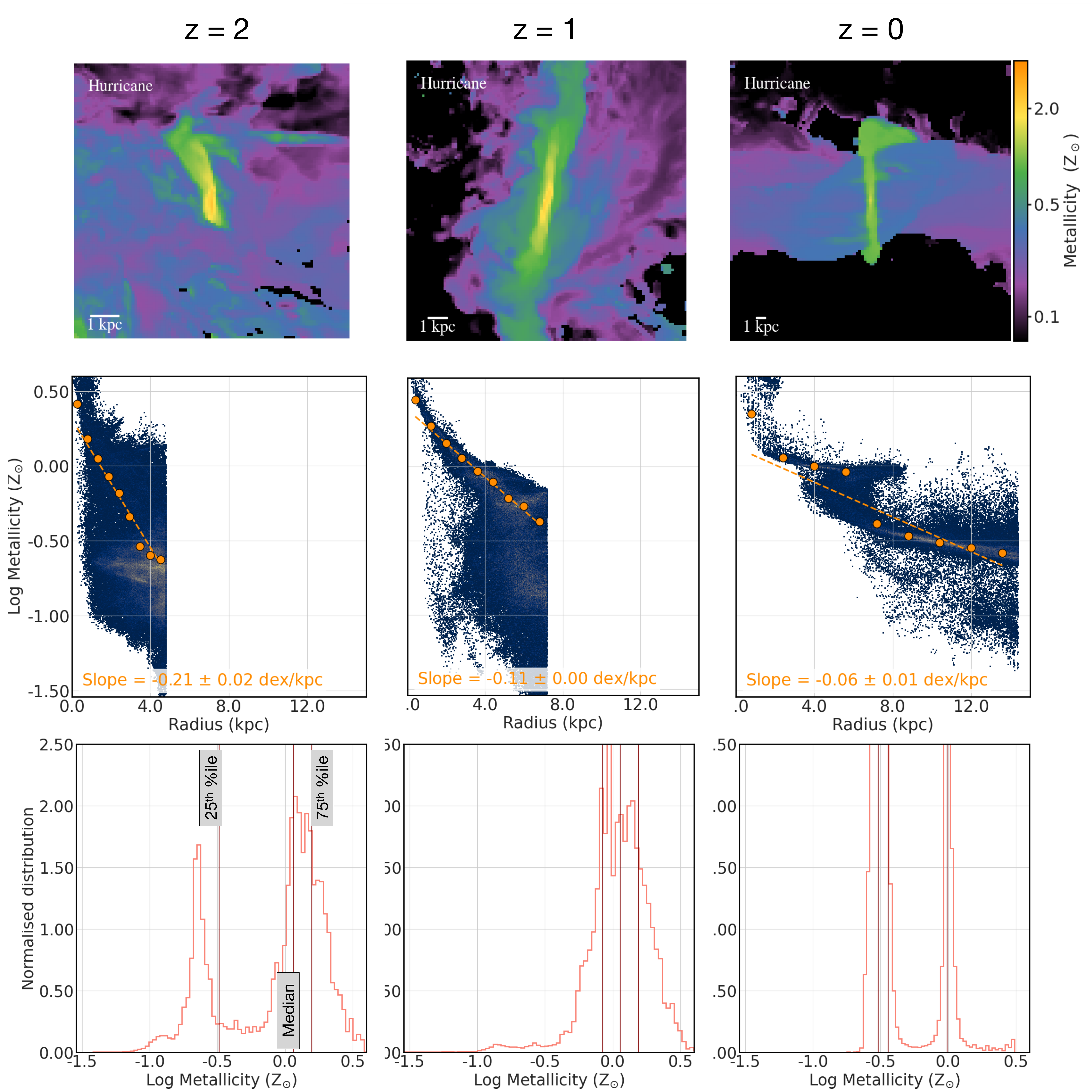}
    \caption{Similar to \autoref{fig:Zgrad_snap_tempest} but for another FOGGIE halo---Hurricane.}
    \label{fig:Zgrad_snap_hurricane}
\end{figure*}

In Figures \ref{fig:Zgrad_snap_tempest}-\ref{fig:Zgrad_snap_hurricane} we present snapshots of mass-weighted metallicity map, radial profile and distribution at redshifts $z=$2, 1, and 0 for each halo. These figures are similar to \autoref{fig:Zgrad_snap} in the main text.

\begin{figure*}
    \centering
    \includegraphics[page=1,width=\linewidth]{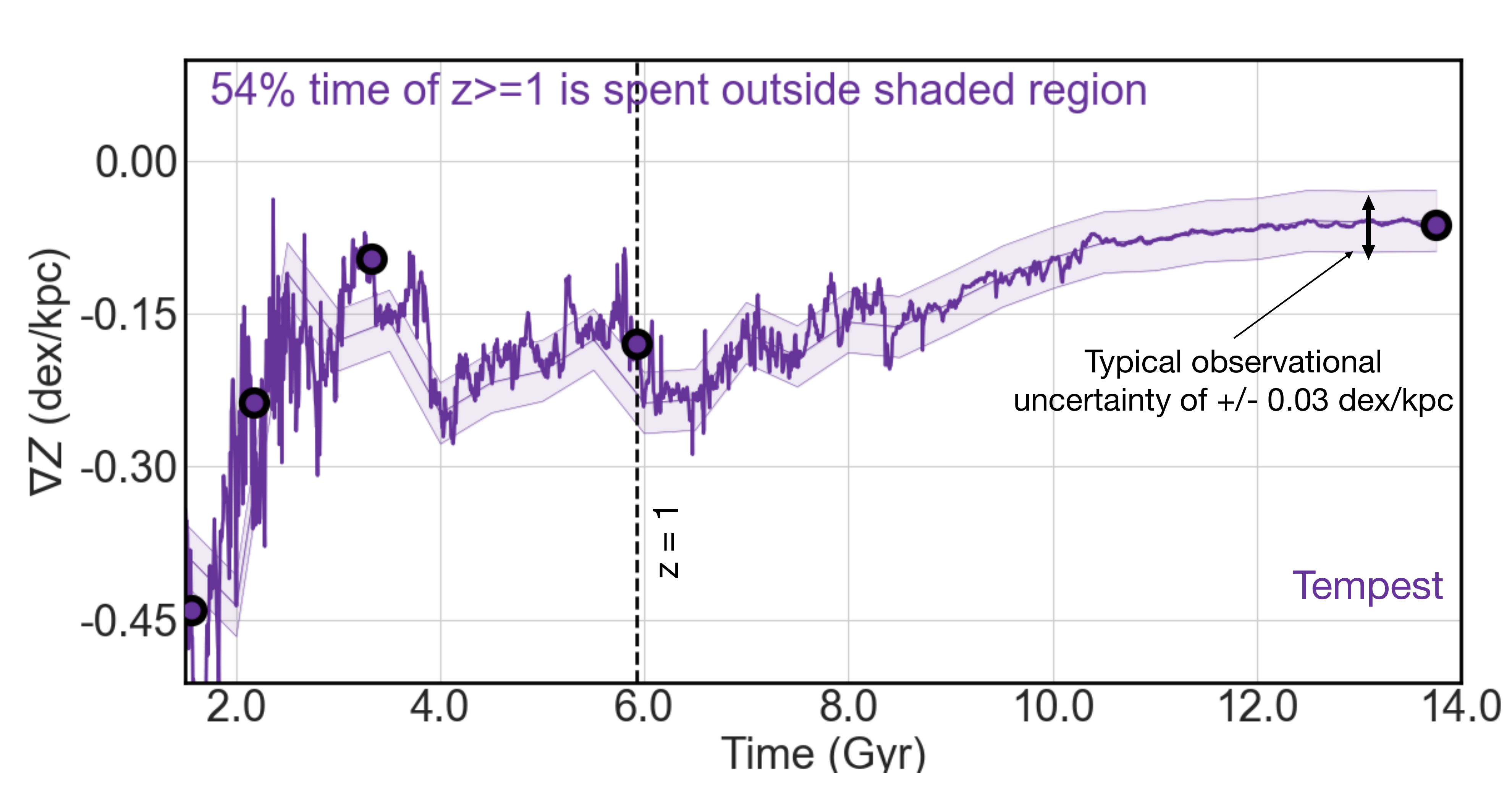}
    \includegraphics[page=2,width=\linewidth]{1_by_1_Zgrad_timefrac.pdf}
    \caption{Similar to \autoref{fig:Zgrad_smooth} but for the other FOGGIE halos---Tempest (top) and Maelstrom (bottom).}
    \label{fig:Zgrad_smooth_TM}
\end{figure*}

\begin{figure*}
    \centering
    \includegraphics[page=3,width=\linewidth]{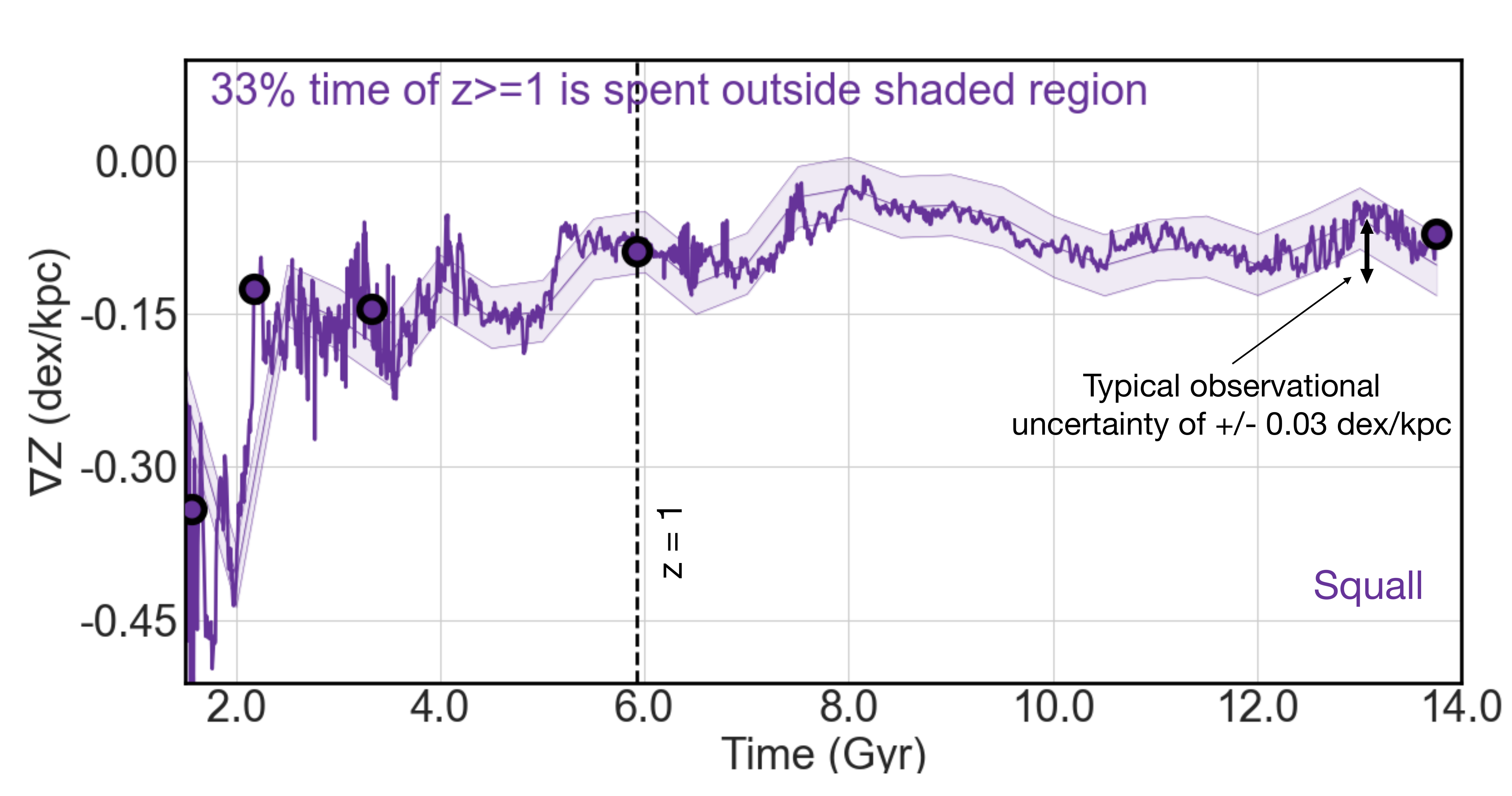}
    \includegraphics[page=4,width=\linewidth]{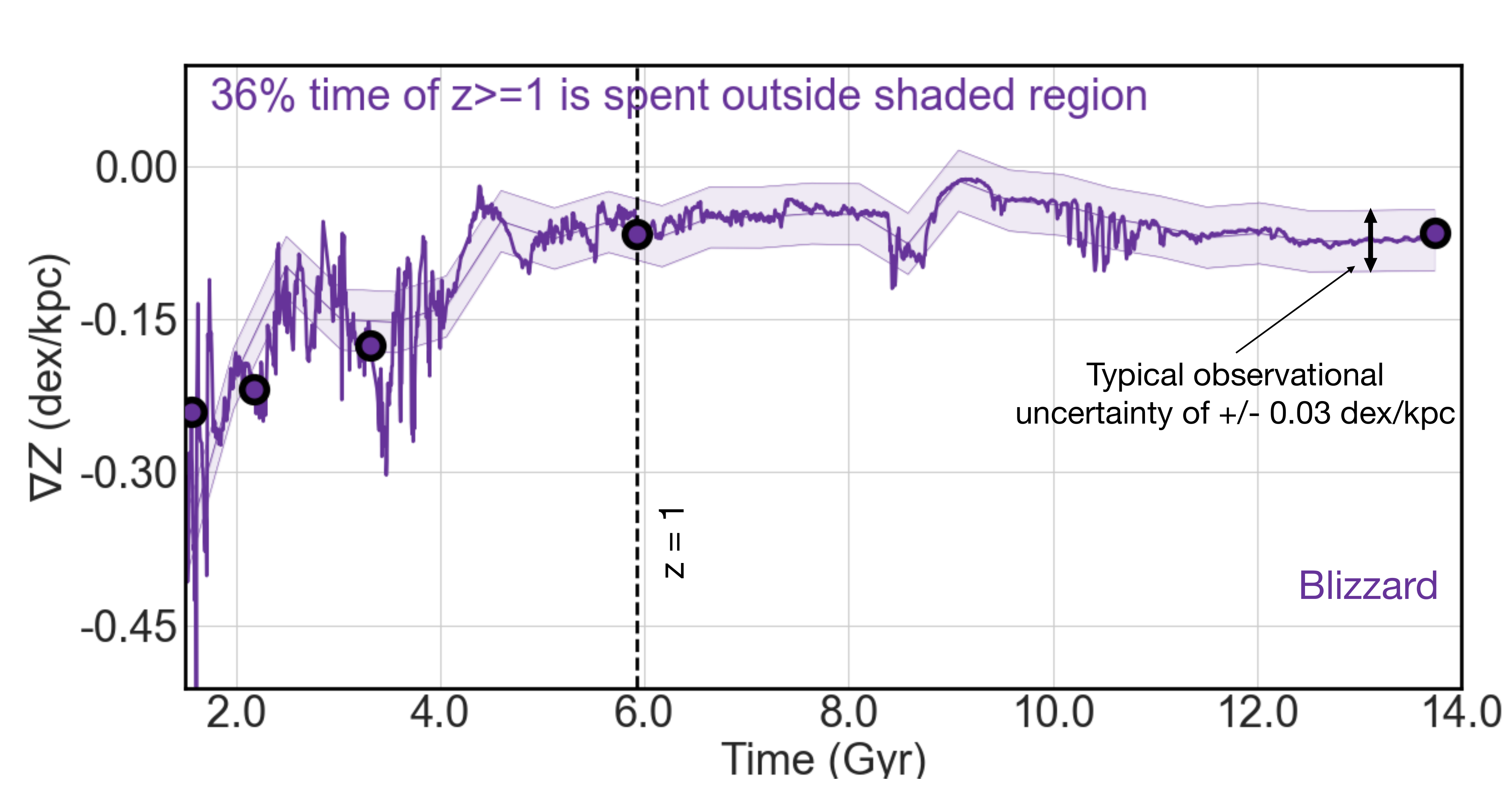}
    \caption{Similar to \autoref{fig:Zgrad_smooth} but for the other FOGGIE halos---Squall (top) and Blizzard (bottom).}
    \label{fig:Zgrad_smooth_SB}
\end{figure*}

\begin{figure*}
    \centering
    \includegraphics[page=5,width=\linewidth]{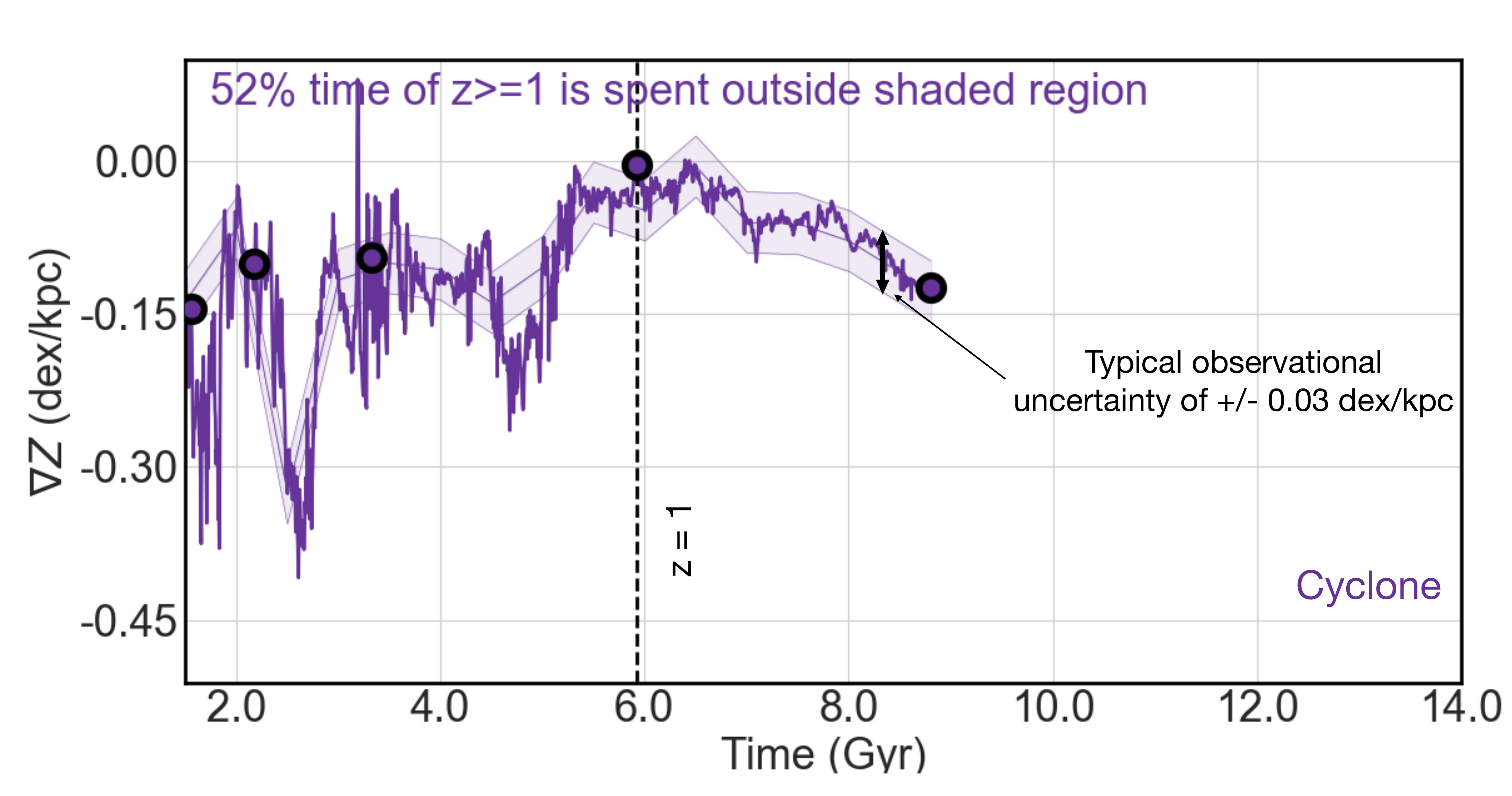}
    \caption{Similar to \autoref{fig:Zgrad_smooth} but for another FOGGIE halo---Cyclone. This FOGGIE halo had been simulated only up to $z=0.5$ at the time of this analysis. So the rightmost purple circle corresponds to $z=0.5$ instead of $z=0$.}
    \label{fig:Zgrad_smooth_C}
\end{figure*}

In Figures \ref{fig:Zgrad_smooth_TM}-\ref{fig:Zgrad_smooth_C} we present the time evolution of the metallicity gradient for each halo, accompanied with the quantification for what fraction of time a given galaxy spends significantly away from its general evolution (see \autoref{sec:variations}). These figures are similar to the top panel of \autoref{fig:Zgrad_smooth} in the main text.

\begin{figure*}
    \centering
    \includegraphics[width=\linewidth]{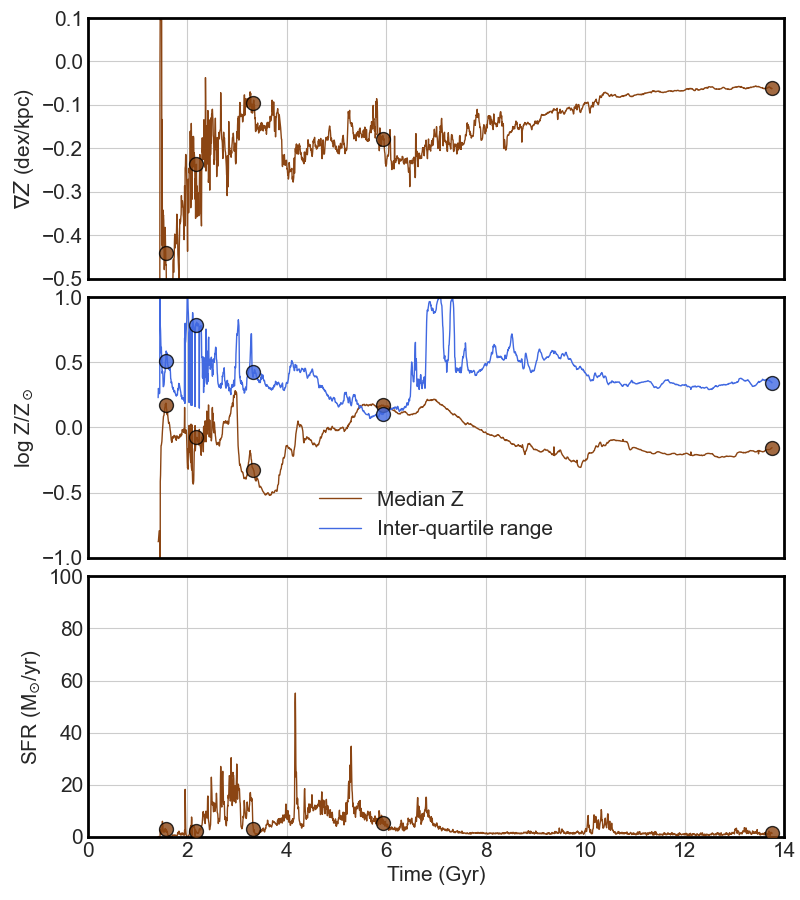}
    \caption{Similar to \autoref{fig:time_series_hurricane} but for a different halo---Tempest.}
    \label{fig:time_series_tempest}
\end{figure*}

\begin{figure*}
    \centering
    \includegraphics[width=\linewidth]{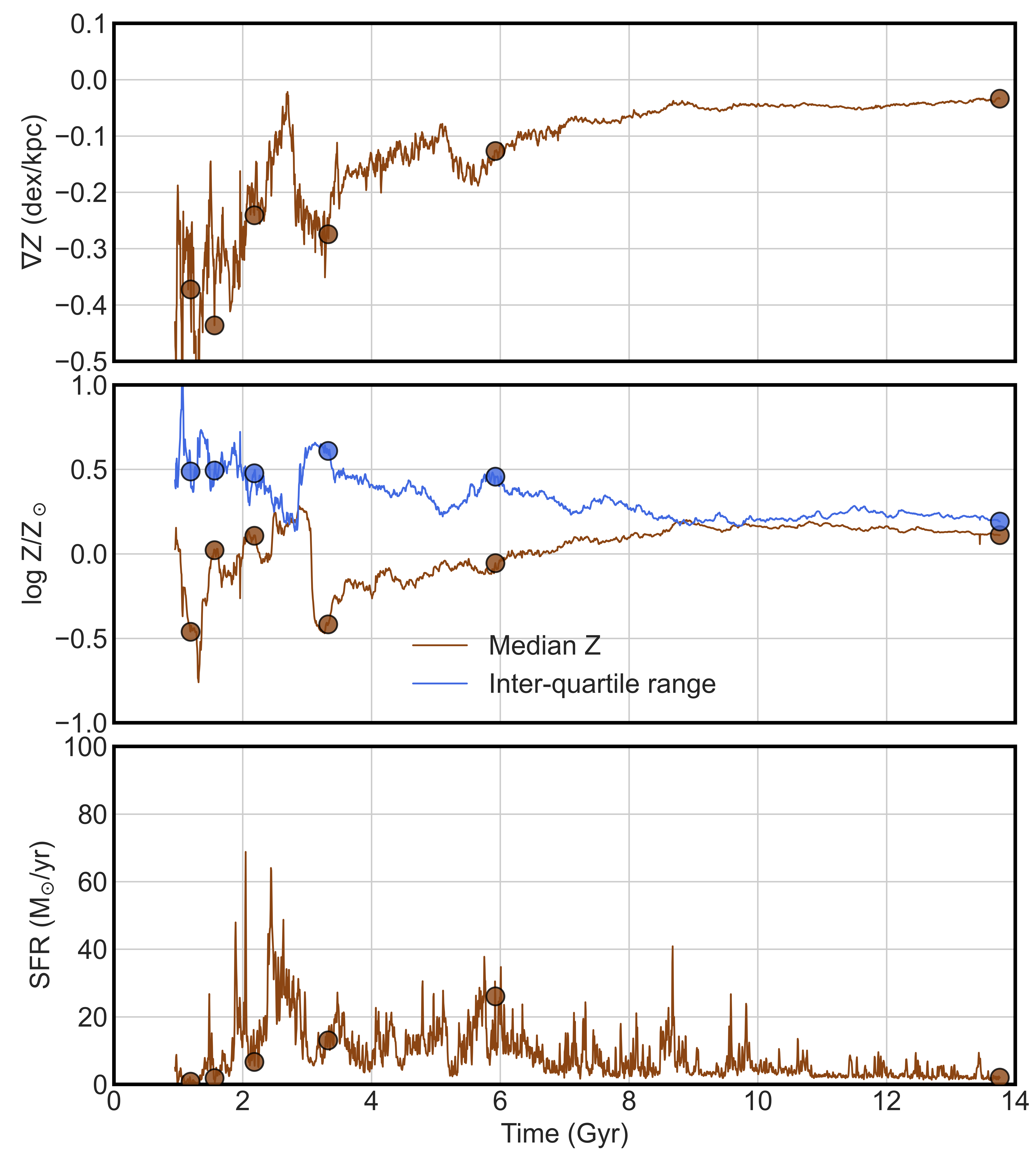}
    \caption{Similar to \autoref{fig:time_series_hurricane} but for a different halo---Maelstrom.}
    \label{fig:time_series_maelstrom}
\end{figure*}

\begin{figure*}
    \centering
    \includegraphics[width=\linewidth]{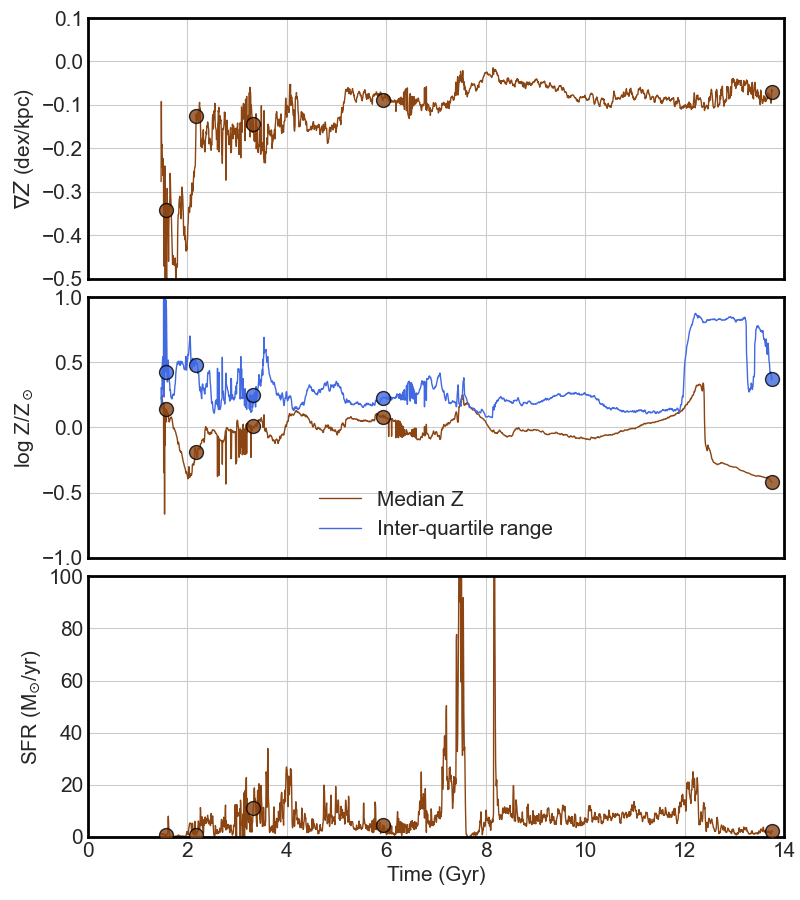}
    \caption{Similar to \autoref{fig:time_series_hurricane} but for a different halo---Squall.}
    \label{fig:time_series_squall}
\end{figure*}

\begin{figure*}
    \centering
    \includegraphics[width=\linewidth]{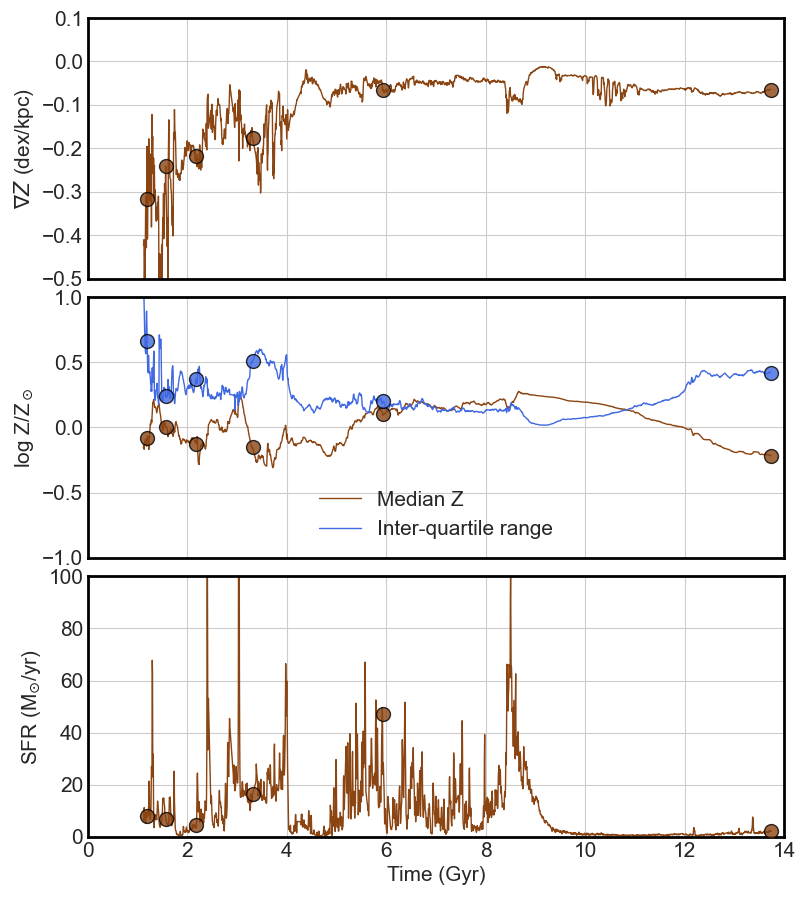}
    \caption{Similar to \autoref{fig:time_series_hurricane} but for a different halo---Blizzard.}
    \label{fig:time_series_blizzard}
\end{figure*}

\begin{figure*}
    \centering
    \includegraphics[width=\linewidth]{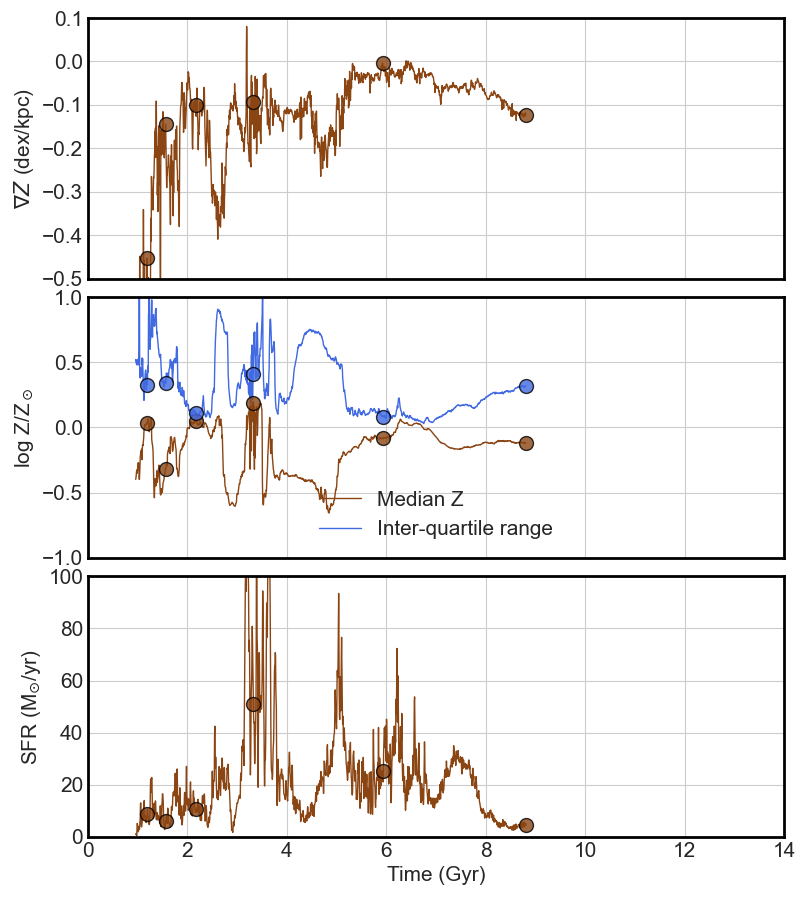}
    \caption{Similar to \autoref{fig:time_series_hurricane} but for a different halo---Cyclone. This FOGGIE halo had been simulated only up to $z=0.5$ at the time of this analysis. So the rightmost circle in each panel corresponds to $z=0.5$ instead of $z=0$.}
    \label{fig:time_series_cyclone}
\end{figure*}

In Figures \ref{fig:time_series_tempest}-\ref{fig:time_series_cyclone} we present evolution of the metallicity gradient as well as the non-parametric quantities---median and IQR of the metallicity distribution---in comparison with the star formation history of each halo. These figures are similar to \autoref{fig:time_series_hurricane} in the main text.


\bibliographystyle{aasjournal}
\bibliography{paper}{}

\begin{thebibliography}{}
\expandafter\ifx\csname natexlab\endcsname\relax\def\natexlab#1{#1}\fi
\providecommand{\url}[1]{\href{#1}{#1}}
\providecommand{\dodoi}[1]{doi:~\href{http://doi.org/#1}{\nolinkurl{#1}}}
\providecommand{\doeprint}[1]{\href{http://ascl.net/#1}{\nolinkurl{http://ascl.net/#1}}}
\providecommand{\doarXiv}[1]{\href{https://arxiv.org/abs/#1}{\nolinkurl{https://arxiv.org/abs/#1}}}

\bibitem[{{Acharyya} {et~al.}(2020){Acharyya}, {Krumholz}, {Federrath},
  {Kewley}, {Goldbaum}, \& {Sharp}}]{Acharyya:2020aa}
{Acharyya}, A., {Krumholz}, M.~R., {Federrath}, C., {et~al.} 2020, \mnras, 495,
  3819, \dodoi{10.1093/mnras/staa1100}

\bibitem[{{Astropy Collaboration} {et~al.}(2013){Astropy Collaboration},
  {Robitaille}, {Tollerud}, {Greenfield}, {Droettboom}, {Bray}, {Aldcroft},
  {Davis}, {Ginsburg}, {Price-Whelan}, {Kerzendorf}, {Conley}, {Crighton},
  {Barbary}, {Muna}, {Ferguson}, {Grollier}, {Parikh}, {Nair}, {Unther},
  {Deil}, {Woillez}, {Conseil}, {Kramer}, {Turner}, {Singer}, {Fox}, {Weaver},
  {Zabalza}, {Edwards}, {Azalee Bostroem}, {Burke}, {Casey}, {Crawford},
  {Dencheva}, {Ely}, {Jenness}, {Labrie}, {Lim}, {Pierfederici}, {Pontzen},
  {Ptak}, {Refsdal}, {Servillat}, \& {Streicher}}]{astropy2013}
{Astropy Collaboration}, {Robitaille}, T.~P., {Tollerud}, E.~J., {et~al.} 2013,
  \aap, 558, A33, \dodoi{10.1051/0004-6361/201322068}

\bibitem[{{Astropy Collaboration} {et~al.}(2018){Astropy Collaboration},
  {Price-Whelan}, {Sip{\H{o}}cz}, {G{\"u}nther}, {Lim}, {Crawford}, {Conseil},
  {Shupe}, {Craig}, {Dencheva}, {Ginsburg}, {VanderPlas}, {Bradley},
  {P{\'e}rez-Su{\'a}rez}, {de Val-Borro}, {Aldcroft}, {Cruz}, {Robitaille},
  {Tollerud}, {Ardelean}, {Babej}, {Bach}, {Bachetti}, {Bakanov}, {Bamford},
  {Barentsen}, {Barmby}, {Baumbach}, {Berry}, {Biscani}, {Boquien}, {Bostroem},
  {Bouma}, {Brammer}, {Bray}, {Breytenbach}, {Buddelmeijer}, {Burke},
  {Calderone}, {Cano Rodr{\'\i}guez}, {Cara}, {Cardoso}, {Cheedella}, {Copin},
  {Corrales}, {Crichton}, {D'Avella}, {Deil}, {Depagne}, {Dietrich}, {Donath},
  {Droettboom}, {Earl}, {Erben}, {Fabbro}, {Ferreira}, {Finethy}, {Fox},
  {Garrison}, {Gibbons}, {Goldstein}, {Gommers}, {Greco}, {Greenfield},
  {Groener}, {Grollier}, {Hagen}, {Hirst}, {Homeier}, {Horton}, {Hosseinzadeh},
  {Hu}, {Hunkeler}, {Ivezi{\'c}}, {Jain}, {Jenness}, {Kanarek}, {Kendrew},
  {Kern}, {Kerzendorf}, {Khvalko}, {King}, {Kirkby}, {Kulkarni}, {Kumar},
  {Lee}, {Lenz}, {Littlefair}, {Ma}, {Macleod}, {Mastropietro}, {McCully},
  {Montagnac}, {Morris}, {Mueller}, {Mumford}, {Muna}, {Murphy}, {Nelson},
  {Nguyen}, {Ninan}, {N{\"o}the}, {Ogaz}, {Oh}, {Parejko}, {Parley}, {Pascual},
  {Patil}, {Patil}, {Plunkett}, {Prochaska}, {Rastogi}, {Reddy Janga},
  {Sabater}, {Sakurikar}, {Seifert}, {Sherbert}, {Sherwood-Taylor}, {Shih},
  {Sick}, {Silbiger}, {Singanamalla}, {Singer}, {Sladen}, {Sooley},
  {Sornarajah}, {Streicher}, {Teuben}, {Thomas}, {Tremblay}, {Turner},
  {Terr{\'o}n}, {van Kerkwijk}, {de la Vega}, {Watkins}, {Weaver}, {Whitmore},
  {Woillez}, {Zabalza}, \& {Astropy Contributors}}]{astropy2018}
{Astropy Collaboration}, {Price-Whelan}, A.~M., {Sip{\H{o}}cz}, B.~M., {et~al.}
  2018, \aj, 156, 123, \dodoi{10.3847/1538-3881/aabc4f}

\bibitem[{{Astropy Collaboration} {et~al.}(2022){Astropy Collaboration},
  {Price-Whelan}, {Lim}, {Earl}, {Starkman}, {Bradley}, {Shupe}, {Patil},
  {Corrales}, {Brasseur}, {N{\"o}the}, {Donath}, {Tollerud}, {Morris},
  {Ginsburg}, {Vaher}, {Weaver}, {Tocknell}, {Jamieson}, {van Kerkwijk},
  {Robitaille}, {Merry}, {Bachetti}, {G{\"u}nther}, {Aldcroft},
  {Alvarado-Montes}, {Archibald}, {B{\'o}di}, {Bapat}, {Barentsen},
  {Baz{\'a}n}, {Biswas}, {Boquien}, {Burke}, {Cara}, {Cara}, {Conroy},
  {Conseil}, {Craig}, {Cross}, {Cruz}, {D'Eugenio}, {Dencheva}, {Devillepoix},
  {Dietrich}, {Eigenbrot}, {Erben}, {Ferreira}, {Foreman-Mackey}, {Fox},
  {Freij}, {Garg}, {Geda}, {Glattly}, {Gondhalekar}, {Gordon}, {Grant},
  {Greenfield}, {Groener}, {Guest}, {Gurovich}, {Handberg}, {Hart},
  {Hatfield-Dodds}, {Homeier}, {Hosseinzadeh}, {Jenness}, {Jones}, {Joseph},
  {Kalmbach}, {Karamehmetoglu}, {Ka{\l}uszy{\'n}ski}, {Kelley}, {Kern},
  {Kerzendorf}, {Koch}, {Kulumani}, {Lee}, {Ly}, {Ma}, {MacBride}, {Maljaars},
  {Muna}, {Murphy}, {Norman}, {O'Steen}, {Oman}, {Pacifici}, {Pascual},
  {Pascual-Granado}, {Patil}, {Perren}, {Pickering}, {Rastogi}, {Roulston},
  {Ryan}, {Rykoff}, {Sabater}, {Sakurikar}, {Salgado}, {Sanghi}, {Saunders},
  {Savchenko}, {Schwardt}, {Seifert-Eckert}, {Shih}, {Jain}, {Shukla}, {Sick},
  {Simpson}, {Singanamalla}, {Singer}, {Singhal}, {Sinha}, {Sip{\H{o}}cz},
  {Spitler}, {Stansby}, {Streicher}, {{\v{S}}umak}, {Swinbank}, {Taranu},
  {Tewary}, {Tremblay}, {Val-Borro}, {Van Kooten}, {Vasovi{\'c}}, {Verma}, {de
  Miranda Cardoso}, {Williams}, {Wilson}, {Winkel}, {Wood-Vasey}, {Xue},
  {Yoachim}, {Zhang}, {Zonca}, \& {Astropy Project Contributors}}]{astropy2022}
{Astropy Collaboration}, {Price-Whelan}, A.~M., {Lim}, P.~L., {et~al.} 2022,
  \apj, 935, 167, \dodoi{10.3847/1538-4357/ac7c74}

\bibitem[{{Avila-Reese} {et~al.}(2018){Avila-Reese}, {Gonz{\'a}lez-Samaniego},
  {Col{\'\i}n}, {Ibarra-Medel}, \&
  {Rodr{\'\i}guez-Puebla}}]{Avila-Reese:2018aa}
{Avila-Reese}, V., {Gonz{\'a}lez-Samaniego}, A., {Col{\'\i}n}, P.,
  {Ibarra-Medel}, H., \& {Rodr{\'\i}guez-Puebla}, A. 2018, \apj, 854, 152,
  \dodoi{10.3847/1538-4357/aaab69}

\bibitem[{{Avila-Reese} {et~al.}(2023){Avila-Reese}, {Ibarra-Medel}, {Lacerna},
  {Rodr{\'\i}guez-Puebla}, {V{\'a}zquez-Mata}, {S{\'a}nchez},
  {Hern{\'a}ndez-Toledo}, \& {Cannarozzo}}]{Avila-Reese:2023aa}
{Avila-Reese}, V., {Ibarra-Medel}, H., {Lacerna}, I., {et~al.} 2023, \mnras,
  523, 4251, \dodoi{10.1093/mnras/stad1638}

\bibitem[{Bednar {et~al.}(2022)Bednar, Crail, Crist-Harif, Rudiger, Brener, B,
  Thomas, Mease, Signell, Liquet, Stevens, Collins, Thorve, Bird, thuydotm,
  esc, kbowen, Abdennur, Smirnov, Hansen, maihde, Hawley, Oriekhov, Ahmadia,
  Jr, Brandt, Tolboom, G., Welch, \& Bourbeau}]{datashader2022}
Bednar, J.~A., Crail, J., Crist-Harif, J., {et~al.} 2022, holoviz/datashader:
  Version 0.14.3, v0.14.3,  Zenodo, \dodoi{10.5281/zenodo.7331952}

\bibitem[{{Belfiore} {et~al.}(2017){Belfiore}, {Maiolino}, {Tremonti},
  {S{\'a}nchez}, {Bundy}, {Bershady}, {Westfall}, {Lin}, {Drory}, {Boquien},
  {Thomas}, \& {Brinkmann}}]{Belfiore:2017aa}
{Belfiore}, F., {Maiolino}, R., {Tremonti}, C., {et~al.} 2017, \mnras, 469,
  151, \dodoi{10.1093/mnras/stx789}

\bibitem[{{Bellardini} {et~al.}(2022){Bellardini}, {Wetzel}, {Loebman}, \&
  {Bailin}}]{Bellardini:2022aa}
{Bellardini}, M.~A., {Wetzel}, A., {Loebman}, S.~R., \& {Bailin}, J. 2022,
  \mnras, 514, 4270, \dodoi{10.1093/mnras/stac1637}

\bibitem[{{Bellardini} {et~al.}(2021){Bellardini}, {Wetzel}, {Loebman},
  {Faucher-Gigu{\`e}re}, {Ma}, \& {Feldmann}}]{Bellardini:2021aa}
{Bellardini}, M.~A., {Wetzel}, A., {Loebman}, S.~R., {et~al.} 2021, \mnras,
  505, 4586, \dodoi{10.1093/mnras/stab1606}

\bibitem[{{Bresolin}(2007)}]{Bresolin:2007aa}
{Bresolin}, F. 2007, \apj, 656, 186, \dodoi{10.1086/510380}

\bibitem[{{Bresolin} {et~al.}(2002){Bresolin}, {Gieren}, {Kudritzki},
  {Pietrzy{\'n}ski}, \& {Przybilla}}]{Bresolin:2002aa}
{Bresolin}, F., {Gieren}, W., {Kudritzki}, R.-P., {Pietrzy{\'n}ski}, G., \&
  {Przybilla}, N. 2002, \apj, 567, 277, \dodoi{10.1086/338505}

\bibitem[{{Brummel-Smith} {et~al.}(2019){Brummel-Smith}, {Bryan}, {Butsky},
  {Corlies}, {Emerick}, {Forbes}, {Fujimoto}, {Goldbaum}, {Grete}, {Hummels},
  {Kim}, {Koh}, {Li}, {Li}, {Li}, {OShea}, {Peeples}, {Regan}, {Salem},
  {Schmidt}, {Simpson}, {Smith}, {Tumlinson}, {Turk}, {Wise}, {Abel},
  {Bordner}, {Cen}, {Collins}, {Crosby}, {Edelmann}, {Hahn}, {Harkness},
  {Harper-Clark}, {Kong}, {Kritsuk}, {Kuhlen}, {Larrue}, {Lee}, {Meece},
  {Norman}, {Oishi}, {Paschos}, {Peruta}, {Razoumov}, {Reynolds}, {Silvia},
  {Skillman}, {Skory}, {So}, {Tasker}, {Wagner}, {Wang}, {Xu}, \&
  {Zhao}}]{Brummel-Smith:2019aa}
{Brummel-Smith}, C., {Bryan}, G., {Butsky}, I., {et~al.} 2019, The Journal of
  Open Source Software, 4, 1636, \dodoi{10.21105/joss.01636}

\bibitem[{{Bryan} {et~al.}(2014){Bryan}, {Norman}, {O'Shea}, {Abel}, {Wise},
  {Turk}, {Reynolds}, {Collins}, {Wang}, {Skillman}, {Smith}, {Harkness},
  {Bordner}, {Kim}, {Kuhlen}, {Xu}, {Goldbaum}, {Hummels}, {Kritsuk}, {Tasker},
  {Skory}, {Simpson}, {Hahn}, {Oishi}, {So}, {Zhao}, {Cen}, {Li}, \& {Enzo
  Collaboration}}]{Bryan:2014aa}
{Bryan}, G.~L., {Norman}, M.~L., {O'Shea}, B.~W., {et~al.} 2014, \apjs, 211,
  19, \dodoi{10.1088/0067-0049/211/2/19}

\bibitem[{{Buck} {et~al.}(2023){Buck}, {Obreja}, {Ratcliffe}, {Lu}, {Minchev},
  \& {Macci{\`o}}}]{Buck:2023aa}
{Buck}, T., {Obreja}, A., {Ratcliffe}, B., {et~al.} 2023, \mnras, 523, 1565,
  \dodoi{10.1093/mnras/stad1503}

\bibitem[{{Bundy} {et~al.}(2015){Bundy}, {Bershady}, {Law}, {Yan}, {Drory},
  {MacDonald}, {Wake}, {Cherinka}, {S{\'a}nchez-Gallego}, {Weijmans}, {Thomas},
  {Tremonti}, {Masters}, {Coccato}, {Diamond-Stanic}, {Arag{\'o}n-Salamanca},
  {Avila-Reese}, {Badenes}, {Falc{\'o}n-Barroso}, {Belfiore}, {Bizyaev},
  {Blanc}, {Bland-Hawthorn}, {Blanton}, {Brownstein}, {Byler}, {Cappellari},
  {Conroy}, {Dutton}, {Emsellem}, {Etherington}, {Frinchaboy}, {Fu}, {Gunn},
  {Harding}, {Johnston}, {Kauffmann}, {Kinemuchi}, {Klaene}, {Knapen},
  {Leauthaud}, {Li}, {Lin}, {Maiolino}, {Malanushenko}, {Malanushenko}, {Mao},
  {Maraston}, {McDermid}, {Merrifield}, {Nichol}, {Oravetz}, {Pan}, {Parejko},
  {Sanchez}, {Schlegel}, {Simmons}, {Steele}, {Steinmetz}, {Thanjavur},
  {Thompson}, {Tinker}, {van den Bosch}, {Westfall}, {Wilkinson}, {Wright},
  {Xiao}, \& {Zhang}}]{Bundy:2015aa}
{Bundy}, K., {Bershady}, M.~A., {Law}, D.~R., {et~al.} 2015, \apj, 798, 7,
  \dodoi{10.1088/0004-637X/798/1/7}

\bibitem[{{Carton} {et~al.}(2018){Carton}, {Brinchmann}, {Contini}, {Epinat},
  {Finley}, {Richard}, {Patr{\'\i}cio}, {Schaye}, {Nanayakkara}, {Weilbacher},
  \& {Wisotzki}}]{Carton:2018aa}
{Carton}, D., {Brinchmann}, J., {Contini}, T., {et~al.} 2018, \mnras, 478,
  4293, \dodoi{10.1093/mnras/sty1343}

\bibitem[{{Cecil} {et~al.}(2014){Cecil}, {Croom}, \& {The SAMI Galaxy Survey
  Team}}]{Cecil:2014aa}
{Cecil}, G.~N., {Croom}, S., \& {The SAMI Galaxy Survey Team}. 2014, in
  American Astronomical Society Meeting Abstracts, Vol. 223, American
  Astronomical Society Meeting Abstracts \#223, 246.03

\bibitem[{{Chiti} {et~al.}(2020){Chiti}, {Frebel}, {Jerjen}, {Kim}, \&
  {Norris}}]{Chiti:2020aa}
{Chiti}, A., {Frebel}, A., {Jerjen}, H., {Kim}, D., \& {Norris}, J.~E. 2020,
  \apj, 891, 8, \dodoi{10.3847/1538-4357/ab6d72}

\bibitem[{{Corlies} {et~al.}(2020){Corlies}, {Peeples}, {Tumlinson}, {O'Shea},
  {Lehner}, {Howk}, {O'Meara}, \& {Smith}}]{Corlies:2020aa}
{Corlies}, L., {Peeples}, M.~S., {Tumlinson}, J., {et~al.} 2020, \apj, 896,
  125, \dodoi{10.3847/1538-4357/ab9310}

\bibitem[{{Curti} {et~al.}(2020){Curti}, {Maiolino}, {Cirasuolo}, {Mannucci},
  {Williams}, {Auger}, {Mercurio}, {Hayden-Pawson}, {Cresci}, {Marconi},
  {Belfiore}, {Cappellari}, {Cicone}, {Cullen}, {Meneghetti}, {Ota}, {Peng},
  {Pettini}, {Swinbank}, \& {Troncoso}}]{Curti:2020aa}
{Curti}, M., {Maiolino}, R., {Cirasuolo}, M., {et~al.} 2020, \mnras, 492, 821,
  \dodoi{10.1093/mnras/stz3379}

\bibitem[{{Durrell} {et~al.}(2001){Durrell}, {Harris}, \&
  {Pritchet}}]{Durrell:2001aa}
{Durrell}, P.~R., {Harris}, W.~E., \& {Pritchet}, C.~J. 2001, \aj, 121, 2557,
  \dodoi{10.1086/320403}

\bibitem[{{F{\"o}rster Schreiber} {et~al.}(2018){F{\"o}rster Schreiber},
  {Renzini}, {Mancini}, {Genzel}, {Bouch{\'e}}, {Cresci}, {Hicks}, {Lilly},
  {Peng}, {Burkert}, {Carollo}, {Cimatti}, {Daddi}, {Davies}, {Genel}, {Kurk},
  {Lang}, {Lutz}, {Mainieri}, {McCracken}, {Mignoli}, {Naab}, {Oesch},
  {Pozzetti}, {Scodeggio}, {Shapiro Griffin}, {Shapley}, {Sternberg},
  {Tacchella}, {Tacconi}, {Wuyts}, \& {Zamorani}}]{Forster-Schreiber:2018aa}
{F{\"o}rster Schreiber}, N.~M., {Renzini}, A., {Mancini}, C., {et~al.} 2018,
  \apjs, 238, 21, \dodoi{10.3847/1538-4365/aadd49}

\bibitem[{{Fu} {et~al.}(2013){Fu}, {Kauffmann}, {Huang}, {Yates}, {Moran},
  {Heckman}, {Dav{\'e}}, {Guo}, \& {Henriques}}]{Fu:2013aa}
{Fu}, J., {Kauffmann}, G., {Huang}, M.-l., {et~al.} 2013, \mnras, 434, 1531,
  \dodoi{10.1093/mnras/stt1117}

\bibitem[{{Garcia} {et~al.}(2023){Garcia}, {Torrey}, {Hemler}, {Hernquist},
  {Kewley}, {Nelson}, {Grasha}, {Zovaro}, \& {Chen}}]{Garcia:2023aa}
{Garcia}, A.~M., {Torrey}, P., {Hemler}, Z.~S., {et~al.} 2023, \mnras, 519,
  4716, \dodoi{10.1093/mnras/stac3749}

\bibitem[{{Garnett} {et~al.}(1997){Garnett}, {Shields}, {Skillman}, {Sagan}, \&
  {Dufour}}]{Garnett:1997aa}
{Garnett}, D.~R., {Shields}, G.~A., {Skillman}, E.~D., {Sagan}, S.~P., \&
  {Dufour}, R.~J. 1997, \apj, 489, 63, \dodoi{10.1086/304775}

\bibitem[{{Gibson} {et~al.}(2013){Gibson}, {Pilkington}, {Brook}, {Stinson}, \&
  {Bailin}}]{Gibson:2013a}
{Gibson}, B.~K., {Pilkington}, K., {Brook}, C.~B., {Stinson}, G.~S., \&
  {Bailin}, J. 2013, A\&A, 554, A47, \dodoi{10.1051/0004-6361/201321239}

\bibitem[{{Graf} {et~al.}(2024){Graf}, {Wetzel}, {Bellardini}, \&
  {Bailin}}]{Graf:2024aa}
{Graf}, R.~L., {Wetzel}, A., {Bellardini}, M.~A., \& {Bailin}, J. 2024, arXiv
  e-prints, arXiv:2402.15614, \dodoi{10.48550/arXiv.2402.15614}

\bibitem[{{Hayden} {et~al.}(2015){Hayden}, {Bovy}, {Holtzman}, {Nidever},
  {Bird}, {Weinberg}, {Andrews}, {Majewski}, {Allende Prieto}, {Anders},
  {Beers}, {Bizyaev}, {Chiappini}, {Cunha}, {Frinchaboy},
  {Garc{\'\i}a-Her{\'n}andez}, {Garc{\'\i}a P{\'e}rez}, {Girardi}, {Harding},
  {Hearty}, {Johnson}, {M{\'e}sz{\'a}ros}, {Minchev}, {O'Connell}, {Pan},
  {Robin}, {Schiavon}, {Schneider}, {Schultheis}, {Shetrone}, {Skrutskie},
  {Steinmetz}, {Smith}, {Wilson}, {Zamora}, \& {Zasowski}}]{Hayden:2015aa}
{Hayden}, M.~R., {Bovy}, J., {Holtzman}, J.~A., {et~al.} 2015, \apj, 808, 132,
  \dodoi{10.1088/0004-637X/808/2/132}

\bibitem[{{Hemler} {et~al.}(2021){Hemler}, {Torrey}, {Qi}, {Hernquist},
  {Vogelsberger}, {Ma}, {Kewley}, {Nelson}, {Pillepich}, {Pakmor}, \&
  {Marinacci}}]{Hemler:2021a}
{Hemler}, Z.~S., {Torrey}, P., {Qi}, J., {et~al.} 2021, MNRAS, 506, 3024,
  \dodoi{10.1093/mnras/stab1803}

\bibitem[{{Hummels} {et~al.}(2019){Hummels}, {Smith}, {Hopkins}, {O'Shea},
  {Silvia}, {Werk}, {Lehner}, {Wise}, {Collins}, \& {Butsky}}]{Hummels:2019aa}
{Hummels}, C.~B., {Smith}, B.~D., {Hopkins}, P.~F., {et~al.} 2019, \apj, 882,
  156, \dodoi{10.3847/1538-4357/ab378f}

\bibitem[{{Hunter}(2007)}]{matplotlib2007}
{Hunter}, J.~D. 2007, Computing in Science and Engineering, 9, 90,
  \dodoi{10.1109/MCSE.2007.55}

\bibitem[{{Jones} {et~al.}(2013{\natexlab{a}}){Jones}, {Ellis}, {Richard}, \&
  {Jullo}}]{Jones:2013ab}
{Jones}, T., {Ellis}, R.~S., {Richard}, J., \& {Jullo}, E. 2013{\natexlab{a}},
  \apj, 765, 48, \dodoi{10.1088/0004-637X/765/1/48}

\bibitem[{{Jones} {et~al.}(2015){Jones}, {Wang}, {Schmidt}, {Treu}, {Brammer},
  {Brada{\v c}}, {Dressler}, {Henry}, {Malkan}, {Pentericci}, \&
  {Trenti}}]{Jones:2015aa}
{Jones}, T., {Wang}, X., {Schmidt}, K.~B., {et~al.} 2015, \aj, 149, 107,
  \dodoi{10.1088/0004-6256/149/3/107}

\bibitem[{{Jones} {et~al.}(2013{\natexlab{b}}){Jones}, {Ellis}, {Schenker}, \&
  {Stark}}]{Jones:2013aa}
{Jones}, T.~A., {Ellis}, R.~S., {Schenker}, M.~A., \& {Stark}, D.~P.
  2013{\natexlab{b}}, \apj, 779, 52, \dodoi{10.1088/0004-637X/779/1/52}

\bibitem[{{Kennicutt} {et~al.}(2003){Kennicutt}, {Bresolin}, \&
  {Garnett}}]{Kennicutt:2003aa}
{Kennicutt}, Jr., R.~C., {Bresolin}, F., \& {Garnett}, D.~R. 2003, \apj, 591,
  801, \dodoi{10.1086/375398}

\bibitem[{{Kewley} \& {Ellison}(2008)}]{Kewley:2008aa}
{Kewley}, L.~J., \& {Ellison}, S.~L. 2008, \apj, 681, 1183,
  \dodoi{10.1086/587500}

\bibitem[{{Kewley} {et~al.}(2010){Kewley}, {Rupke}, {Zahid}, {Geller}, \&
  {Barton}}]{Kewley:2010aa}
{Kewley}, L.~J., {Rupke}, D., {Zahid}, H.~J., {Geller}, M.~J., \& {Barton},
  E.~J. 2010, \apj, 721, L48, \dodoi{10.1088/2041-8205/721/1/L48}

\bibitem[{{Kirby} {et~al.}(2011){Kirby}, {Lanfranchi}, {Simon}, {Cohen}, \&
  {Guhathakurta}}]{Kirby:2011aa}
{Kirby}, E.~N., {Lanfranchi}, G.~A., {Simon}, J.~D., {Cohen}, J.~G., \&
  {Guhathakurta}, P. 2011, \apj, 727, 78, \dodoi{10.1088/0004-637X/727/2/78}

\bibitem[{{Krabbe} {et~al.}(2008){Krabbe}, {Pastoriza}, {Winge}, {Rodrigues},
  \& {Ferreiro}}]{Krabbe:2008aa}
{Krabbe}, A.~C., {Pastoriza}, M.~G., {Winge}, C., {Rodrigues}, I., \&
  {Ferreiro}, D.~L. 2008, \mnras, 389, 1593,
  \dodoi{10.1111/j.1365-2966.2008.13701.x}

\bibitem[{{Leethochawalit} {et~al.}(2016){Leethochawalit}, {Jones}, {Ellis},
  {Stark}, \& {Zitrin}}]{Leethochawalit:2016aa}
{Leethochawalit}, N., {Jones}, T.~A., {Ellis}, R.~S., {Stark}, D.~P., \&
  {Zitrin}, A. 2016, \apj, 831, 152, \dodoi{10.3847/0004-637X/831/2/152}

\bibitem[{{Li} {et~al.}(2022){Li}, {Wang}, {Cai}, {Shi}, {Fan}, {Zheng},
  {Malkan}, {Teplitz}, {Henry}, {Bian}, \& {Colbert}}]{Li:2022aa}
{Li}, Z., {Wang}, X., {Cai}, Z., {et~al.} 2022, \apjl, 929, L8,
  \dodoi{10.3847/2041-8213/ac626f}

\bibitem[{{Lochhaas} {et~al.}(2021){Lochhaas}, {Tumlinson}, {O'Shea},
  {Peeples}, {Smith}, {Werk}, {Augustin}, \& {Simons}}]{Lochhaas:2021aa}
{Lochhaas}, C., {Tumlinson}, J., {O'Shea}, B.~W., {et~al.} 2021, \apj, 922,
  121, \dodoi{10.3847/1538-4357/ac2496}

\bibitem[{{Lochhaas} {et~al.}(2023){Lochhaas}, {Tumlinson}, {Peeples},
  {O'Shea}, {Werk}, {Simons}, {Juno}, {Kopenhafer}, {Augustin}, {Wright},
  {Acharyya}, \& {Smith}}]{Lochhaas:2023aa}
{Lochhaas}, C., {Tumlinson}, J., {Peeples}, M.~S., {et~al.} 2023, \apj, 948,
  43, \dodoi{10.3847/1538-4357/acbb06}

\bibitem[{{Loebman} {et~al.}(2016){Loebman}, {Debattista}, {Nidever}, {Hayden},
  {Holtzman}, {Clarke}, {Ro{\v{s}}kar}, \& {Valluri}}]{Loebman:2016aa}
{Loebman}, S.~R., {Debattista}, V.~P., {Nidever}, D.~L., {et~al.} 2016, \apjl,
  818, L6, \dodoi{10.3847/2041-8205/818/1/L6}

\bibitem[{{Ma} {et~al.}(2017){Ma}, {Hopkins}, {Feldmann}, {Torrey},
  {Faucher-Gigu{\`e}re}, \& {Kere{\v{s}}}}]{Ma:2017aa}
{Ma}, X., {Hopkins}, P.~F., {Feldmann}, R., {et~al.} 2017, \mnras, 466, 4780,
  \dodoi{10.1093/mnras/stx034}

\bibitem[{{Marino} {et~al.}(2016){Marino}, {Gil de Paz}, {S{\'a}nchez},
  {S{\'a}nchez-Bl{\'a}zquez}, {Cardiel}, {Castillo-Morales}, {Pascual},
  {V{\'\i}lchez}, {Kehrig}, {Moll{\'a}}, {Mendez-Abreu},
  {Catal{\'a}n-Torrecilla}, {Florido}, {Perez}, {Ruiz-Lara}, {Ellis},
  {L{\'o}pez-S{\'a}nchez}, {Gonz{\'a}lez Delgado}, {de Lorenzo-C{\'a}ceres},
  {Garc{\'\i}a-Benito}, {Galbany}, {Zibetti}, {Cortijo}, {Kalinova}, {Mast},
  {Iglesias-P{\'a}ramo}, {Papaderos}, {Walcher}, \&
  {Bland-Hawthorn}}]{Marino:2016aa}
{Marino}, R.~A., {Gil de Paz}, A., {S{\'a}nchez}, S.~F., {et~al.} 2016, \aap,
  585, A47, \dodoi{10.1051/0004-6361/201526986}

\bibitem[{{Mingozzi} {et~al.}(2020){Mingozzi}, {Belfiore}, {Cresci}, {Bundy},
  {Bershady}, {Bizyaev}, {Blanc}, {Boquien}, {Drory}, {Fu}, {Maiolino},
  {Riffel}, {Schaefer}, {Storchi-Bergmann}, {Telles}, {Tremonti}, {Zakamska},
  \& {Zhang}}]{Mingozzi:2020aa}
{Mingozzi}, M., {Belfiore}, F., {Cresci}, G., {et~al.} 2020, \aap, 636, A42,
  \dodoi{10.1051/0004-6361/201937203}

\bibitem[{{Minniti} \& {Zoccali}(2008)}]{Minniti:2008aa}
{Minniti}, D., \& {Zoccali}, M. 2008, in Formation and Evolution of Galaxy
  Bulges, ed. M.~{Bureau}, E.~{Athanassoula}, \& B.~{Barbuy}, Vol. 245,
  323--332, \dodoi{10.1017/S1743921308018048}

\bibitem[{{Miralles-Caballero} {et~al.}(2014){Miralles-Caballero}, {D{\'\i}az},
  {Rosales-Ortega}, {P{\'e}rez-Montero}, \& {S{\'a}nchez}}]{Miralles:2014aa}
{Miralles-Caballero}, D., {D{\'\i}az}, A.~I., {Rosales-Ortega}, F.~F.,
  {P{\'e}rez-Montero}, E., \& {S{\'a}nchez}, S.~F. 2014, \mnras, 440, 2265,
  \dodoi{10.1093/mnras/stu435}

\bibitem[{{Molina} {et~al.}(2017){Molina}, {Ibar}, {Swinbank}, {Sobral},
  {Best}, {Smail}, {Escala}, \& {Cirasuolo}}]{Molina:2017aa}
{Molina}, J., {Ibar}, E., {Swinbank}, A.~M., {et~al.} 2017, \mnras, 466, 892,
  \dodoi{10.1093/mnras/stw3120}

\bibitem[{{Moustakas} {et~al.}(2010){Moustakas}, {Kennicutt}, {Tremonti},
  {Dale}, {Smith}, \& {Calzetti}}]{Moustakas:2010aa}
{Moustakas}, J., {Kennicutt}, Robert~C., J., {Tremonti}, C.~A., {et~al.} 2010,
  The Astrophysical Journal Supplement Series, 190, 233,
  \dodoi{10.1088/0067-0049/190/2/233}

\bibitem[{{Mu{\~n}oz-Elgueta} {et~al.}(2018){Mu{\~n}oz-Elgueta},
  {Torres-Flores}, {Amram}, {Hernandez-Jimenez}, {Urrutia-Viscarra}, {Mendes de
  Oliveira}, \& {G{\'o}mez-L{\'o}pez}}]{Munoz:2018aa}
{Mu{\~n}oz-Elgueta}, N., {Torres-Flores}, S., {Amram}, P., {et~al.} 2018,
  \mnras, 480, 3257, \dodoi{10.1093/mnras/sty1934}

\bibitem[{{Peeples} {et~al.}(2019){Peeples}, {Corlies}, {Tumlinson}, {O'Shea},
  {Lehner}, {O'Meara}, {Howk}, {Earl}, {Smith}, {Wise}, \&
  {Hummels}}]{Peeples:2019aa}
{Peeples}, M.~S., {Corlies}, L., {Tumlinson}, J., {et~al.} 2019, \apj, 873,
  129, \dodoi{10.3847/1538-4357/ab0654}

\bibitem[{{Planck Collaboration} {et~al.}(2014){Planck Collaboration}, {Ade},
  {Aghanim}, {Armitage-Caplan}, {Arnaud}, {Ashdown}, {Atrio-Barandela},
  {Aumont}, {Baccigalupi}, {Banday}, {Barreiro}, {Bartlett}, {Battaner},
  {Benabed}, {Beno{\^\i}t}, {Benoit-L{\'e}vy}, {Bernard}, {Bersanelli},
  {Bielewicz}, {Bobin}, {Bock}, {Bonaldi}, {Bond}, {Borrill}, {Bouchet},
  {Bridges}, {Bucher}, {Burigana}, {Butler}, {Calabrese}, {Cappellini},
  {Cardoso}, {Catalano}, {Challinor}, {Chamballu}, {Chary}, {Chen}, {Chiang},
  {Chiang}, {Christensen}, {Church}, {Clements}, {Colombi}, {Colombo},
  {Couchot}, {Coulais}, {Crill}, {Curto}, {Cuttaia}, {Danese}, {Davies},
  {Davis}, {de Bernardis}, {de Rosa}, {de Zotti}, {Delabrouille}, {Delouis},
  {D{\'e}sert}, {Dickinson}, {Diego}, {Dolag}, {Dole}, {Donzelli}, {Dor{\'e}},
  {Douspis}, {Dunkley}, {Dupac}, {Efstathiou}, {Elsner}, {En{\ss}lin},
  {Eriksen}, {Finelli}, {Forni}, {Frailis}, {Fraisse}, {Franceschi}, {Gaier},
  {Galeotta}, {Galli}, {Ganga}, {Giard}, {Giardino}, {Giraud-H{\'e}raud},
  {Gjerl{\o}w}, {Gonz{\'a}lez-Nuevo}, {G{\'o}rski}, {Gratton}, {Gregorio},
  {Gruppuso}, {Gudmundsson}, {Haissinski}, {Hamann}, {Hansen}, {Hanson},
  {Harrison}, {Henrot-Versill{\'e}}, {Hern{\'a}ndez-Monteagudo}, {Herranz},
  {Hildebrandt}, {Hivon}, {Hobson}, {Holmes}, {Hornstrup}, {Hou}, {Hovest},
  {Huffenberger}, {Jaffe}, {Jaffe}, {Jewell}, {Jones}, {Juvela},
  {Keih{\"a}nen}, {Keskitalo}, {Kisner}, {Kneissl}, {Knoche}, {Knox}, {Kunz},
  {Kurki-Suonio}, {Lagache}, {L{\"a}hteenm{\"a}ki}, {Lamarre}, {Lasenby},
  {Lattanzi}, {Laureijs}, {Lawrence}, {Leach}, {Leahy}, {Leonardi},
  {Le{\'o}n-Tavares}, {Lesgourgues}, {Lewis}, {Liguori}, {Lilje},
  {Linden-V{\o}rnle}, {L{\'o}pez-Caniego}, {Lubin}, {Mac{\'\i}as-P{\'e}rez},
  {Maffei}, {Maino}, {Mandolesi}, {Maris}, {Marshall}, {Martin},
  {Mart{\'\i}nez-Gonz{\'a}lez}, {Masi}, {Massardi}, {Matarrese}, {Matthai},
  {Mazzotta}, {Meinhold}, {Melchiorri}, {Melin}, {Mendes}, {Menegoni},
  {Mennella}, {Migliaccio}, {Millea}, {Mitra}, {Miville-Desch{\^e}nes},
  {Moneti}, {Montier}, {Morgante}, {Mortlock}, {Moss}, {Munshi}, {Murphy},
  {Naselsky}, {Nati}, {Natoli}, {Netterfield}, {N{\o}rgaard-Nielsen},
  {Noviello}, {Novikov}, {Novikov}, {O'Dwyer}, {Osborne}, {Oxborrow}, {Paci},
  {Pagano}, {Pajot}, {Paladini}, {Paoletti}, {Partridge}, {Pasian},
  {Patanchon}, {Pearson}, {Pearson}, {Peiris}, {Perdereau}, {Perotto},
  {Perrotta}, {Pettorino}, {Piacentini}, {Piat}, {Pierpaoli}, {Pietrobon},
  {Plaszczynski}, {Platania}, {Pointecouteau}, {Polenta}, {Ponthieu}, {Popa},
  {Poutanen}, {Pratt}, {Pr{\'e}zeau}, {Prunet}, {Puget}, {Rachen}, {Reach},
  {Rebolo}, {Reinecke}, {Remazeilles}, {Renault}, {Ricciardi}, {Riller},
  {Ristorcelli}, {Rocha}, {Rosset}, {Roudier}, {Rowan-Robinson},
  {Rubi{\~n}o-Mart{\'\i}n}, {Rusholme}, {Sandri}, {Santos}, {Savelainen},
  {Savini}, {Scott}, {Seiffert}, {Shellard}, {Spencer}, {Starck}, {Stolyarov},
  {Stompor}, {Sudiwala}, {Sunyaev}, {Sureau}, {Sutton}, {Suur-Uski}, {Sygnet},
  {Tauber}, {Tavagnacco}, {Terenzi}, {Toffolatti}, {Tomasi}, {Tristram},
  {Tucci}, {Tuovinen}, {T{\"u}rler}, {Umana}, {Valenziano}, {Valiviita}, {Van
  Tent}, {Vielva}, {Villa}, {Vittorio}, {Wade}, {Wandelt}, {Wehus}, {White},
  {White}, {Wilkinson}, {Yvon}, {Zacchei}, \& {Zonca}}]{Planck:2014aa}
{Planck Collaboration}, {Ade}, P.~A.~R., {Aghanim}, N., {et~al.} 2014, \aap,
  571, A16, \dodoi{10.1051/0004-6361/201321591}

\bibitem[{{Poetrodjojo} {et~al.}(2018){Poetrodjojo}, {Groves}, {Kewley},
  {Medling}, {Sweet}, {van de Sande}, {Sanchez}, {Bland-Hawthorn}, {Brough},
  {Bryant}, {Cortese}, {Croom}, {L{\'o}pez-S{\'a}nchez}, {Richards}, {Zafar},
  {Lawrence}, {Lorente}, {Owers}, \& {Scott}}]{Poetrodjojo:2018aa}
{Poetrodjojo}, H., {Groves}, B., {Kewley}, L.~J., {et~al.} 2018, \mnras, 479,
  5235, \dodoi{10.1093/mnras/sty1782}

\bibitem[{{Poetrodjojo} {et~al.}(2021){Poetrodjojo}, {Groves}, {Kewley},
  {Sweet}, {Sanchez}, {Medling}, {L{\'o}pez-S{\'a}nchez}, {Brough}, {Cortese},
  {van de Sande}, {Vaughan}, {Richards}, {Bryant}, {Croom}, {Bland-Hawthorn},
  {Goodwin}, {Lawrence}, {Owers}, \& {Scott}}]{Poetrodjojo:2021aa}
---. 2021, \mnras, 502, 3357, \dodoi{10.1093/mnras/stab205}

\bibitem[{{Rich} {et~al.}(2012){Rich}, {Torrey}, {Kewley}, {Dopita}, \&
  {Rupke}}]{Rich:2012aa}
{Rich}, J.~A., {Torrey}, P., {Kewley}, L.~J., {Dopita}, M.~A., \& {Rupke},
  D.~S.~N. 2012, \apj, 753, 5, \dodoi{10.1088/0004-637X/753/1/5}

\bibitem[{{Rosas-Guevara} {et~al.}(2022){Rosas-Guevara}, {Tissera}, {Lagos},
  {Paillas}, \& {Padilla}}]{Rosas-Guevara:202aa}
{Rosas-Guevara}, Y., {Tissera}, P., {Lagos}, C. d.~P., {Paillas}, E., \&
  {Padilla}, N. 2022, \mnras, 517, 712, \dodoi{10.1093/mnras/stac2583}

\bibitem[{{S{\'a}nchez} {et~al.}(2014){S{\'a}nchez}, {Rosales-Ortega},
  {Iglesias-P{\'a}ramo}, {Moll{\'a}}, {Barrera-Ballesteros}, {Marino},
  {P{\'e}rez}, {S{\'a}nchez-Blazquez}, {Gonz{\'a}lez Delgado}, {Cid Fernandes},
  {de Lorenzo-C{\'a}ceres}, {Mendez-Abreu}, {Galbany}, {Falcon-Barroso},
  {Miralles-Caballero}, {Husemann}, {Garc{\'{\i}}a-Benito}, {Mast}, {Walcher},
  {Gil de Paz}, {Garc{\'{\i}}a-Lorenzo}, {Jungwiert}, {V{\'{\i}}lchez},
  {J{\'{\i}}lkov{\'a}}, {Lyubenova}, {Cortijo-Ferrero}, {D{\'{\i}}az},
  {Wisotzki}, {M{\'a}rquez}, {Bland-Hawthorn}, {Ellis}, {van de Ven}, {Jahnke},
  {Papaderos}, {Gomes}, {Mendoza}, \& {L{\'o}pez-S{\'a}nchez}}]{Sanchez:2014aa}
{S{\'a}nchez}, S.~F., {Rosales-Ortega}, F.~F., {Iglesias-P{\'a}ramo}, J.,
  {et~al.} 2014, \aap, 563, A49, \dodoi{10.1051/0004-6361/201322343}

\bibitem[{{S{\'a}nchez-Menguiano} {et~al.}(2016){S{\'a}nchez-Menguiano},
  {S{\'a}nchez}, {P{\'e}rez}, {Garc{\'\i}a-Benito}, {Husemann}, {Mast},
  {Mendoza}, {Ruiz-Lara}, {Ascasibar}, {Bland-Hawthorn}, {Cavichia},
  {D{\'\i}az}, {Florido}, {Galbany}, {G{\'o}nzalez Delgado}, {Kehrig},
  {Marino}, {M{\'a}rquez}, {Masegosa}, {M{\'e}ndez-Abreu}, {Moll{\'a}}, {Del
  Olmo}, {P{\'e}rez}, {S{\'a}nchez-Bl{\'a}zquez}, {Stanishev}, {Walcher},
  {L{\'o}pez-S{\'a}nchez}, \& {Califa
  Collaboration}}]{Sanchez-Menguiano:2016aa}
{S{\'a}nchez-Menguiano}, L., {S{\'a}nchez}, S.~F., {P{\'e}rez}, I., {et~al.}
  2016, \aap, 587, A70, \dodoi{10.1051/0004-6361/201527450}

\bibitem[{{S{\'a}nchez-Menguiano} {et~al.}(2018){S{\'a}nchez-Menguiano},
  {S{\'a}nchez}, {P{\'e}rez}, {Ruiz-Lara}, {Galbany}, {Anderson},
  {Kr{\"u}hler}, {Kuncarayakti}, \& {Lyman}}]{Sanchez-Menguiano:2018aa}
---. 2018, \aap, 609, A119, \dodoi{10.1051/0004-6361/201731486}

\bibitem[{{Sharda} {et~al.}(2021{\natexlab{a}}){Sharda}, {Krumholz},
  {Wisnioski}, {Forbes}, {Federrath}, \& {Acharyya}}]{Sharda:2021aa}
{Sharda}, P., {Krumholz}, M.~R., {Wisnioski}, E., {et~al.} 2021{\natexlab{a}},
  \mnras, 502, 5935, \dodoi{10.1093/mnras/stab252}

\bibitem[{{Sharda} {et~al.}(2021{\natexlab{b}}){Sharda}, {Wisnioski},
  {Krumholz}, \& {Federrath}}]{Sharda:2021ac}
{Sharda}, P., {Wisnioski}, E., {Krumholz}, M.~R., \& {Federrath}, C.
  2021{\natexlab{b}}, \mnras, 506, 1295, \dodoi{10.1093/mnras/stab1836}

\bibitem[{{Simons} {et~al.}(2020){Simons}, {Peeples}, {Tumlinson}, {O'Shea},
  {Smith}, {Corlies}, {Lochhaas}, {Zheng}, {Augustin}, {Prasad}, {Snyder}, \&
  {Tollerud}}]{Simons:2020aa}
{Simons}, R.~C., {Peeples}, M.~S., {Tumlinson}, J., {et~al.} 2020, \apj, 905,
  167, \dodoi{10.3847/1538-4357/abc5b8}

\bibitem[{{Simons} {et~al.}(2021){Simons}, {Papovich}, {Momcheva}, {Trump},
  {Brammer}, {Estrada-Carpenter}, {Backhaus}, {Cleri}, {Finkelstein},
  {Giavalisco}, {Ji}, {Jung}, {Matharu}, \& {Weiner}}]{Simons:2021aa}
{Simons}, R.~C., {Papovich}, C., {Momcheva}, I., {et~al.} 2021, \apj, 923, 203,
  \dodoi{10.3847/1538-4357/ac28f4}

\bibitem[{{Simons} {et~al.}(2024){Simons}, {Peeples}, {Tumlinson}, {O'Shea},
  {Lochhaas}, {Wright}, {Acharyya}, {Augustin}, {Hamilton-Campos}, {Smith},
  {Lehner}, {Werk}, \& {Zheng}}]{Simons:2024aa}
{Simons}, R.~C., {Peeples}, M.~S., {Tumlinson}, J., {et~al.} 2024, arXiv
  e-prints, arXiv:2409.17244, \dodoi{10.48550/arXiv.2409.17244}

\bibitem[{{Suess} {et~al.}(2019{\natexlab{a}}){Suess}, {Kriek}, {Price}, \&
  {Barro}}]{Suess:2019aa}
{Suess}, K.~A., {Kriek}, M., {Price}, S.~H., \& {Barro}, G. 2019{\natexlab{a}},
  \apj, 877, 103, \dodoi{10.3847/1538-4357/ab1bda}

\bibitem[{{Suess} {et~al.}(2019{\natexlab{b}}){Suess}, {Kriek}, {Price}, \&
  {Barro}}]{Suess:2019ab}
---. 2019{\natexlab{b}}, \apjl, 885, L22, \dodoi{10.3847/2041-8213/ab4db3}

\bibitem[{{Swinbank} {et~al.}(2012){Swinbank}, {Smail}, {Sobral}, {Theuns},
  {Best}, \& {Geach}}]{Swinbank:2012aa}
{Swinbank}, A.~M., {Smail}, I., {Sobral}, D., {et~al.} 2012, \apj, 760, 130,
  \dodoi{10.1088/0004-637X/760/2/130}

\bibitem[{{Tacchella} {et~al.}(2016){Tacchella}, {Dekel}, {Carollo},
  {Ceverino}, {DeGraf}, {Lapiner}, {Mandelker}, \&
  {Primack}}]{Tacchella:2016ab}
{Tacchella}, S., {Dekel}, A., {Carollo}, C.~M., {et~al.} 2016, \mnras, 458,
  242, \dodoi{10.1093/mnras/stw303}

\bibitem[{{Tissera} {et~al.}(2016){Tissera}, {Pedrosa}, {Sillero}, \&
  {Vilchez}}]{Tissera:2016aa}
{Tissera}, P.~B., {Pedrosa}, S.~E., {Sillero}, E., \& {Vilchez}, J.~M. 2016,
  \mnras, 456, 2982, \dodoi{10.1093/mnras/stv2736}

\bibitem[{{Tissera} {et~al.}(2022){Tissera}, {Rosas-Guevara}, {Sillero},
  {Pedrosa}, {Theuns}, \& {Bignone}}]{Tissera:2022aa}
{Tissera}, P.~B., {Rosas-Guevara}, Y., {Sillero}, E., {et~al.} 2022, \mnras,
  511, 1667, \dodoi{10.1093/mnras/stab3644}

\bibitem[{{Turk} {et~al.}(2011{\natexlab{a}}){Turk}, {Smith}, {Oishi}, {Skory},
  {Skillman}, {Abel}, \& {Norman}}]{Turk:2011aa}
{Turk}, M.~J., {Smith}, B.~D., {Oishi}, J.~S., {et~al.} 2011{\natexlab{a}},
  \apjs, 192, 9, \dodoi{10.1088/0067-0049/192/1/9}

\bibitem[{{Turk} {et~al.}(2011{\natexlab{b}}){Turk}, {Smith}, {Oishi}, {Skory},
  {Skillman}, {Abel}, \& {Norman}}]{yt2011}
---. 2011{\natexlab{b}}, The Astrophysical Journal Supplement Series, 192, 9,
  \dodoi{10.1088/0067-0049/192/1/9}

\bibitem[{{Vallini} {et~al.}(2024){Vallini}, {Witstok}, {Sommovigo},
  {Pallottini}, {Ferrara}, {Carniani}, {Kohandel}, {Smit}, {Gallerani}, \&
  {Gruppioni}}]{Vallini:2024aa}
{Vallini}, L., {Witstok}, J., {Sommovigo}, L., {et~al.} 2024, \mnras, 527, 10,
  \dodoi{10.1093/mnras/stad3150}

\bibitem[{{van de Voort} {et~al.}(2019){van de Voort}, {Springel}, {Mandelker},
  {van den Bosch}, \& {Pakmor}}]{vandeVoort:2019aa}
{van de Voort}, F., {Springel}, V., {Mandelker}, N., {van den Bosch}, F.~C., \&
  {Pakmor}, R. 2019, \mnras, 482, L85, \dodoi{10.1093/mnrasl/sly190}

\bibitem[{{Venturi} {et~al.}(2024){Venturi}, {Carniani}, {Parlanti},
  {Kohandel}, {Curti}, {Pallottini}, {Vallini}, {Arribas}, {Bunker}, {Cameron},
  {Castellano}, {Ferrara}, {Fontana}, {Gallerani}, {Gelli}, {Maiolino},
  {Ntormousi}, {Pacifici}, {Pentericci}, {Salvadori}, \&
  {Vanzella}}]{Venturi:2024aa}
{Venturi}, G., {Carniani}, S., {Parlanti}, E., {et~al.} 2024, arXiv e-prints,
  arXiv:2403.03977.
\newblock \doarXiv{2403.03977}

\bibitem[{{Vila-Costas} \& {Edmunds}(1992)}]{Vila-Costas:1992aa}
{Vila-Costas}, M.~B., \& {Edmunds}, M.~G. 1992, \mnras, 259, 121,
  \dodoi{10.1093/mnras/259.1.121}

\bibitem[{{Virtanen} {et~al.}(2020){Virtanen}, {Gommers}, {Oliphant},
  {Haberland}, {Reddy}, {Cournapeau}, {Burovski}, {Peterson}, {Weckesser},
  {Bright}, {van der Walt}, {Brett}, {Wilson}, {Millman}, {Mayorov}, {Nelson},
  {Jones}, {Kern}, {Larson}, {Carey}, {Polat}, {Feng}, {Moore}, {VanderPlas},
  {Laxalde}, {Perktold}, {Cimrman}, {Henriksen}, {Quintero}, {Harris},
  {Archibald}, {Ribeiro}, {Pedregosa}, {van Mulbregt}, \& {SciPy 1. 0
  Contributors}}]{scipy2020}
{Virtanen}, P., {Gommers}, R., {Oliphant}, T.~E., {et~al.} 2020, Nature
  Methods, 17, 261, \dodoi{10.1038/s41592-019-0686-2}

\bibitem[{{Vogt} {et~al.}(2015){Vogt}, {Dopita}, {Borthakur},
  {Verdes-Montenegro}, {Heckman}, {Yun}, \& {Chambers}}]{Vogt:2015aa}
{Vogt}, F. P.~A., {Dopita}, M.~A., {Borthakur}, S., {et~al.} 2015, \mnras, 450,
  2593, \dodoi{10.1093/mnras/stv749}

\bibitem[{{Wadsley} {et~al.}(2008){Wadsley}, {Veeravalli}, \&
  {Couchman}}]{Wadsley:2008aa}
{Wadsley}, J.~W., {Veeravalli}, G., \& {Couchman}, H.~M.~P. 2008, \mnras, 387,
  427, \dodoi{10.1111/j.1365-2966.2008.13260.x}

\bibitem[{Walt {et~al.}(2011)Walt, Colbert, \& Varoquaux}]{numpy2011}
Walt, S. v.~d., Colbert, S.~C., \& Varoquaux, G. 2011, Computing in Science \&
  Engineering, 13, 22

\bibitem[{{Wang} {et~al.}(2017){Wang}, {Jones}, {Treu}, {Morishita},
  {Abramson}, {Brammer}, {Huang}, {Malkan}, {Schmidt}, {Fontana}, {Grillo},
  {Henry}, {Karman}, {Kelly}, {Mason}, {Mercurio}, {Rosati}, {Sharon},
  {Trenti}, \& {Vulcani}}]{Wang:2017aa}
{Wang}, X., {Jones}, T.~A., {Treu}, T., {et~al.} 2017, \apj, 837, 89,
  \dodoi{10.3847/1538-4357/aa603c}

\bibitem[{{Wang} {et~al.}(2019){Wang}, {Jones}, {Treu}, {Hirtenstein},
  {Brammer}, {Daddi}, {Meng}, {Morishita}, {Abramson}, {Henry}, {Peng},
  {Schmidt}, {Sharon}, {Trenti}, \& {Vulcani}}]{Wang:2019aa}
---. 2019, \apj, 882, 94, \dodoi{10.3847/1538-4357/ab3861}

\bibitem[{Wang {et~al.}(2020)Wang, Jones, Treu, Daddi, Brammer, Sharon,
  Morishita, Abramson, Colbert, Henry, Hopkins, Malkan, Schmidt, Teplitz, \&
  Vulcani}]{Wang:2020aa}
Wang, X., Jones, T.~A., Treu, T., {et~al.} 2020, The Astrophysical Journal,
  900, 183, \dodoi{10.3847/1538-4357/abacce}

\bibitem[{{Wang} {et~al.}(2022){Wang}, {Jones}, {Vulcani}, {Treu}, {Morishita},
  {Roberts-Borsani}, {Malkan}, {Henry}, {Brammer}, {Strait}, {Brada{\v{c}}},
  {Boyett}, {Calabr{\`o}}, {Castellano}, {Fontana}, {Glazebrook}, {Kelly},
  {Leethochawalit}, {Marchesini}, {Santini}, {Trenti}, \& {Yang}}]{Wang:2022aa}
{Wang}, X., {Jones}, T., {Vulcani}, B., {et~al.} 2022, \apjl, 938, L16,
  \dodoi{10.3847/2041-8213/ac959e}

\bibitem[{{Werk} {et~al.}(2011){Werk}, {Putman}, {Meurer}, \&
  {Santiago-Figueroa}}]{Werk:2011aa}
{Werk}, J.~K., {Putman}, M.~E., {Meurer}, G.~R., \& {Santiago-Figueroa}, N.
  2011, \apj, 735, 71, \dodoi{10.1088/0004-637X/735/2/71}

\bibitem[{{Wright} {et~al.}(2023){Wright}, {Tumlinson}, {Peeples}, {O'Shea},
  {Lochhaas}, {Corlies}, {Smith}, {Binh}, {Augustin}, \&
  {Simons}}]{Wright:2024aa}
{Wright}, A.~C., {Tumlinson}, J., {Peeples}, M.~S., {et~al.} 2023, arXiv
  e-prints, arXiv:2309.10039, \dodoi{10.48550/arXiv.2309.10039}

\bibitem[{{Wuyts} {et~al.}(2016){Wuyts}, {Wisnioski}, {Fossati}, {F{\"o}rster
  Schreiber}, {Genzel}, {Davies}, {Mendel}, {Naab}, {R{\"o}ttgers}, {Wilman},
  {Wuyts}, {Bandara}, {Beifiori}, {Belli}, {Bender}, {Brammer}, {Burkert},
  {Chan}, {Galametz}, {Kulkarni}, {Lang}, {Lutz}, {Momcheva}, {Nelson},
  {Rosario}, {Saglia}, {Seitz}, {Tacconi}, {Tadaki}, {{\"U}bler}, \& {van
  Dokkum}}]{Wuyts:2016aa}
{Wuyts}, E., {Wisnioski}, E., {Fossati}, M., {et~al.} 2016, \apj, 827, 74,
  \dodoi{10.3847/0004-637X/827/1/74}

\bibitem[{{Yuan} {et~al.}(2013{\natexlab{a}}){Yuan}, {Kewley}, \&
  {Rich}}]{Yuan:2013aa}
{Yuan}, T.-T., {Kewley}, L.~J., \& {Rich}, J. 2013{\natexlab{a}}, \apj, 767,
  106, \dodoi{10.1088/0004-637X/767/2/106}

\bibitem[{{Yuan} {et~al.}(2013{\natexlab{b}}){Yuan}, {Kewley}, \&
  {Richard}}]{Yuan:2013ab}
{Yuan}, T.-T., {Kewley}, L.~J., \& {Richard}, J. 2013{\natexlab{b}}, \apj, 763,
  9, \dodoi{10.1088/0004-637X/763/1/9}

\bibitem[{{Zheng} {et~al.}(2020){Zheng}, {Peeples}, {O'Shea}, {Simons},
  {Lochhaas}, {Corlies}, {Tumlinson}, {Smith}, \& {Augustin}}]{Zheng:2020aa}
{Zheng}, Y., {Peeples}, M.~S., {O'Shea}, B.~W., {et~al.} 2020, \apj, 896, 143,
  \dodoi{10.3847/1538-4357/ab960a}

\end{thebibliography}

\end{document}